\documentclass[11pt,a4paper,twoside,groupcitations]{article}
\usepackage[T1]{fontenc}
\usepackage[ansinew]{inputenc}
\usepackage[english]{babel}
\usepackage{amsfonts}
\usepackage{amsmath}
\usepackage{array}
\usepackage{amsthm}
\usepackage{amssymb}
\usepackage{graphicx}
\usepackage{subfigure}
\usepackage{braket}
\usepackage{eucal}
\usepackage{verbatim}
\usepackage[table]{xcolor}
\usepackage{caption}
\usepackage{cite}
\usepackage{textcomp}
\usepackage{multicol}
\usepackage{tikz}
\usetikzlibrary{positioning,arrows}
\usetikzlibrary{decorations.pathmorphing}
\usetikzlibrary{decorations.markings}
\usetikzlibrary{calc,decorations.markings}
\usetikzlibrary{arrows,shapes}
\usetikzlibrary{matrix,arrows}
\usepackage{pgfplots}
\usepackage{xparse}
\definecolor{jade}{HTML}{00A86B}
\newcommand{\be}{\begin{eqnarray}}
\newcommand{\ee}{\end{eqnarray}}

\renewcommand{\d}{\mbox{${\rm d}$}} 

\newcommand{\gn}{G_{\rm N}}
\newcommand{\rh}{r_{\rm H}}
\newcommand{\Rh}{R_{\rm H}}

\title{\bf Bootstrapped Newtonian stars and black holes}
\author{Roberto~Casadio$^{ab}$\thanks{E-mail: casadio@bo.infn.it},
$\ $
Michele~Lenzi$^{ab}$\thanks{E-mail: michele.lenzi@studio.unibo.it}
$\ $
and
Octavian Micu$^c$\thanks{E-mail: octavian.micu@spacescience.ro}
\\
\\
$^a${\em Dipartimento di Fisica e Astronomia, Universit\`a di Bologna}
\\
{\em via Irnerio~46, 40126 Bologna, Italy}
\\
\\
$^b${\em I.N.F.N., Sezione di Bologna, I.S.~FLAG}
\\
{\em viale B.~Pichat~6/2, 40127 Bologna, Italy}
\\
\\
{\em $^b$Institute of Space Science, Bucharest, Romania}
\\
{\em P.O. Box MG-23, RO-077125 Bucharest-Magurele, Romania}
}
\begin{document}
\maketitle
\begin{abstract}
We study equilibrium configurations of a homogenous ball of matter in a bootstrapped description
of gravity which includes a gravitational self-interaction term beyond the Newtonian coupling.
Both matter density and pressure are accounted for as sources of the gravitational potential
for test particles.
Unlike the general relativistic case, no Buchdahl limit is found and the pressure can 
in principle support a star of arbitrarily large compactness.
By defining the horizon as the location where the escape velocity of test particles equals the
speed of light, like in Newtonian gravity, we find a minimum value of the compactness
for which this occurs.
The solutions for the gravitational potential here found could effectively
describe the interior of macroscopic black holes in the quantum theory, as well as
predict consequent deviations from general relativity in the strong field regime of very
compact objects.
\par
\null
\par
\noindent
\textit{PACS - 04.70.Dy, 04.70.-s, 04.60.-m}
\end{abstract}
\section{Introduction and motivation}
\setcounter{equation}{0}
\label{Sintro}
The true nature of black holes is already problematic in the classical description given 
by general relativity and notoriously more so once one tries to incorporate the unavoidable
quantum physics.
Once a trapping surface appears, singularity theorems of general relativity require an object to collapse
all the way into a region of vanishing volume and infinite density~\cite{HE}.
At the same time, a point-like source is well known to be classically unacceptable~\cite{geroch}.
One would therefore hope that quantum physics cures this problem, the same way it makes
the hydrogen atom stable, by affecting the gravitational dynamics, at least in the strong
field regime (where the description of matter likely requires physics beyond the standard
model as well~\cite{brustein}).
\par
In light of the above observations, in Ref.~\cite{BootN} we studied an effective equation for the
gravitational potential of a static source which contains a gravitational self-interaction term besides
the usual Newtonian coupling with the matter density.
Following an idea from Ref.~\cite{Casadio:2016zpl}, this equation was derived in details
from a Fierz-Pauli Lagrangian in Ref.~\cite{Casadio:2017cdv}, and it can therefore be viewed
as stemming from the truncation of the relativistic theory at some ``post-Newtonian'' order
(for the standard post-Newtonian formalism, see Ref.~\cite{PPNlr}).
However, since the ``post-Newtonian'' correction $V_{\rm PN}\sim M^2/r^2$ is positive and grows
faster than the Newtonian potential $V_{\rm N}\sim M/r$ near the surface of the source, one is allowed
to consider only matter sources with radius $R\gg\Rh$ in this approximation (where we remark that $M$
is the total ADM mass~\cite{adm} of the system and
\be
\Rh\equiv 2\,\gn\,M
\label{def:Rh}
\ee
is the gravitational radius of the source.)
This consistency condition clearly excludes the possibility to study very compact matter sources and,
in particular, those with $R\simeq \Rh$ which are on the verge of forming a black hole.
For the ultimate purpose of including such cases and gain some hindsight about the fate of matter
which collapses inside a black hole, in Ref.~\cite{BootN} we studied the non-linear equation
of the effective theory derived in Ref.~\cite{Casadio:2017cdv} at face value, without requiring that the
corrections it introduces with respect to the Newtonian potential remain small.
\par
In Ref.~\cite{BootN}, we showed that the qualitative behaviour of the complete solutions to that non-linear
equation resembles rather closely the Newtonian counterpart.
This result, which essentially stems from including a gravitational self-interaction in the Poisson equation, 
is what we call ``bootstrapping'' the Newtonian gravity.
In fact, including those specific non-linear terms could be viewed as the first step in the perturbative
reconstruction of classical general relativity (see, {\em e.g.}~Refs.~\cite{carballo}).
However, we could also entertain the idea that these terms are meaningful to determine the (mean field)
gravitational potential of extremely compact objects if the {\em quantum\/} break-down of classical
general relativity occurs at {\em macroscopic\/} scales and general relativity therefore fails at describing
(the interior of) black holes~\cite{DvaliGomez}.
In this case, although its origin lies in the quantum nature of gravity (and matter), if
this effective potential applies for macroscopic sources, it does not need to contain
explicitly a dependence on the Planck constant $\hbar$ such that general relativistic configurations
are (formally) recovered for vanishing $\hbar$~\footnote{Terms explicitly proportional to
$\hbar$ are usually obtained as perturbative corrections to classical solutions and they can therefore
be trusted only as long as they remain smaller than the quantities they perturb (see, {\em e.g.}~Ref.~\cite{hbar}).}.
On the other hand, Newtonian physics is recovered (by construction) for sources with small compactness
$\gn\,M/R\sim \Rh/R\ll 1$, and one can therefore consider that the compactness $\gn\,M/R$ is the parameter
measuring deviations from general relativity in the bootstrapped potential~\footnote{A detailed study of orbits
in the region outside the source is underway.}.
To be more specific, we expect that the bootstrapped potential admits a description in terms of
a quantum state of bound gravitons, like the coherent state that can be used to reproduce the
Newtonian potential~\cite{Casadio:2016zpl,Casadio:2017cdv}.
The quantum features of the system would hence become apparent only after such a quantum
state is constructed explicitly (which we leave for future investigations).
Moreover, by studying static sources of uniform density $\rho$ in Ref.~\cite{BootN}, we found 
a finite matter pressure $p$ could support sources of arbitrarily large compactenss $\gn\,M/R\sim \Rh/R\gg 1$,
so that there is no equivalent of the Buchdahl limit~\cite{buchdahl} of general relativity in the bootstrapped
Newtonian gravity.
Of course, the pressure becomes the dominant source of energy when $\gn\,M/R\gg 1$ and,
although the strong energy condition $\rho+p>0$ still holds, the dominant energy condition $\rho\ge |p|$
is violated in this regime (see, {\em e.g.}~Ref.\cite{Wald}).
This  suggests that the source of highly compact configurations, such as black holes, must be matter in
a quantum state with no purely classical analogue (like Bose-Einstein condensates~\cite{DvaliGomez}
or degenerate neutron stars).
This result is again consistent with the fact that classical general relativistic configurations are expected
to become physically relevant only for astrophysical objects with small compactness $\Rh/R\ll 1$.
\par
Since we are mainly interested in investigating static sources which we found can be very compact,
a pressure term which prevents the gravitational collapse needs to be included from the onset. 
For this reason, we here modify the effective theory used in Ref.~\cite{BootN} in order to consistently
supplement the matter density with the pressure as sources of the gravitational potential.
We then study systems with generic compactness $\gn\,M/R\sim \Rh/R$, from the regime $R\gg\Rh$,
in which we recover the standard post-Newtonian picture, to $R\ll\Rh$ where we find the source is enclosed within
a horizon.
The latter is defined according to the Newtonian view as the location at which the escape velocity of
test particles equals the speed of light.
Of course, it should be possible to treat the single microscopic constituents of the source in this 
test particle approximation and the presence of an horizon therefore refers to their inability to escape
the gravitational pull.
\par
Like in Refs.~\cite{BootN,Casadio:2017cdv}, we shall just consider (static) spherically symmetric systems,
so that all quantities depend only on the radial coordinate $r$, and the matter density $\rho=\rho(r)$ will
also be assumed homogenous inside the source ($r\le R$) for the sake of simplicity.
The pressure will instead be determined consistently from the condition of staticity.
The paper is organised as follows:
in Section~\ref{S:action}, we briefly review the derivation of the equation for the potential with the inclusion
of a pressure term;
in Section~\ref{S:solution}, we solve for the outer and inner potential generated by the homogenous source
using appropriate analytical methods for the diverse regimes.
In particular, we study intermediate and large compact sources with $R\lesssim \Rh$ as
possible candidates for effectively describing collapsed objects which should act as black holes according
to general relativity;
their horizon structure is then analysed in Section~\ref{S:horizon};
we finally comment about our results and possible outlooks in Section~\ref{S:conc}.
\section{Bootstrapped theory for the gravitational potential}
\label{S:action}
\setcounter{equation}{0}
From Ref.~\cite{Casadio:2017cdv}, we recall that the non-linear equation for the potential $V=V(r)$
describing the gravitational pull on test particles generated by a matter density $\rho=\rho(r)$ can be
obtained starting from the Newtonian Lagrangian
\be
L_{\rm N}[V]
=
-4\,\pi
\int_0^\infty
r^2 \,\d r
\left[
\frac{\left(V'\right)^2}{8\,\pi\,\gn}
+\rho\,V
\right]
\ ,
\label{LagrNewt}
\ee
where $f'\equiv\d f/\d r$, and the corresponding equation of motion is the Poisson equation
\be
r^{-2}\left(r^2\,V'\right)'
\equiv
\triangle V
=
4\,\pi\,\gn\,\rho
\label{EOMn}
\ee
for the Newtonian potential $V=V_{\rm N}$.
We can then include the effects of gravitational self-interaction by noting that the Hamiltonian
\be
H_{\rm N}[V]
=
-L_{\rm N}[V]
=
4\,\pi
\int_0^\infty
r^2\,\d r
\left(
-\frac{V\,\triangle V}{8\,\pi\,\gn}
+\rho\,V
\right)
\ ,
\label{NewtHam}
\ee 
computed on-shell by means of Eq.~\eqref{EOMn}, yields the total Newtonian potential energy
\be
U_{\rm N}[V]
&\!\!=\!\!&
2\,\pi
\int_0^\infty
{r}^2\,\d {r}
\,\rho\, V
\nonumber
\\
&\!\!=\!\!&
\frac{1}{2\,\gn}\,
\int_0^\infty
{r}^2 \,\d {r}\,
V\,\triangle V
\nonumber
\\
&\!\!=\!\!&
-4\,\pi
\int_0^\infty
{r}^2 \,\d {r}\,
\frac{\left(V'\right)^2}{{8\,\pi\,\gn}}
\ ,
\label{Unn}
\ee
where we used Eq.~\eqref{EOMn} in the second line and assumed that boundary terms vanish at
$r=0$ and $r=\infty$ as usual in the last line (for an alternative but equivalent derivation, see Appendix~\ref{A:J_V}).
One can therefore view the above $U_{\rm N}$ as given by the interaction of the matter distribution 
with the gravitational field or, following Ref.~\cite{Casadio:2016zpl} (see also Ref.~\cite{dadhich1}), 
as the volume integral of the gravitational current proportional to the gravitational energy $U_{\rm N}$
per unit volume $\delta\mathcal{V}=4\,\pi\,r^2\,\delta r$, that is~\footnote{The factor of $4$ in the
expression~\eqref{JV} of $J_V$ is chosen in order to recover the expected first post-Newtonian correction in the
vacuum potential for the coupling constant $q_V=1$ (see Section~\ref{S:vacuum}).} 
\be
J_V
\simeq
4\,\frac{\delta U_{\rm N}}{\delta \mathcal{V}} 
=
-\frac{\left[V'(r)\right]^2}{2\,\pi\,\gn}
\ .
\label{JV}
\ee
\par
As mentioned in the Introduction, in Ref.~\cite{BootN} we found that the pressure $p$ which prevents 
the system from collapsing becomes very large for compact sources
with a size $R\lesssim\Rh$, where $\Rh$ is the gravitational radius of Eq.~\eqref{def:Rh}.
We must therefore add a corresponding potential energy $U_{\rm B}$, associated with the work done
by the force responsible for the pressure $p$, such that
\be
p
\simeq
-\frac{\delta U_{\rm B}}{\delta \mathcal{V}}
= 
J_{\rm B}
\ .
\label{JP}
\ee
We will accordingly have to couple the potential field with the energy densities $J_V$ and $J_{\rm B}$
and then add the analogous higher order term $J_\rho=-2\,V^2$
which couples with the matter sector, {\em i.e.}~with the total matter energy density $\rho + p$.
Upon including these new source terms, we obtain the total Lagrangian~\cite{Casadio:2017cdv}  
\be
L[V]
&\!\!=\!\!&
L_{\rm N}[V]
-4\,\pi
\int_0^\infty
r^2\,\d r
\left[
q_V\,J_V\,V
+
q_{\rm B}\,J_{\rm B}\,V
+
q_\rho\,J_\rho\left(\rho+p\right)
\right]
\nonumber
\\
&\!\!=\!\!&
-4\,\pi
\int_0^\infty
r^2\,\d r
\left[
\frac{\left(V'\right)^2}{8\,\pi\,\gn}
\left(1-4\,q_V\,V\right)
+V\left(\rho+q_{\rm B}\,p\right)
-2\,q_{\rho} V^2\,\left(\rho+p\right)
\right]
\ .
\label{LagrV}
\ee
The parameters $q_V$, $q_{\rm B}$ and $q_{\rho}$ play the role of coupling
constants~\footnote{Different values of $q_V$, $q_{\rm B}$ and $q_{\rho}$
can be implemented in order to obtain the approximate potentials for different motions of test particles
in general relativity and describe different interiors.}
for the three different currents $J_V$, $J_{\rm B}$ and $J_\rho$ respectively.
They also allow us to control the origin of non-linearities, as we recover the Newtonian Lagrangian~\eqref{LagrNewt}
by setting all of them equal to zero.
\par
The associated effective Hamiltonian is simply given by
\be
\label{HamV}
H[V]
=
-L[V]
\ ,
\ee
and the Euler-Lagrange equation for $V$ reads
\be
\left(1 - 4\,q_V\,V \right) \triangle V 
=
4\,\pi\,\gn\left(\rho+q_{\rm B}\,p\right)
-
16\,\pi\,\gn\,q_{\rho}\,V\,(\rho + p)
+
2\,q_V\left(V'\right)^2
\ .
\label{EOMVcoup}
\ee
The latter must be supplemented with the conservation equation that determines the pressure,
\be
p'
=
-V'\left(\rho+p\right)
\ ,
\label{eqP}
\ee
which can be seen as a correction to the usual Newtonian formula that accounts for the contribution 
of the pressure to the energy density, or as an approximation for the Tolman-Oppenheimer-Volkoff equation
of general relativity.
\par
Although we showed the three parameters $q_\Phi$, $q_{\rm B}$ and $q_{\rho}$ explicitly, 
we shall only consider $q_\Phi=q_{\rm B}=q_{\rho}=1$ in the following for the sake of simplicity.
In this case, Eq.~\eqref{EOMVcoup} reduces to
\be
\triangle V
=
4\,\pi\,\gn\left(\rho+p\right)
+
\frac{2\,\left(V'\right)^2}
{1-4\,V}
\ ,
\label{EOMV}
\ee
from which we see that the differences with respect to the Poisson Eq.~\eqref{EOMn} are given 
by the inclusion of the pressure $p$ and the derivative self-interaction term in the right hand side.
\section{Homogeneous ball in vacuum}
\label{S:solution}
\setcounter{equation}{0}
Since we are interested in compact sources, we will consider the simplest case in which the
matter density is homogeneous and vanishes outside the sphere of radius $r=R$, that is
\be
\rho
=
\rho_0
\equiv
\frac{3\, M_0}{4\,\pi\, R^3}\, 
\Theta(R-r)
\ ,
\label{HomDens}
\ee
where $\Theta$ is the Heaviside step function, and
\be
M_0
=
4\,\pi
\int_0^R
r^2\,\d r\,\rho(r)
\ .
\ee
Of course, the uniform density~\eqref{HomDens} is not expected to be compatible with an equation of state,
since the pressure $p=p(r)$ must depend on the radial position so as to maintain equilibrium~\cite{BootN}. 
We also remark that uniform density is not very realistic and is here used just for mathematical
convenience and because it is the source for the exact interior Schwarzschild solution~\cite{schwIn}
in general relativity~\footnote{More realistic energy densities with physically motivated equations of state
will be considered in future developments.}.
\par
The potential must also satisfy the regularity condition in the centre
\be
V_{\rm in}'(0)=0
\label{b0}
\ee
and be smooth across the surface $r=R$, that is
\be
V_{\rm in}(R)
&\!\!=\!\!&
V_{\rm out}(R)
\equiv
V_R
\label{bR}
\\
\notag
\\
V'_{\rm in}(R)
&\!\!=\!\!&
V'_{\rm out}(R)
\equiv 
V'_R
\ ,
\label{dbR}
\ee
where we defined $V_{\rm in}=V(0\le r\le R)$ and $V_{\rm out}=V(R\le r)$.
\subsection{Outer vacuum solution}
\label{S:vacuum}
In the vacuum, where $\rho=p=0$, Eq.~\eqref{eqP} is trivially satisfied and
Eq.~\eqref{EOMV} with $q_\Phi=1$ reads
\be
\triangle V
=
\frac{2 \left(V'\right)^2}{1-4\,V}
\ ,
\label{EOMV0}
\ee
which is exactly solved by
\be
V_{\rm out}
=
\frac{1}{4}
\left[
1-\left(1+\frac{6\,\gn\,M}{r}\right)^{2/3}
\right]
\ .
\label{sol0}
\ee
where two integration constants were fixed by requiring the expected
Newtonian behaviour in terms of the ADM-like mass $M$ for large $r$. 
In fact, the large $r$ expansion now reads
\be
V_{\rm out}
\underset{r\to\infty}{\simeq}
-\frac{\gn\,M}{r}
+\frac{\gn^2\,M^2}{r^2}
-\frac{8\,\gn^3\,M^3}{3\,r^3}
\ ,
\label{Vlarge}
\ee
and contains the expected post-Newtonian term $V_{\rm PN}$ of order $\gn^2$ without any further
assumptions~\cite{Casadio:2017cdv}.
\par
From Eq.~\eqref{sol0}, we also obtain 
\be
V_R
=
V_{\rm out}(R)
=
\frac{1}{4}
\left[
1-\left(1+\frac{6\,\gn\,M}{R}\right)^{2/3}
\right]
\ ,
\label{VR} 
\ee
and
\be
V'_R
=
V_{\rm out}'(R)
=
\frac{\gn\,M}
{R^{2}\left(1+{6\,\gn\,M}/R\right)^{1/3}}
\ ,
\label{dVR}
\ee
which we will often use since they appear in the boundary conditions~\eqref{bR} and \eqref{dbR}.
\begin{figure}[t]
\centering
\includegraphics[width=10cm]{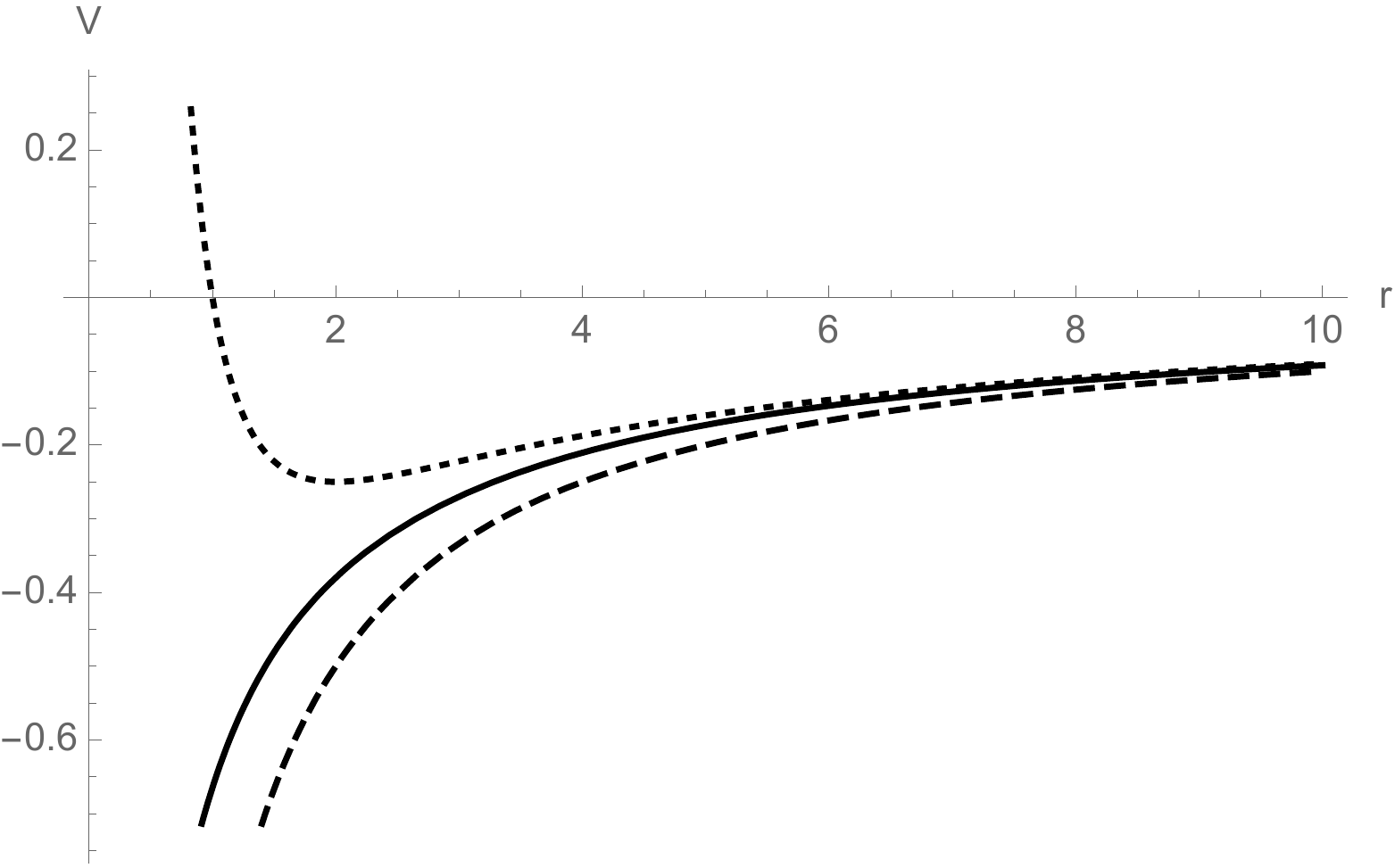}
\caption{Potential $V_{\rm out}$ (solid line) {\em vs\/} Newtonian potential (dashed line)
{\em vs\/} order $\gn^2$ expansion of $V_{\rm out}$ (dotted line) for $r>0$
(all quantities are in units of $\gn\,M$).
}
\label{V0}
\end{figure}
\subsection{The inner pressure}
We first consider the conservation Eq.~\eqref{eqP} and notice that, for $0\le r\le R$, we can write it as
\be
\frac{\left(\rho_0+p\right)'}{\rho_0+p}
=
-V'
\ ,
\ee
which allows us to express the total effective energy density as
\be
\rho_0+p
=
\alpha\,e^{-V}
\ .
\ee
The integration constant can be determined by imposing the usual
boundary condition
\be
p(R)=0
\ ,
\ee
which finally yields
\be \label{pressure}
p
=
\rho_0
\left[
e^{V_R-V}
-1
\right]
\ ,
\ee
where $V_R$ is given in Eq.~\eqref{VR}.
\subsection{The inner potential}
\label{SS:innerV}
The field equation~\eqref{EOMV} for $0\le r\le R$ and $q_\Phi=1$ becomes
\be
\triangle V
&\!\!=\!\!&
4\,\pi\,\gn\,
\rho_0\,e^{V_R-V}
+
\frac{2\left(V'\right)^2}
{1-4\,V}
\nonumber
\\
&\!\!=\!\!&
\frac{3\,\gn\,M_0}{R^3}\,e^{V_R-V}
+
\frac{2\left(V'\right)^2}
{1-4\,V}
\ ,
\label{eomV}
\ee
and we notice that $\rho_0\,e^{V_R}<\rho_0$ since $V_R<0$. 
The relevant solutions $V_{\rm in}$ to Eq.~\eqref{eomV} must also satisfy the regularity condition~\eqref{b0}
and the matching conditions~\eqref{bR} and~\eqref{dbR}, with $V_R$ and $V'_R$ respectively given
in Eq.~\eqref{VR} and \eqref{dVR}.
Since Eq.~\eqref{eomV} is a second order (ordinary) differential equation, 
the three boundary conditions~\eqref{b0}, \eqref{bR} and~\eqref{dbR} will not only fix the potential $V_{\rm in}$
uniquely, but also the ratio of the proper mass parameter $\gn\,M_0/R$ for any given value of the compactness
$\gn\,M/R$.
\par
It is hard to find the complete solution of the above problem for general compactness.
An approximate analytic solution to Eq.~\eqref{eomV} can be found quite straightforwardly only in the regimes
of low and intermediate compactness ({\em i.e.}~for $\gn\,M/R \ll 1$ and $\gn\,M/R \simeq 1$).
On the other hand, for $\gn\,M\gg R$, the non-linearity of Eq.~\eqref{eomV} and the interplay between $M_0$
and the boundary conditions~\eqref{b0}, \eqref{bR} and~\eqref{dbR} make it very difficult to find any
(approximate or numerical) solutions.
In fact, even a slight error in the estimate of $M_0=M_0(M,R)$ can spoil the solution completely.
For this reason, we will take advantage of the comparison method~\cite{BVPexistence, BVPexistence1,ode}
which essentially consists in finding two bounding functions $V_\pm$ (upper and lower approximate solutions)
such that $E_+(r)<0$ and $E_-(r)>0$ for $0\le r\le R$, where
\be
E_\pm
\equiv
\triangle V_\pm
-
\frac{3\,\gn\,M_0^{\pm}(M)}{R^3}\,e^{V_R-V_\pm}
-\frac{2\left(V_\pm'\right)^2}{1-4\,V_\pm}
\ .
\ee
Comparison theorems then guarantee that the proper solution will lie in between the two bounding functions
(see Appendix~\ref{A:comparison} for more details~\footnote{We just remark here that the comparison
theorems do not require that the approximate solutions $V_\pm$ have the same functional forms of
the exact solution $V_{\rm in}$.}), that is
\be
\label{vminvmax}
V_-<V_{\rm in}<V_+
\ .
\ee
The advantage of this method is twofold. 
It will serve as a tool for finding approximate solutions in the regime of large compactness
and will also allow us to check the accuracy of the approximate analytic solution for low
and intermediate compactness.
\subsubsection{Small and intermediate compactness}
\label{SS:lowC}
For the radius $R$ of the source much larger or of the order of $\gn\,M$, an analytic approximation
$V_{\rm s}$ for the solution $V_{\rm in}$ can be found by simply expanding around $r=0$, and turns out to be
\be
V_{\rm s}
=
V_0
+
\frac{\gn\,M_0}{2\,R^3}
\,e^{V_R-V_0}\,r^2
\ .
\label{Vs0}
\ee
where $V_0\equiv V_{\rm in}(0)<0$ and $V_R$ is given in Eq.~\eqref{VR}. 
We remark that the regularity condition~\eqref{b0} requires that all terms of odd order in $r$
in the Taylor expansion about $r=0$  must vanish.
\par
We can immediately notice that the above form is qualitatively similar to the
Newtonian solution recalled in Appendix~\ref{A:newton}.
Like the latter, the present case does not show any singularity in the potential for $r=0$
and the pressure, 
\be
p
\simeq
\rho_0
\left[
e^{V_R-V_0-B\,r^2}
-1
\right]
\ ,
\ee
is also regular in $r=0$,
\be
p(0)
=
\rho_0
\left[
e^{-(V_0-V_R)}
-1
\right]
>0
\ ,
\ee
since $V_0<V_R<0$.
\par
The two matching conditions at $r=R$ can now be written as
\be \label{match}
\left\{
\begin{array}{l}
2\,R\,(V_R-V_0)
\simeq
{\gn\,M_0}\,e^{V_R-V_0}
\\
\\
R^2\,V_R'
\simeq
{\gn\,M_0}\,e^{V_R-V_0}
\ ,
\end{array}
\right.
\label{syslow}
\ee
One can solve the second equation of the system above for $V_0$ to obtain
\be
V_0
=
\frac{1}{4}\left[1- \left( 1+6\,\gn\,M/R\right)^{2/3}\right]
+
\ln\left[\frac{M_0}{M}\left(1+6\,\gn\,M/R\right)^{1/3}\right]\ ,
\label{V0si}
\ee 
which is written in terms of $M_0$ and $M$.
Using the first equation in~\eqref{syslow}, one then finds
\be
M_0
=
\frac{M\,e^{-\frac{\gn\,M}{2\,R\left(1+6\,\gn\,M/R\right)^{1/3}}}}
{\left(1+6\,\gn\,M/R\right)^{1/3}}
\ .
\label{M0}
\ee
This last expression, along with the one for $V_0$, can be used to write the approximate 
solution~\eqref{Vs0} in terms of $M$ only as
\be
\label{Vins}
V_{\rm s}
=
\frac{R^3\left[\left(1+6\,\gn\,M/R\right)^{1/3}-1\right]
+2\,\gn\,M\left(r^2-4\,R^2\right)}
{4\,R^3\left(1+6\,\gn\,M/R\right)^{1/3}}
\ ,
\label{Vintermediate}
\ee
where we remark that this expression contains only the terms of the first two orders in the series
expansion about $r=0$.
\begin{figure}[t]
\centering
\includegraphics[width=5cm,height=4cm]{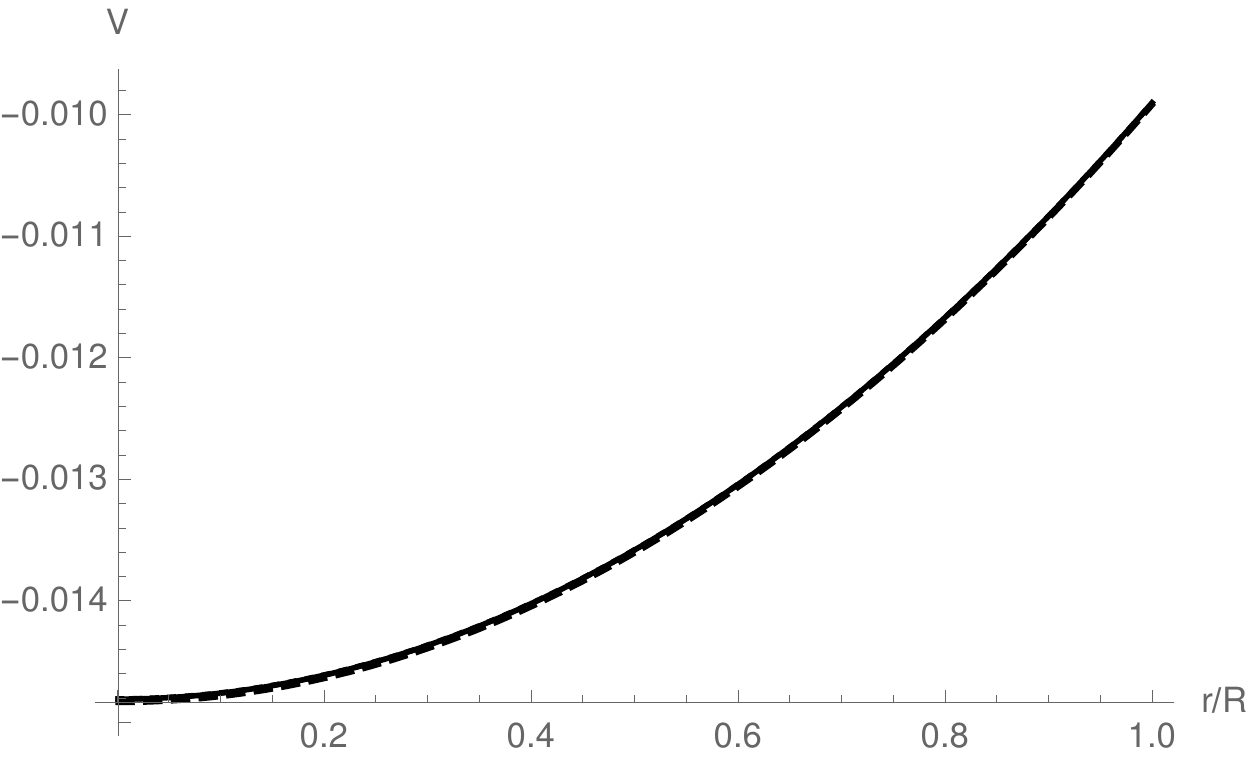}
$\ $
\includegraphics[width=5cm,height=4cm]{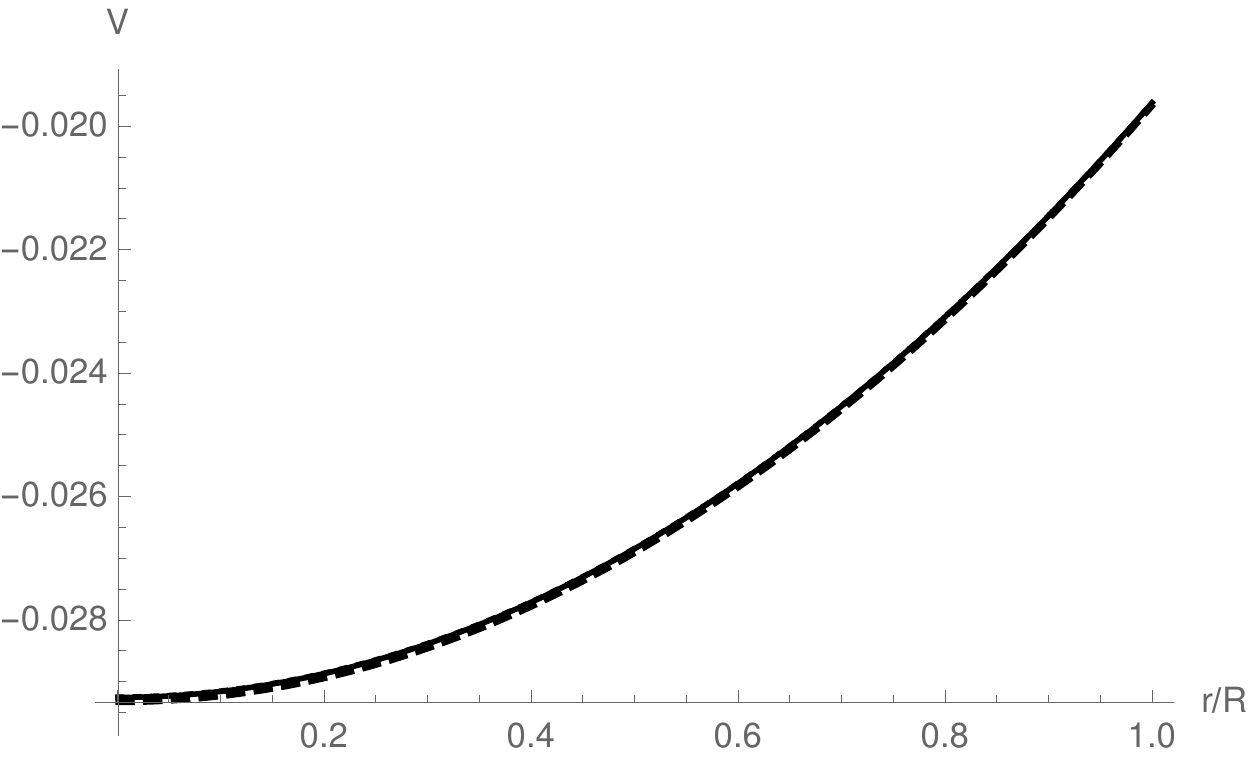}
$\ $
\includegraphics[width=5cm,height=4cm]{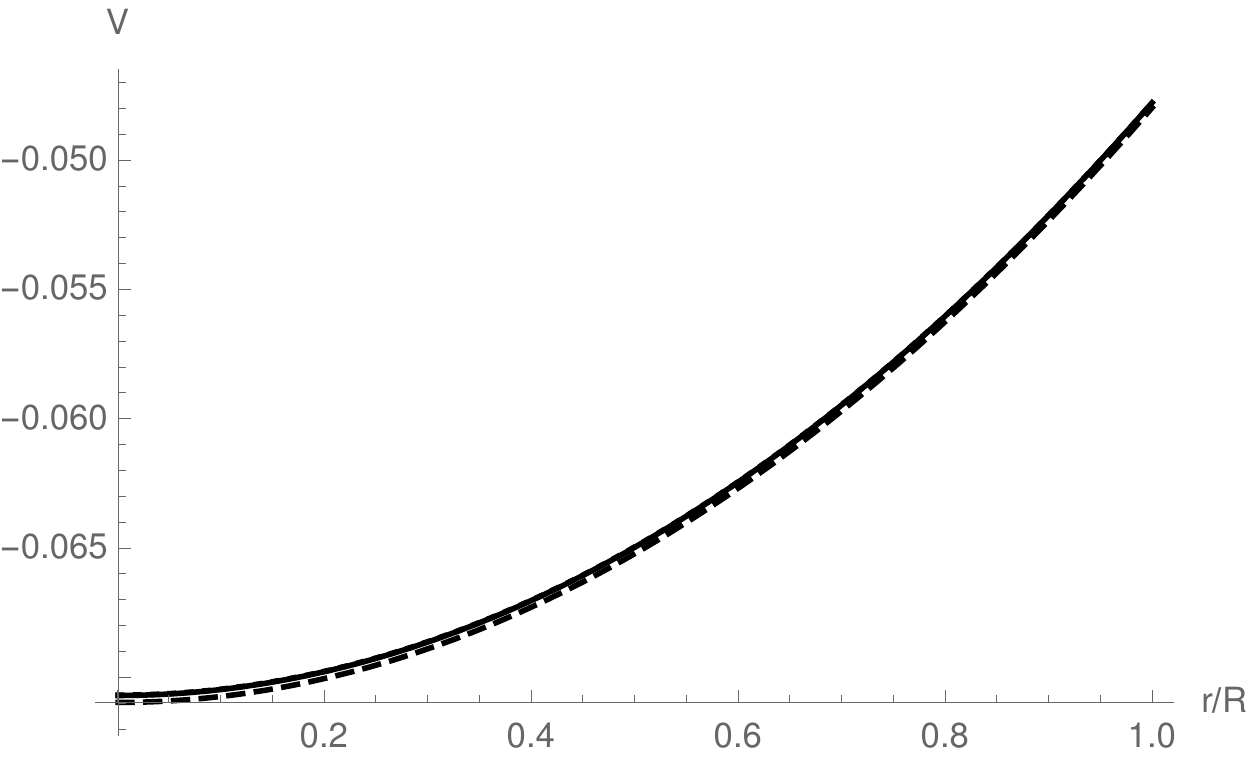}
\\
\includegraphics[width=5cm,height=4cm]{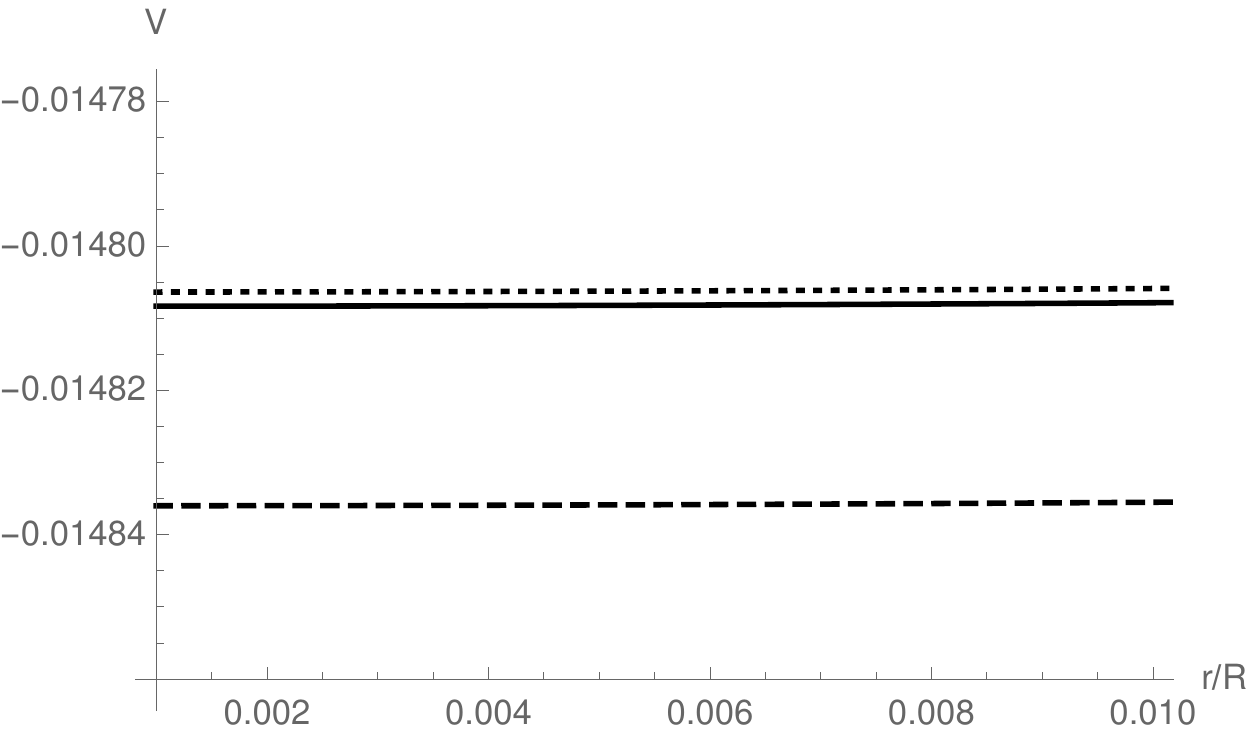}
$\ $
\includegraphics[width=5cm,height=4cm]{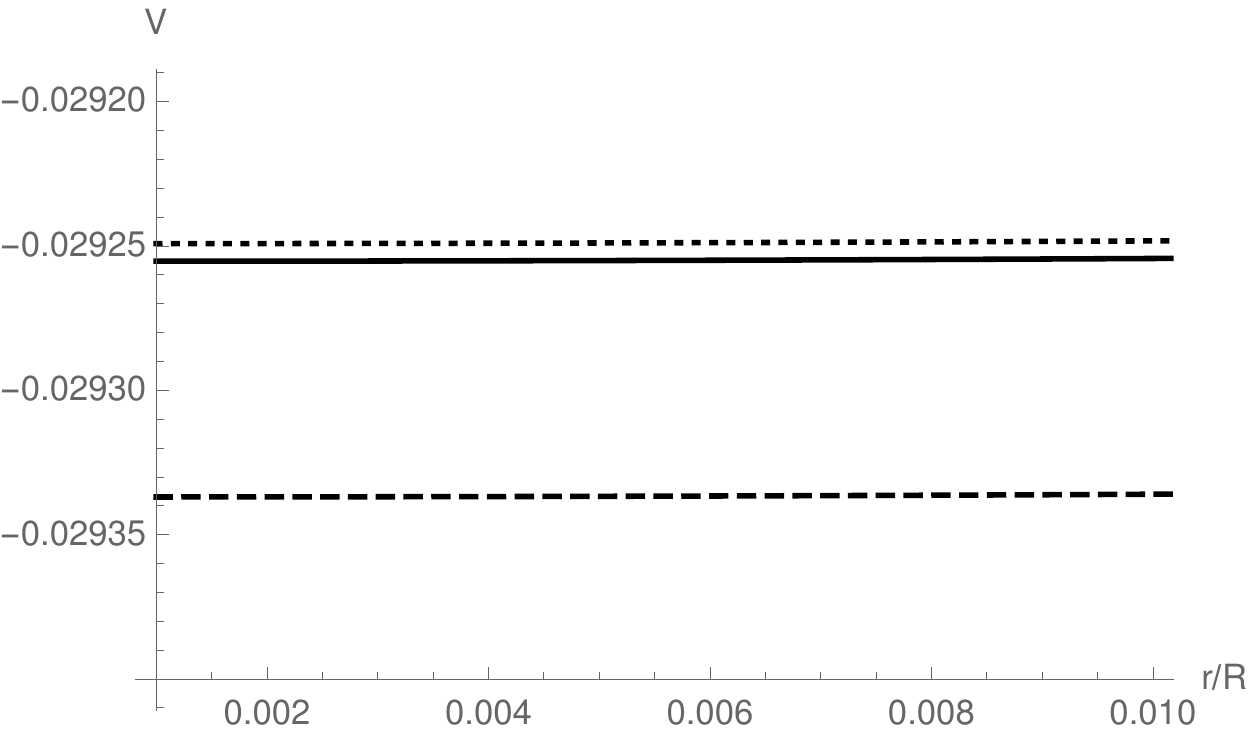}
$\ $
\includegraphics[width=5cm,height=4cm]{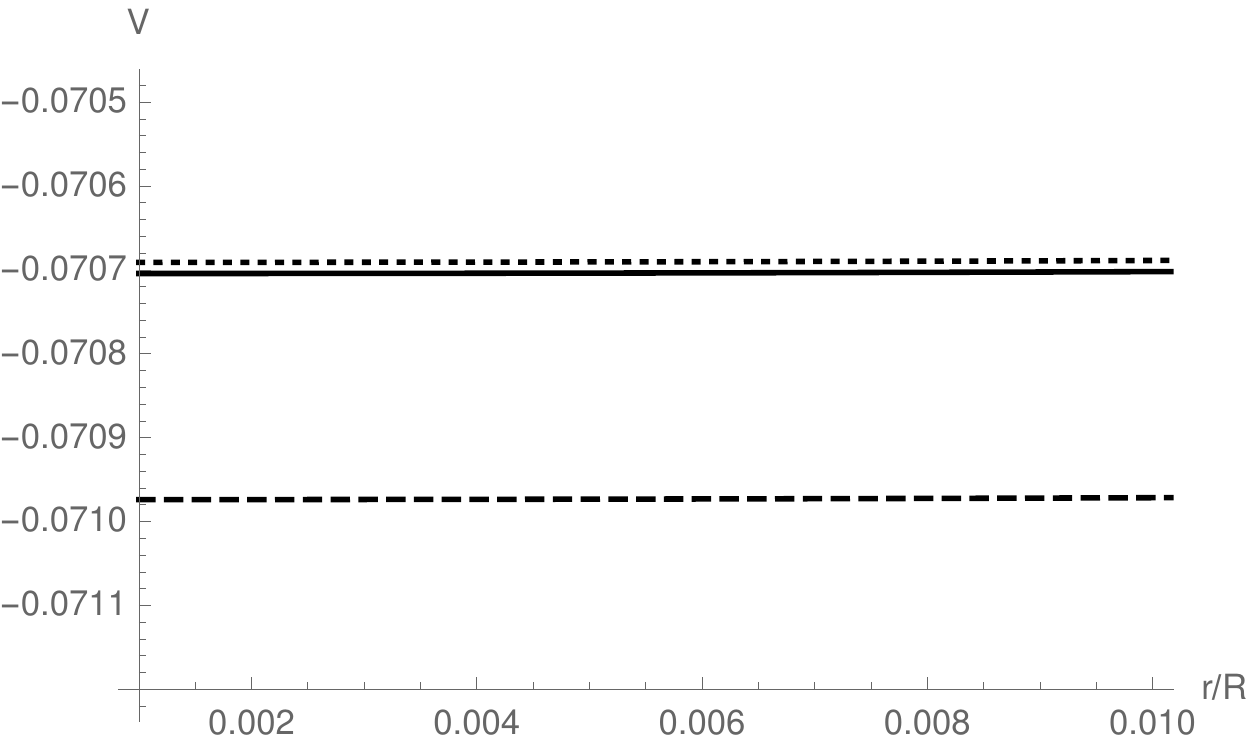}
\caption{Numerical solution to Eq.~\eqref{eomV} (solid line) {\em vs\/} approximate solution $V_{\rm s}=V_+$
in Eq.~\eqref{Vins} (dotted line) {\em vs\/} lower bounding function $V_-=C\, V_{\rm s}$ (dashed line),
for $\gn\,M /R = 1/100$ (top left panel, with $C=1.002$),
$\gn\,M / R= 1/50$ (top central panel, with $C=1.003$)
and $\gn\,M / R= 1/20$ (top right panel, with $C=1.004$).
The bottom panels show the region $0 \le r \le R/100$ where the difference between the three potentials is
the largest.
\label{VinVSVzoomSmall}}
\end{figure}
\begin{figure}[t]
\centering
\includegraphics[width=5cm,height=4cm]{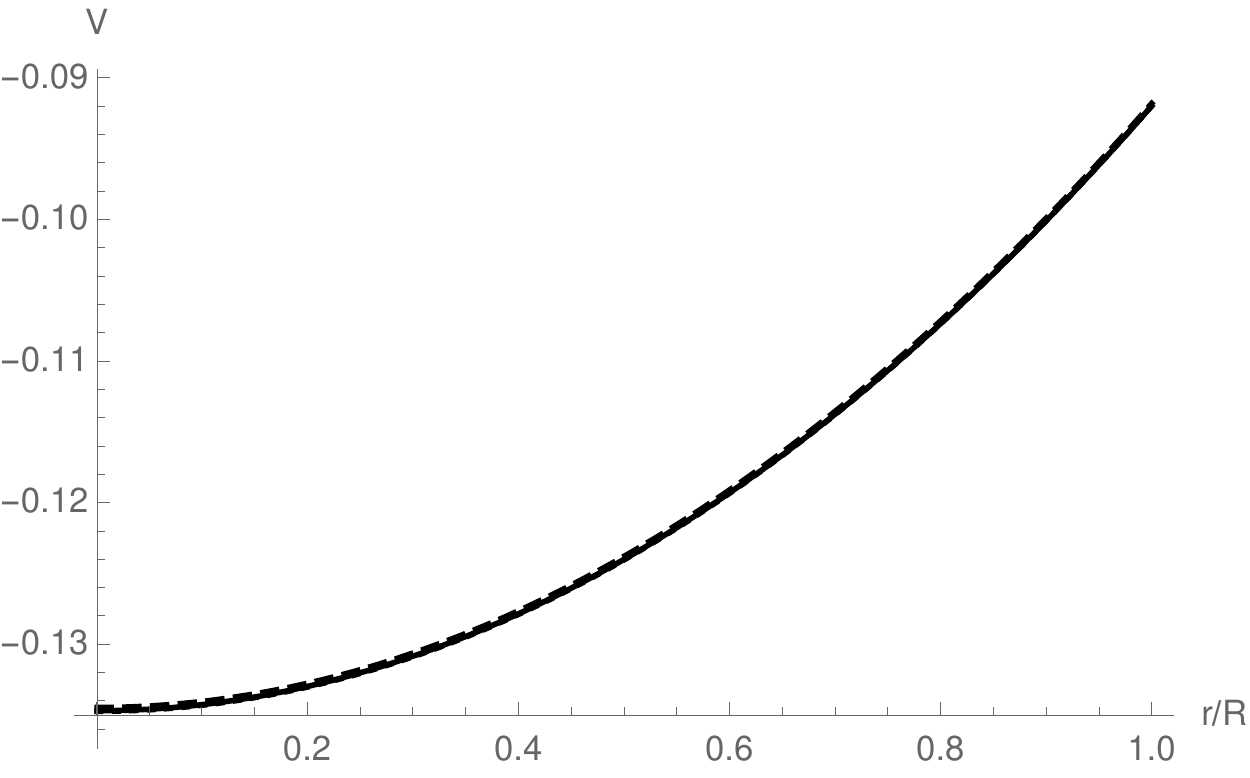}
$\ $
\includegraphics[width=5cm,height=4cm]{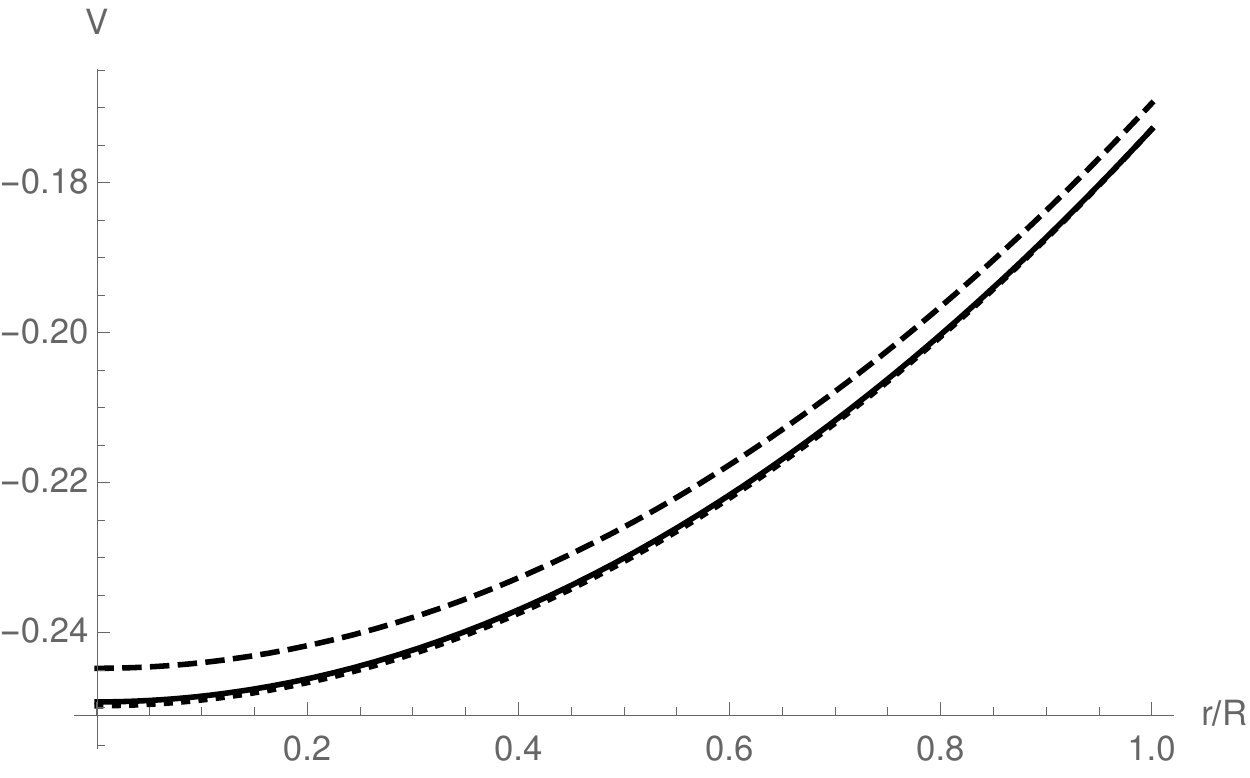}
$\ $
\includegraphics[width=5cm,height=4cm]{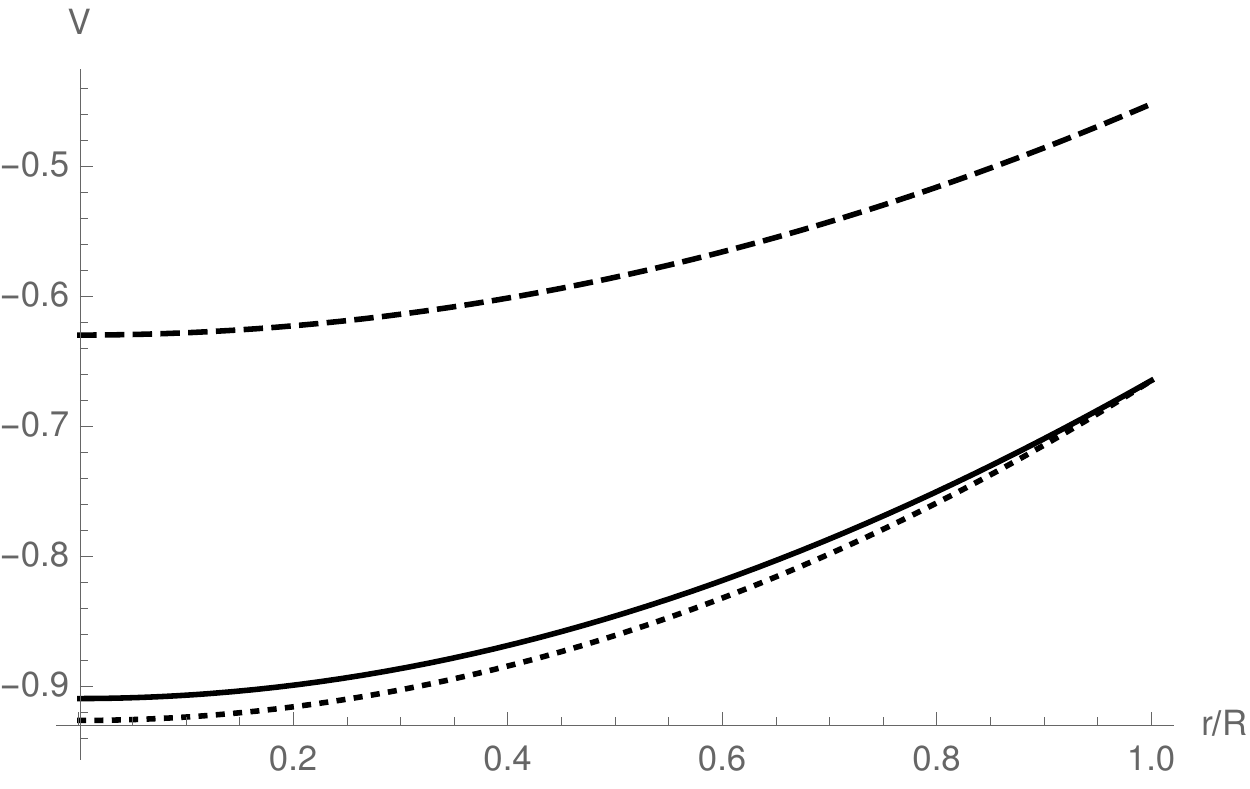}
\\
\includegraphics[width=5cm,height=4cm]{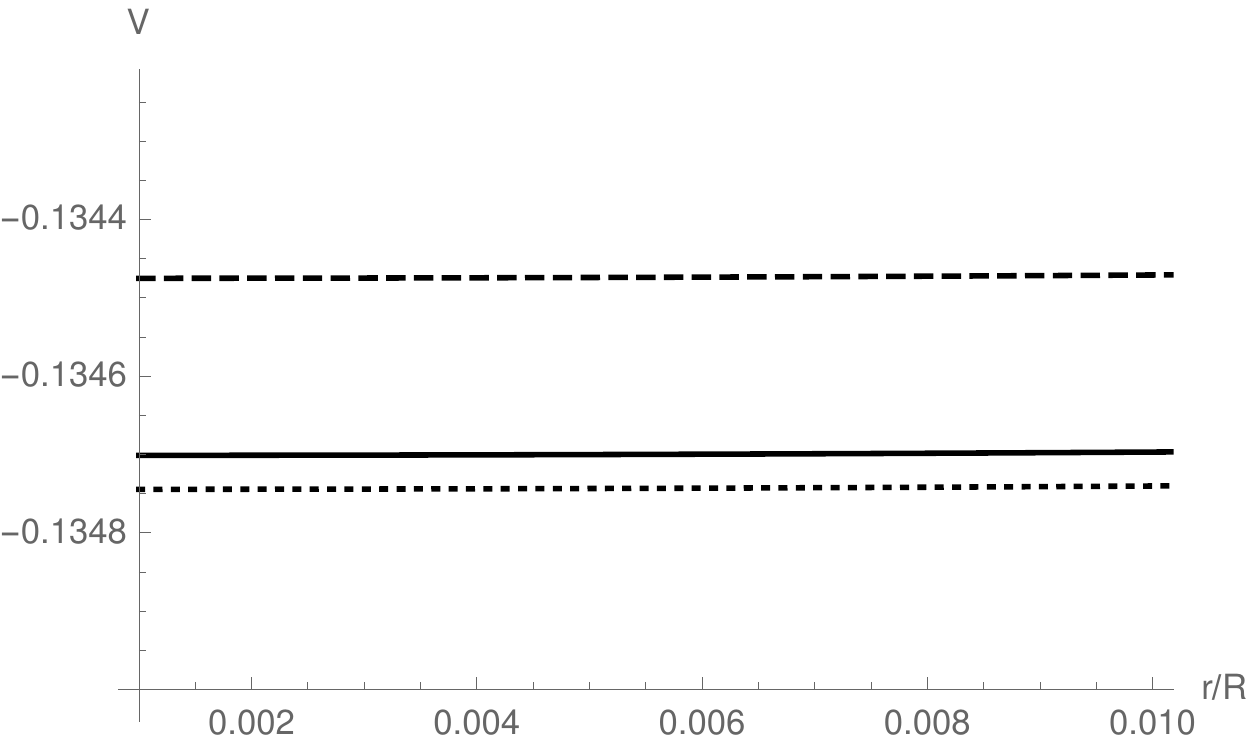}
$\ $
\includegraphics[width=5cm,height=4cm]{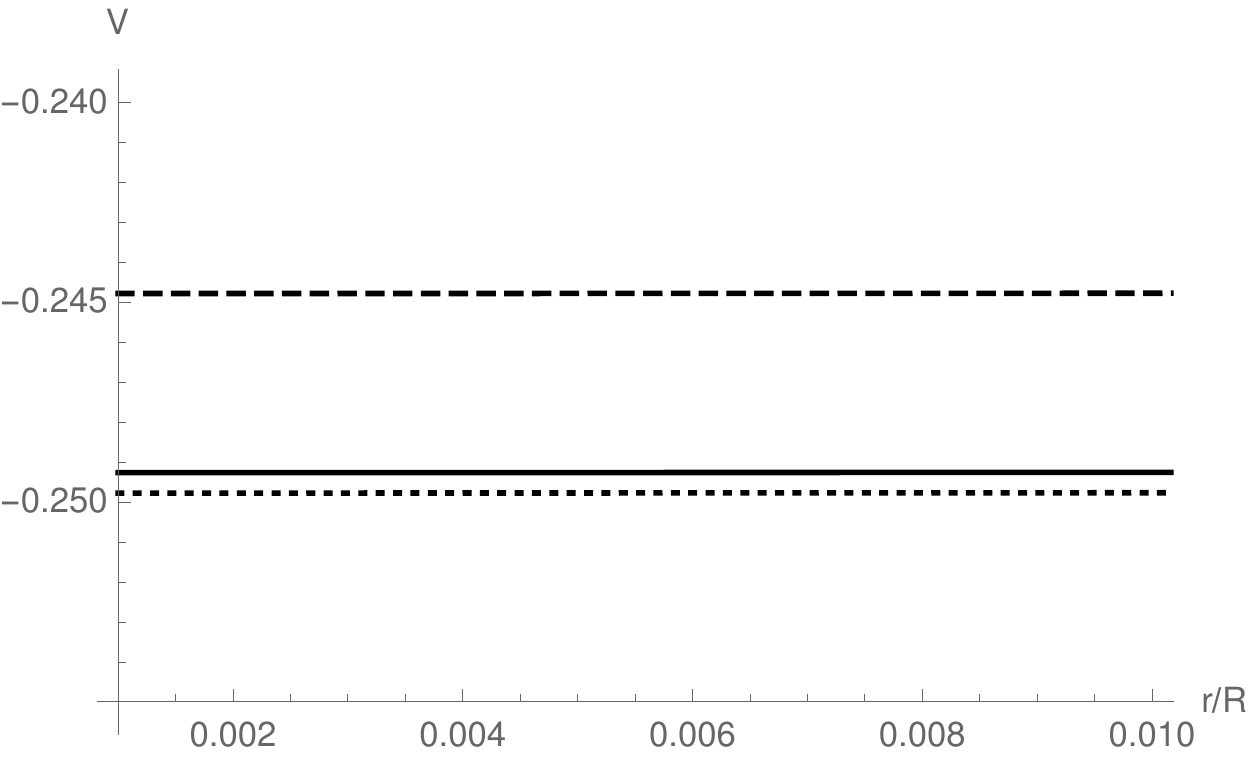}
$\ $
\includegraphics[width=5cm,height=4cm]{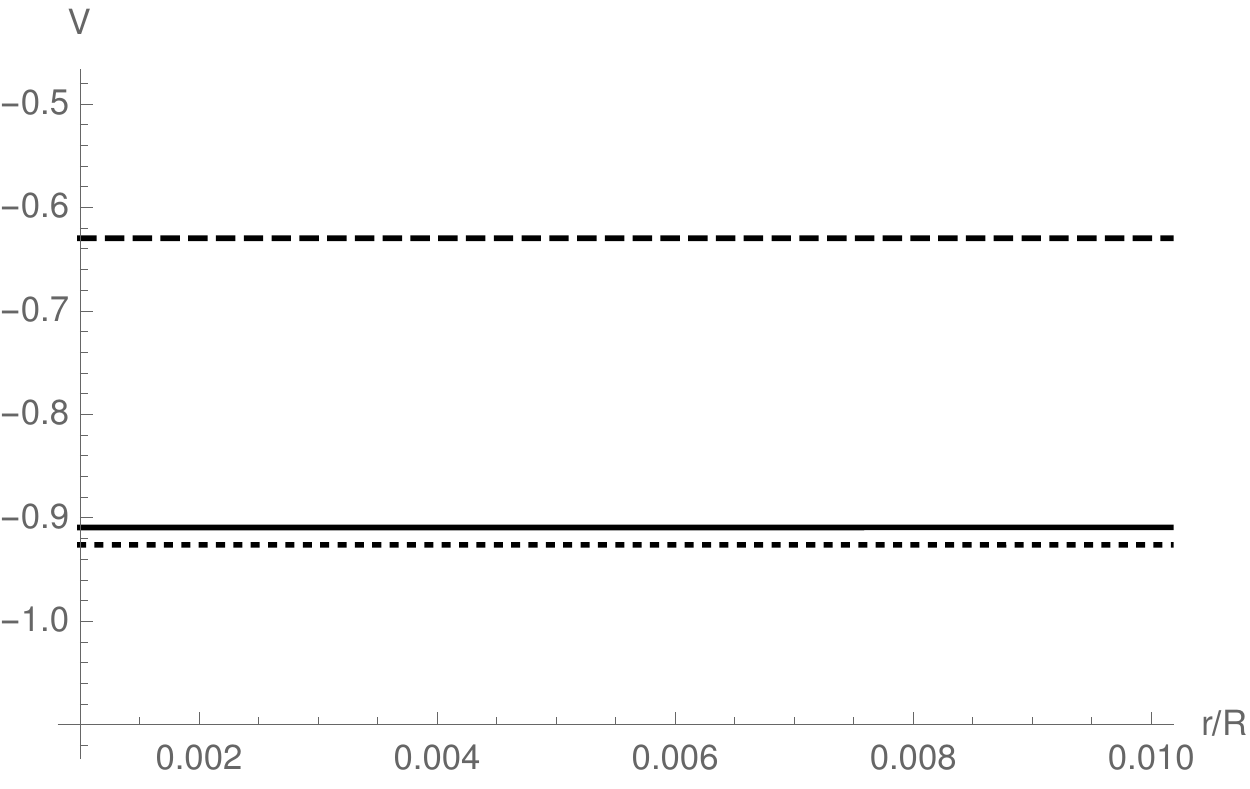}
\caption{Numerical solution to Eq.~\eqref{eomV} (solid line) {\em vs\/} approximate solution $V_{\rm s}=V_-$
in Eq.~\eqref{Vins} (dotted line) {\em vs\/} upper bounding function $V_+=C\, V_{\rm s}$ (dashed line),
for $\gn\,M /R = 1/10$ (top left panel, with $C=0.998$),
$\gn\,M / R= 1/5$ (top central panel, with $C=0.980$)
and $\gn\,M / R= 1$ (top right panel, with $C=0.680$).
The bottom panels show the region $0 < r < R/100$ where the difference between the three expressions
is the largest. 
}
\label{VinVSVzoomInter}
\end{figure}
\par
\begin{figure}[t]
\centering
\includegraphics[width=5cm,height=4cm]{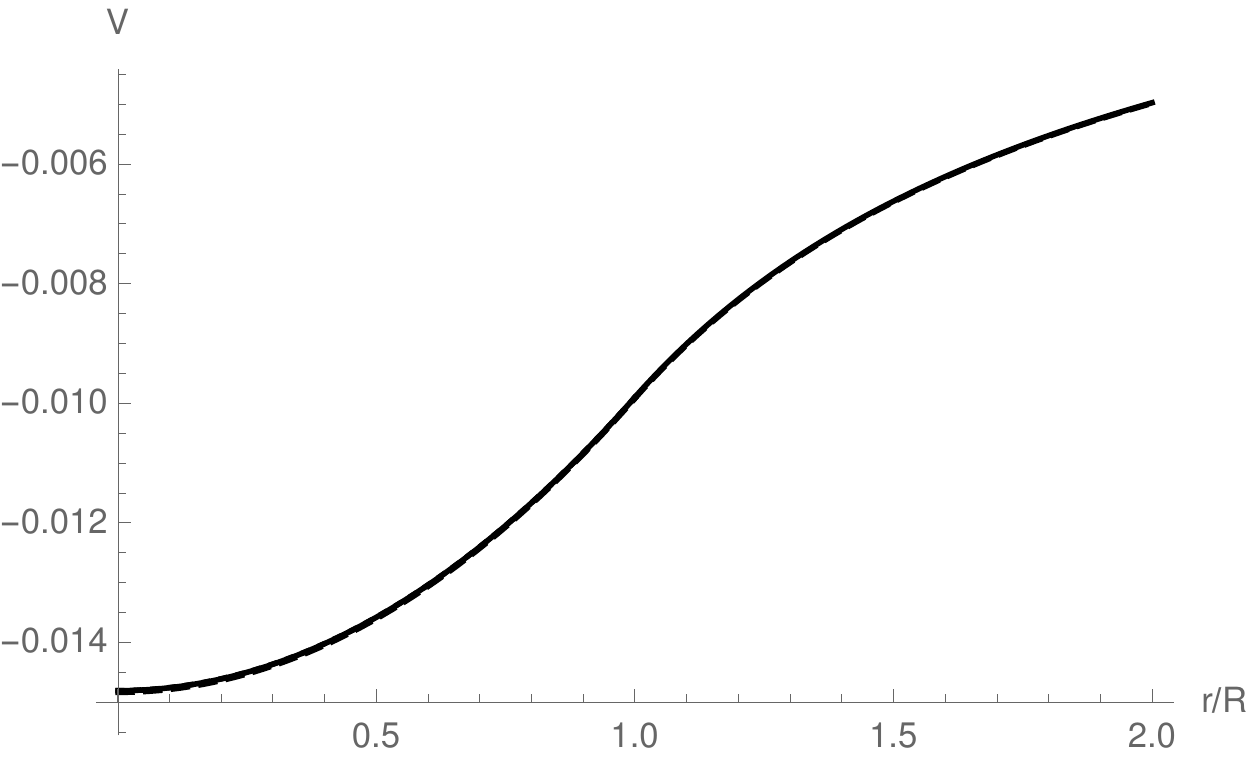}
$\ $
\includegraphics[width=5cm,height=4cm]{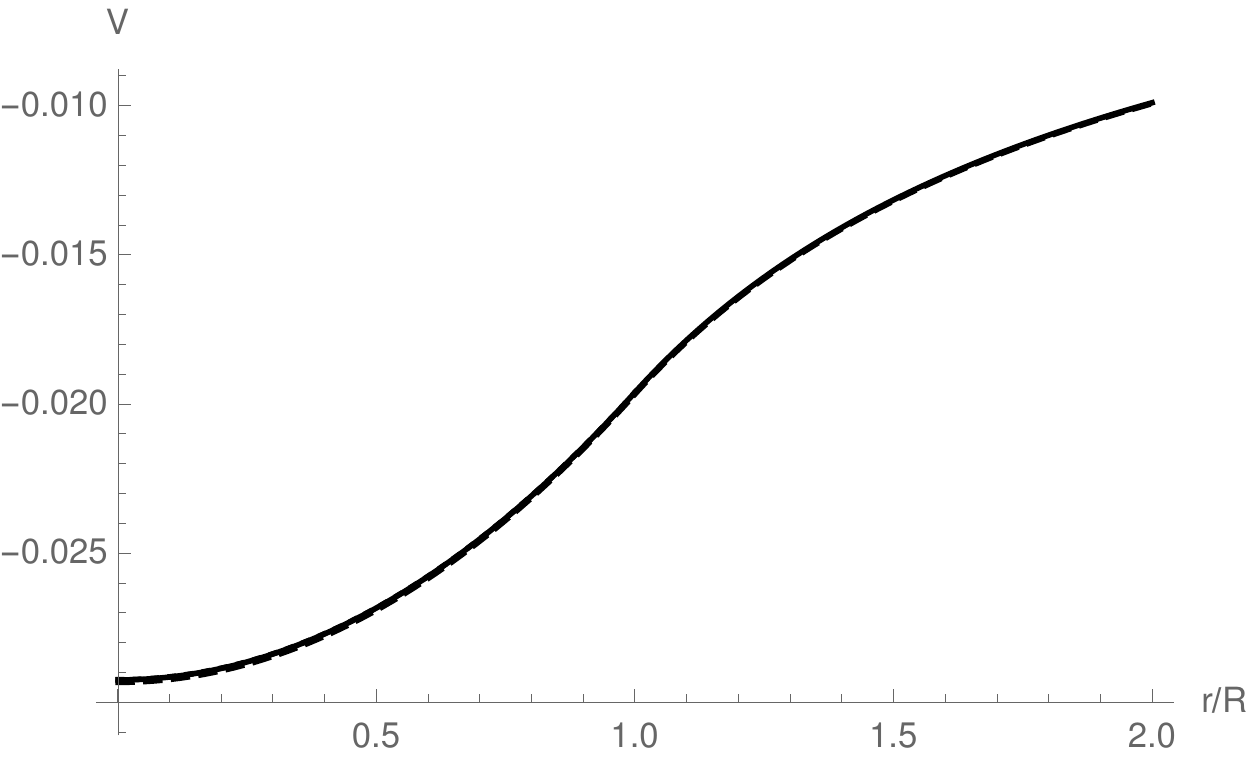}
$\ $
\includegraphics[width=5cm,height=4cm]{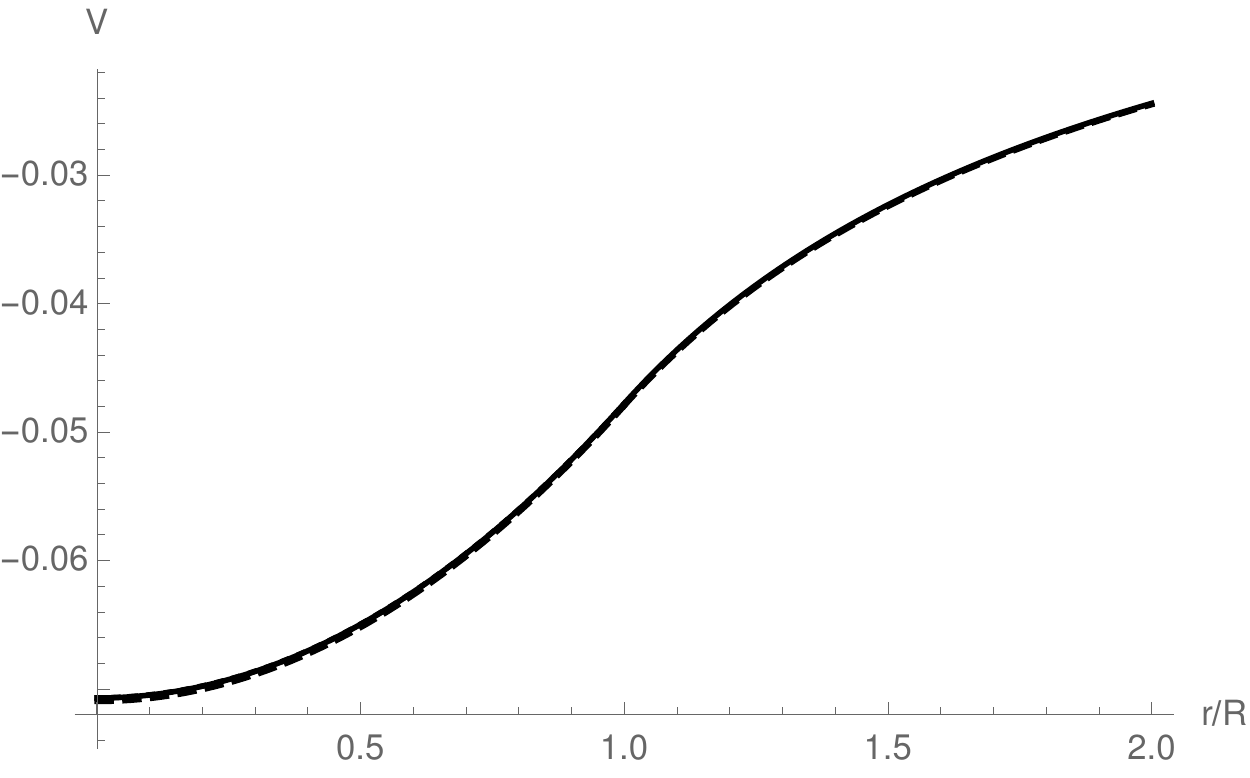}
\\
\includegraphics[width=5cm,height=4cm]{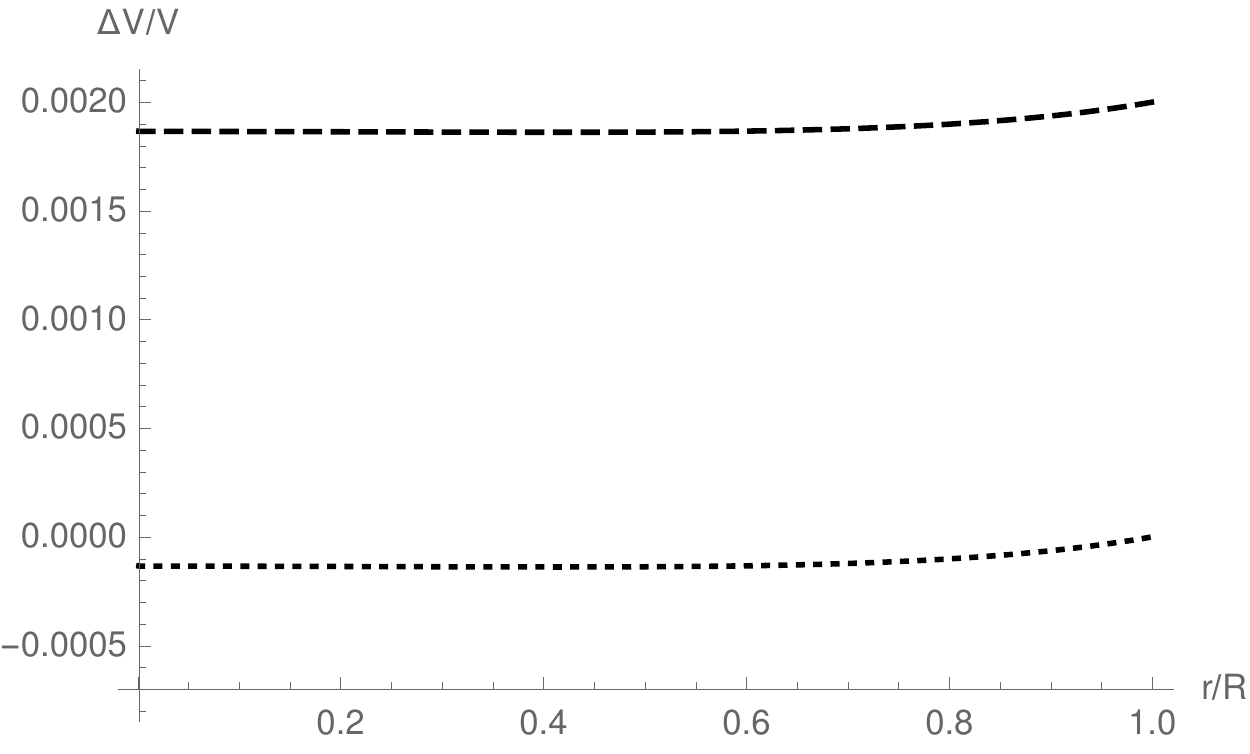}
$\ $
\includegraphics[width=5cm,height=4cm]{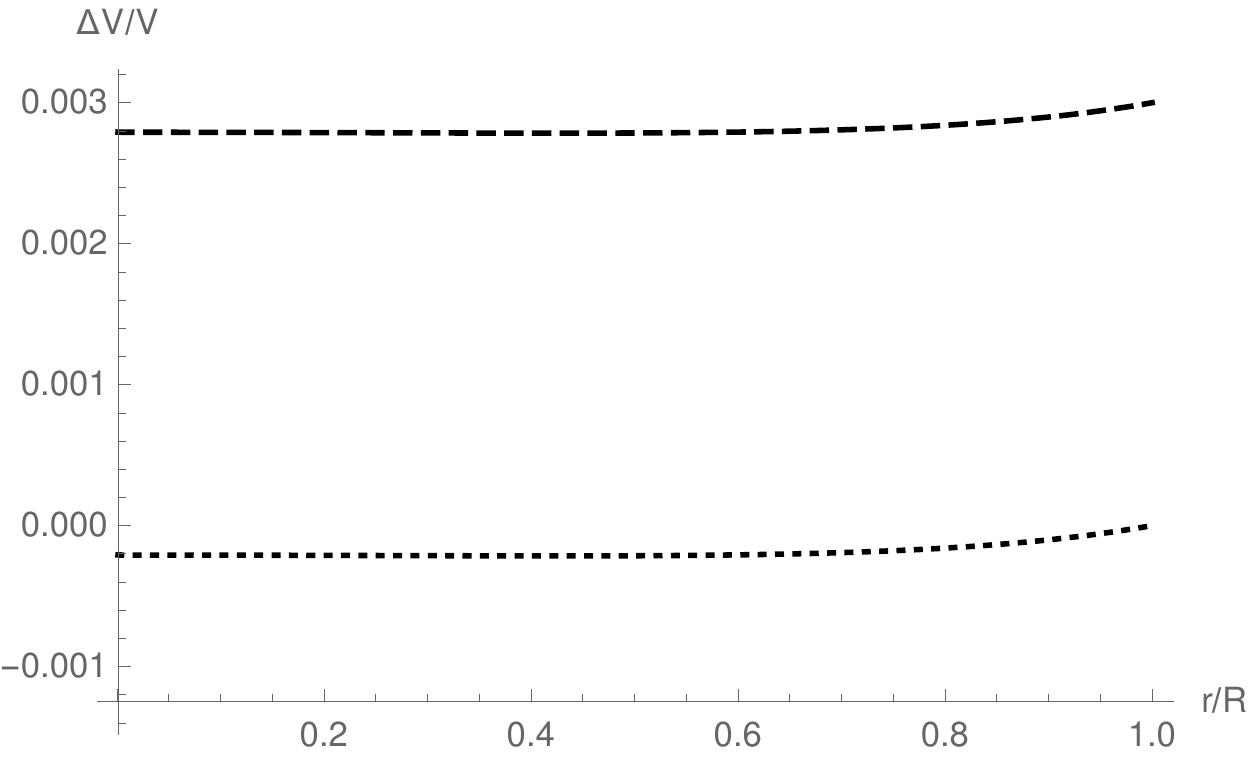}
$\ $
\includegraphics[width=5cm,height=4cm]{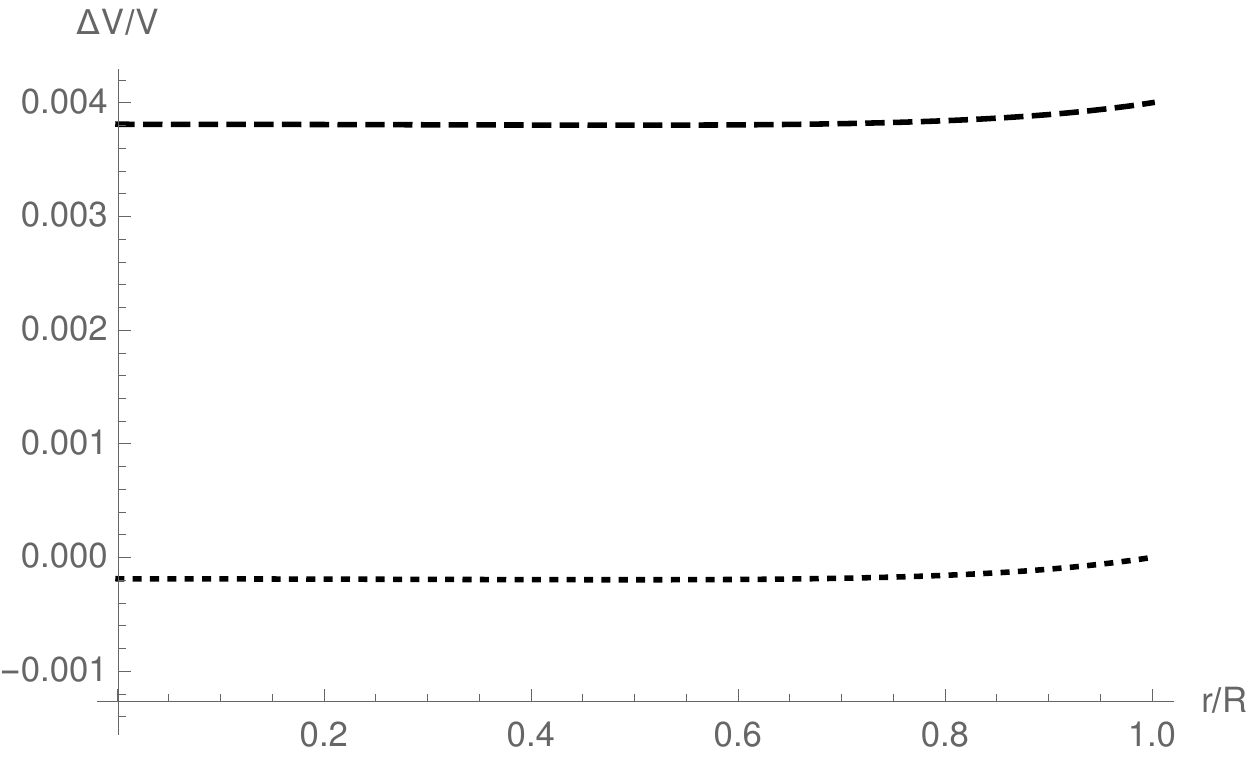}
\caption{
Upper panels:
numerical solution $V_{\rm n}$ to Eq.~\eqref{eomV} matched to the exact outer solution~\eqref{sol0}
(solid line) {\em vs\/} approximate solution $V_{\rm s}=V_+$ in Eq.~\eqref{Vins} (dotted line)
{\em vs\/} lower bounding function $V_-$ (dashed line) for $\gn\,M /R = 1/100$ (top left),
$\gn\,M / R= 1/50$ (top middle) and $\gn\,M / R= 1/20$ (top right).
Bottom panels: 
relative difference $(V_{\rm s}-V_{\rm n})/V_{\rm n}$ (dotted line) {\em vs\/}  
$(V_--V_{\rm n})/V_{\rm n}$ (dashed line) in the interior region
for $\gn\,M /R = 1/100$ (bottom left),
$\gn\,M / R= 1/50$ (bottom middle) and
$\gn\,M / R= 1/20$ (bottom right).
The negative sign of $(V_{\rm s}-V_{\rm n})/V_{\rm n}$ shows that the approximate solution is an
upper bounding function $V_{\rm s}=V_+$ in this range of compactness.
}
\label{Vtot}
\end{figure}
\begin{figure}[t]
\centering
\includegraphics[width=5cm,height=4cm]{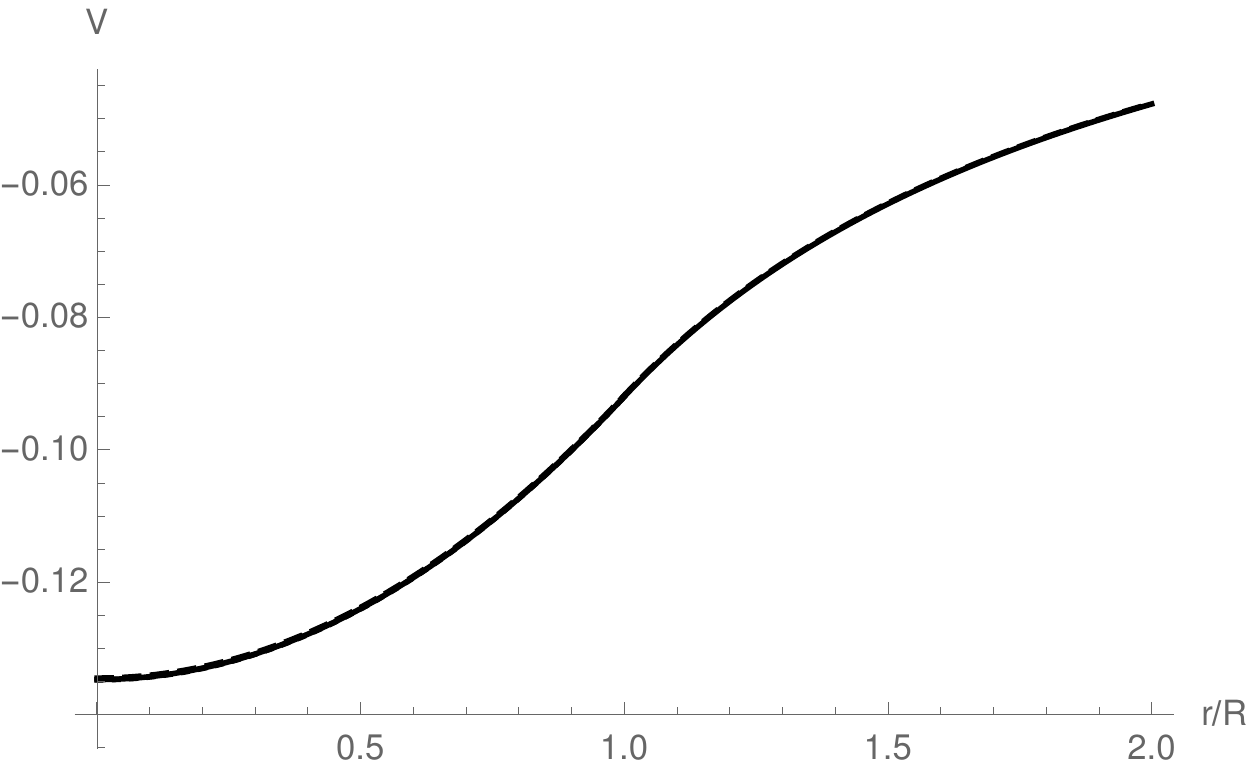}
$\ $
\includegraphics[width=5cm,height=4cm]{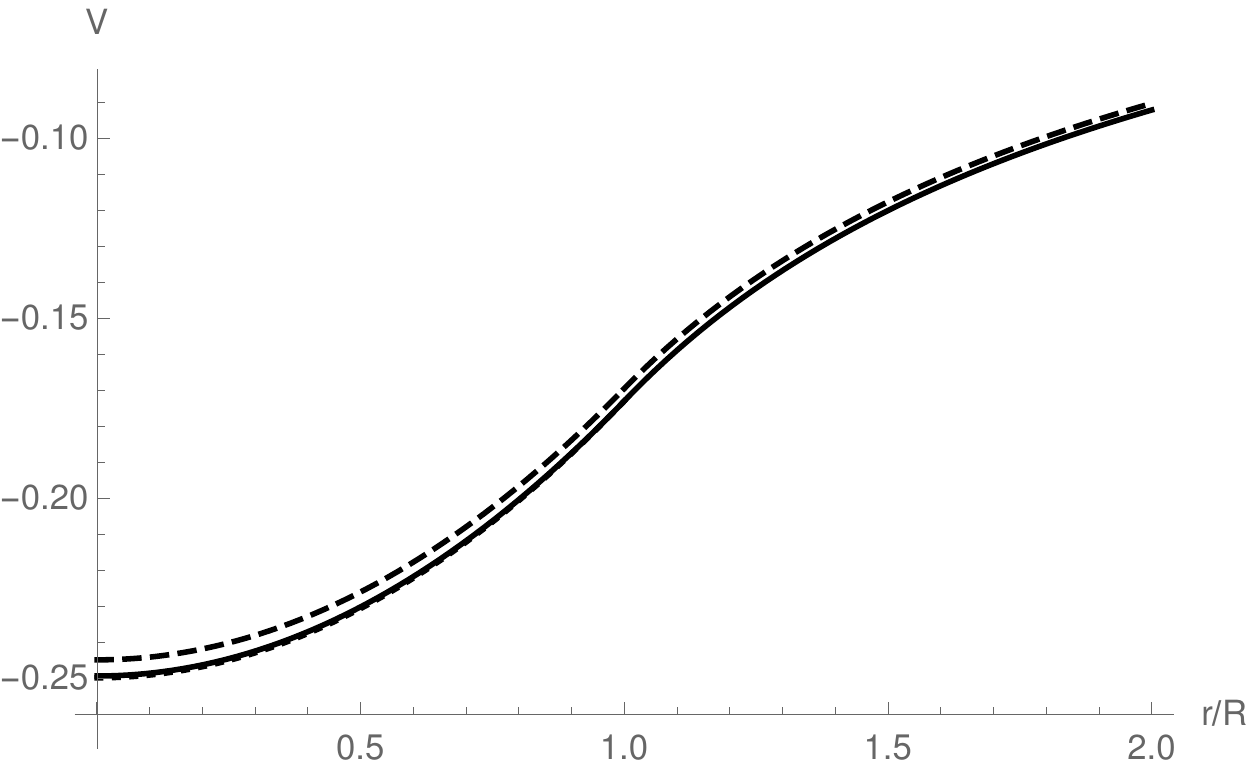}
$\ $
\includegraphics[width=5cm,height=4cm]{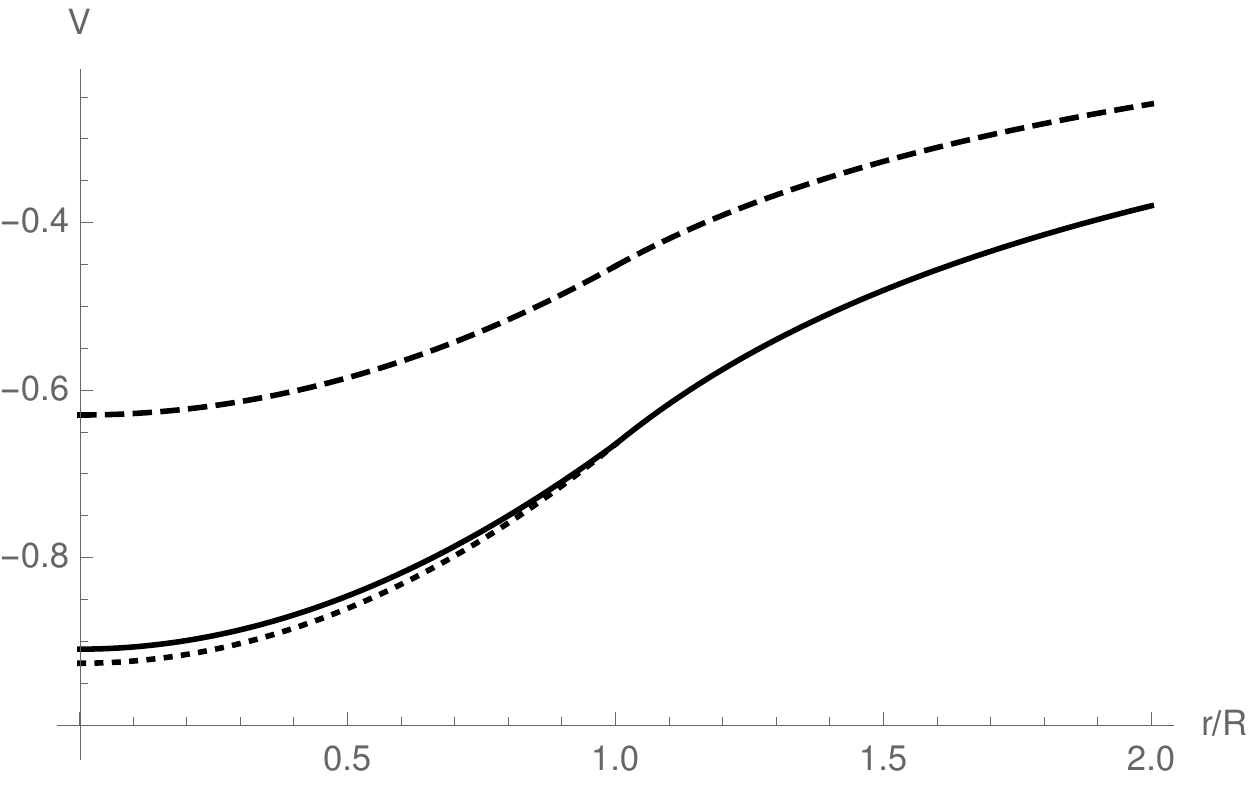}
\\
\includegraphics[width=5cm,height=4cm]{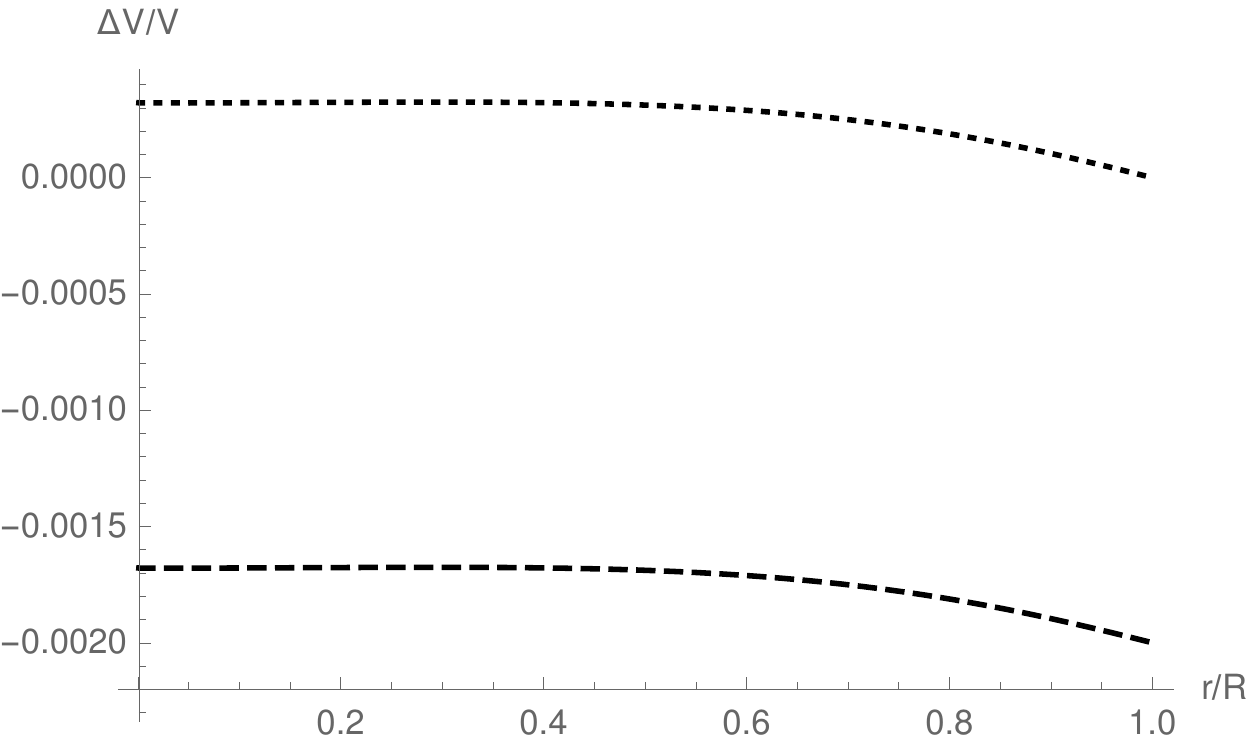}
$\ $
\includegraphics[width=5cm,height=4cm]{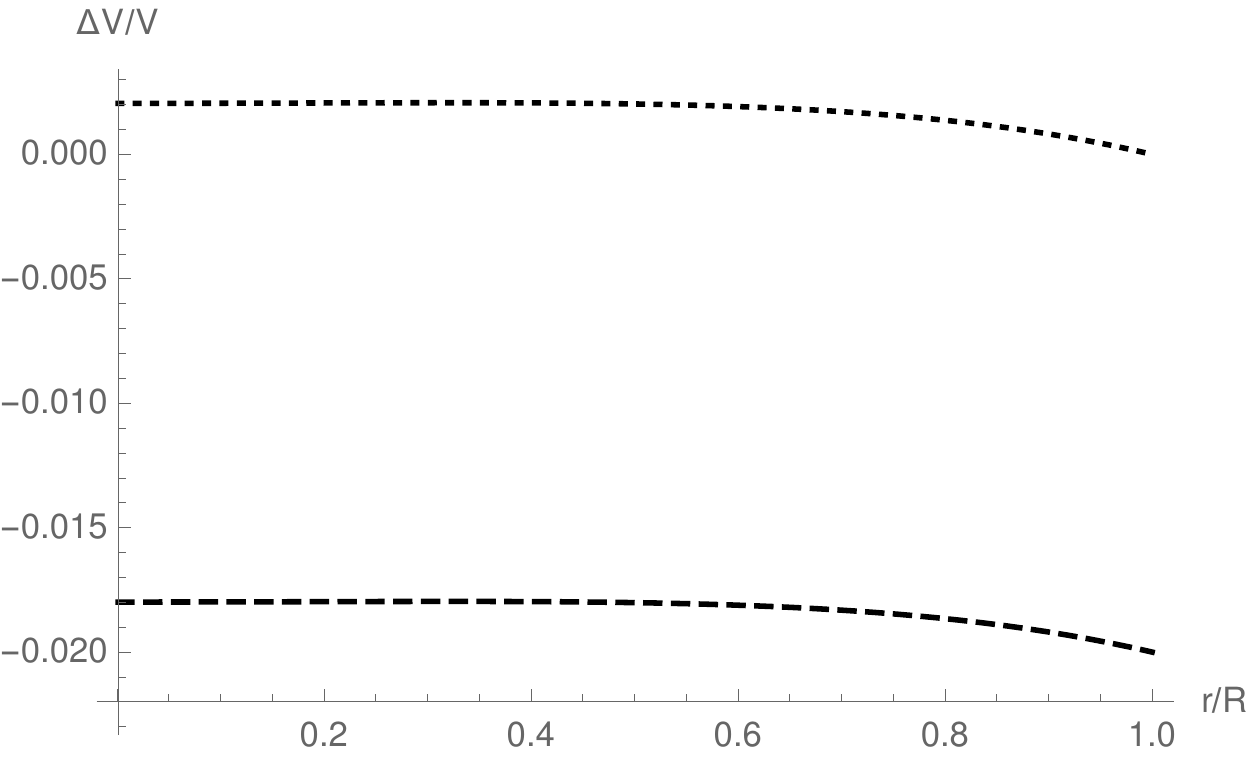}
$\ $
\includegraphics[width=5cm,height=4cm]{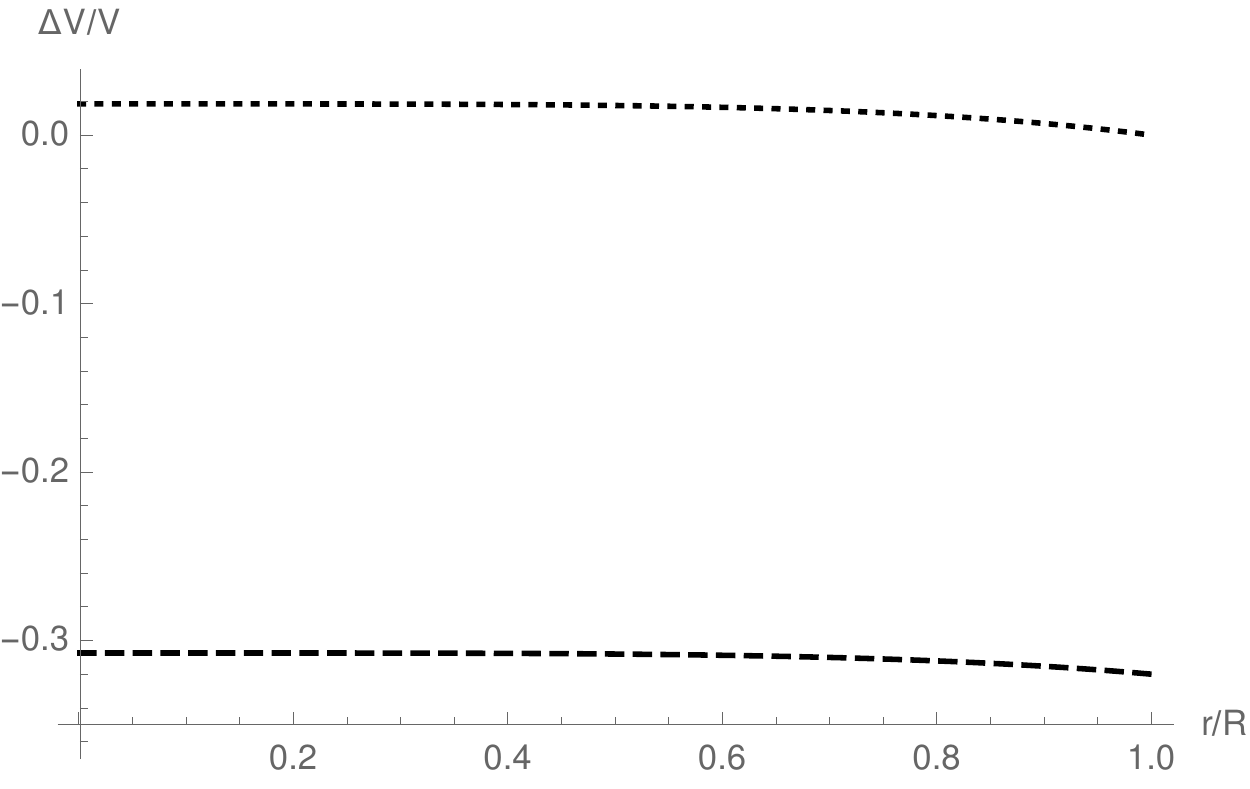}
\caption{
Upper panels:
numerical solution $V_{\rm n}$ to Eq.~\eqref{eomV} matched to the exact outer solution~\eqref{sol0}
(solid line) {\em vs\/} approximate solution $V_{\rm s}=V_-$ in Eq.~\eqref{Vins} (dotted line)
{\em vs\/} upper bounding function $V_+$ (dashed line) for
$\gn\,M /R = 1/10$ (top left),
$\gn\,M / R= 1/5$ (top middle) and
$\gn\,M / R= 1$ (top right).
Bottom panels:
relative difference
$(V_{\rm s}-V_{\rm n})/V_{\rm n}$ (dotted line) {\em vs\/} $(V_+-V_{\rm n})/V_{\rm n}$ (dashed line) in the interior region
for $\gn\,M /R = 1/10$ (bottom left),
$\gn\,M / R= 1/5$ (bottom middle) and
$\gn\,M / R= 1$ (bottom right).
The negative sign of $(V_{\rm s}-V_{\rm n})/V_{\rm n}$ shows that the approximate solution is a lower bounding
function $V_{\rm s}=V_-$ in this range of compactness.
The rapid growth in modulus of $(V_+-V_{\rm n})/V_{\rm n}$ with the compactness
signals the need of a better estimate of $M=M(M_0)$ for a more accurate description.
}
\label{VtotA}
\end{figure}

We can now estimate the accuracy of the approximation~\eqref{Vs0} by means of the comparison
method. 
The plots in Fig.~\ref{VinVSVzoomSmall} and~\ref{VinVSVzoomInter} show that $V_{\rm s}$ is already
in good agreement with the numerical solution for both small and intermediate compactness and
the smaller the ratio $\gn\,M/R$, the less $V_{\rm s}$ differs from the numerical solution.
Indeed, the approximate solution $V_{\rm s}$ fails in the large compactness regime,
which will be studied in the next subsection.
The same plots also tell us that $V_{\rm s}$ is actually an upper bounding function $V_+$ up to
$\gn\,M /R \simeq 1/20$, but becomes a lower bounding function $V_-$ for higher compactness
(this can be verified by showing that it satisfies the required conditions described in Appendix~\ref{A:comparison}).
The other bounding function ($V_-$ or $V_+$) can be found by simply multiplying $V_{\rm s}$ by a suitable
constant factor $C$ determined according to the theorem in Appendix~\ref{A:comparison}
(with $C>1$ for small compactness and $C<1$ for intermediate compactness).
This means that the approximate solution~\eqref{Vs0} overestimates the expected true potential $V_{\rm in}$
for low compactness, whereas it underestimates $V_{\rm in}$ when the compactness grows beyond 
$\gn\,M / R \simeq 1/20$.
We also note that the gap between the above $V_{-}$ and $V_{+}$ increases for increasing compactness,
which signals the need for a better estimate of $M_0=M_0(M)$ in order to narrow this gap and 
gain more precision for describing the intermediate compactness.
The latter regime is particularly useful for understanding objects that have collapsed 
to a size of the order of their gravitational radius~\footnote{The uniform density profile~\eqref{HomDens} can also be viewed
as a crude approximation of the density in the corpuscular model of black holes,
in which the energy is distributed throughout the entire inner
volume~\cite{DvaliGomez,Casadio:2015bna,Mueck,Foit:2015wqa,Kuhnel,kuhnelBaryons,Dvali:2016mur,Casadio:2014vja,Giusti:2019wdx}.}. 
We should remark that, in this analysis, we actually employed the comparison method in the whole range
$0\le r< \infty$ by defining $V_\pm=C_\pm\,V_{\rm out}$, for $r>R$, where $V_{\rm out}$ is the exact solution
in Eq.~\eqref{sol0} (see Figs.~\ref{Vtot} and~\ref{VtotA}).
This means that we did not require that the lower function $V_-$ (for $\gn\,M/R \lesssim 1/20$) and the
upper function $V_+$ (for  $\gn\,M/R \gtrsim 1/20$) satisfy the boundary conditions~\eqref{bR}
and~\eqref{dbR} at $r=R$.
However, since we have the analytical form for $V_{\rm out}$ in its entire range of applicability,
all that is needed to ensure that $V_\pm$ are the upper and lower bounding functions is for the constants
$C_\pm$ which multiply the expression for $V_{\rm out}$ to be smaller, respectively larger than one. 
\par
As stated earlier, the analytic approximation~\eqref{Vins} works best in the regime of small compactness,
in which we can further Taylor expand all quantities to second order in $\gn\,M/R\ll 1$ to obtain
\be
V_0
\simeq
-\frac{3\, \gn\, M}{2\, R}
\left(1 
-
\frac{4\, \gn\, M}{3\,R}
\right)
\ ,
\label{V0s}
\ee 
and finally use Eq.~\eqref{M0} to obtain
\be
M_0
\simeq
M
\left(
1
-
\frac{5\, \gn\, M}{2\, R}
\right)
\ ,
\label{M0small}
\ee
in qualitative agreement with the result of Ref.~\cite{BootN}, where however the effect of the pressure
on the potential was neglected.
\par
\begin{figure}[t]
\centering
\includegraphics[width=5cm,height=4cm]{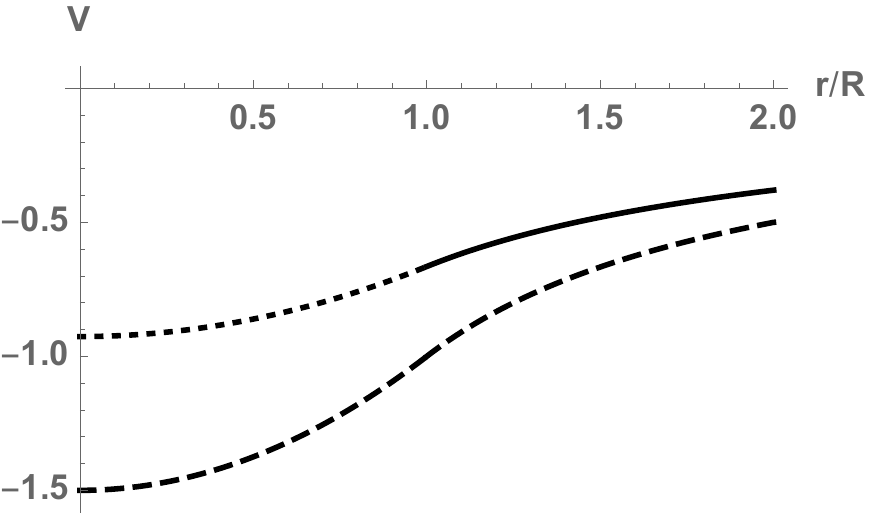}
$\ $
\includegraphics[width=5cm,height=4cm]{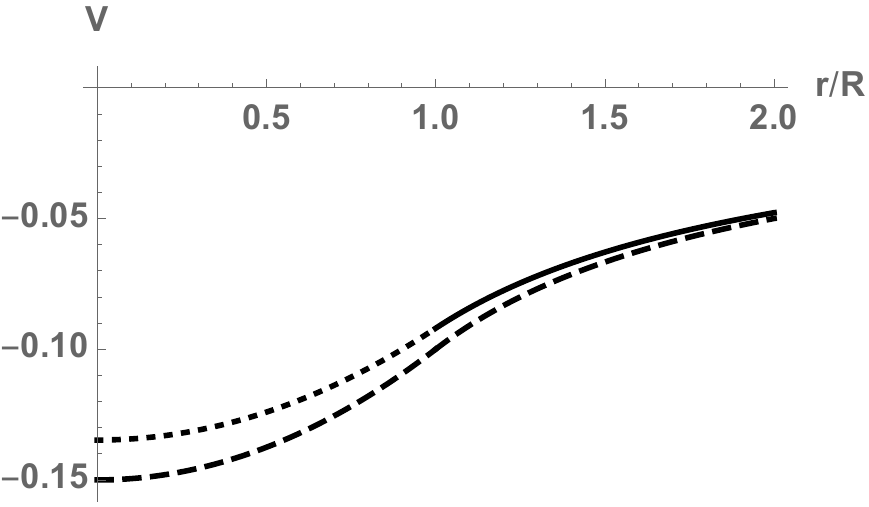}
$\ $
\includegraphics[width=5cm,height=4cm]{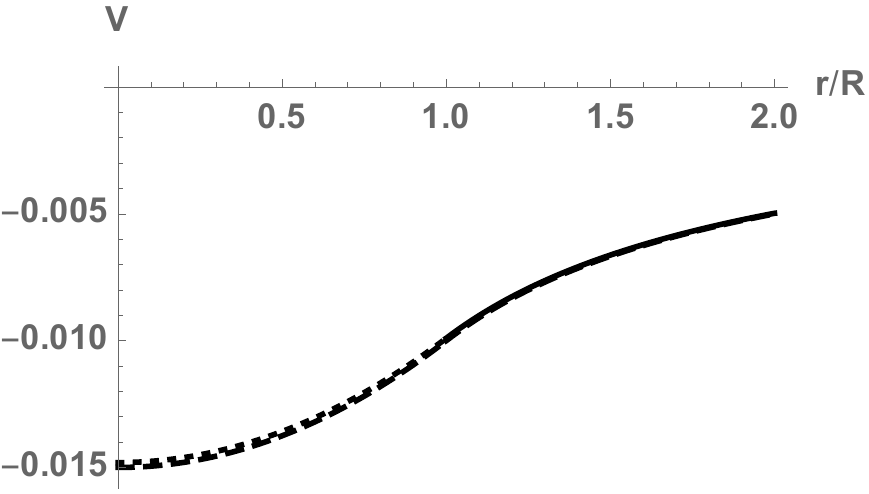}
\caption{Potential $V_{\rm out}$ (solid line) {\em vs\/} approximate solution~\eqref{Vins} (dotted line)
{\em vs\/} Newtonian potential (dashed line), for $\gn\,M/R =1$ (left panel), $\gn\,M/R = 1/10$
(center panel) and $\gn\,M/R = 1/100$ (right panel).
}
\label{Vsmall}
\end{figure}

The above expressions for $M_0$ and $V_0$ can be used to write the inner potential~\eqref{Vs0} in a much
simpler form in terms of $M$ as
\be
\label{Vins1}
V_{\rm in}
\simeq
-\frac{3\, \gn\, M}{2\, R} 
+
\frac{2\, \gn^2\, M^2}{R^2}
+
\frac{\gn\, M\,(R - 2\, \gn\, M)}{2\, R^4}\, r^2 
\ .
\ee
As expected, the solution for small compactness, which can be useful for describing stars with a radius
orders of magnitude larger in size than their gravitational radius, qualitatively tracks the Newtonian case.
This can also be seen from Fig.~\ref{Vsmall}.
The limitations of the small compactness approximation can be inferred from Eq.~\eqref{Vins1}.
For $2\,\gn\,M\equiv\Rh\sim R$ the last term vanishes and $V_{\rm in}$ becomes a constant.
\par
\begin{figure}[t]
\centering
\includegraphics[width=5cm,height=4cm]{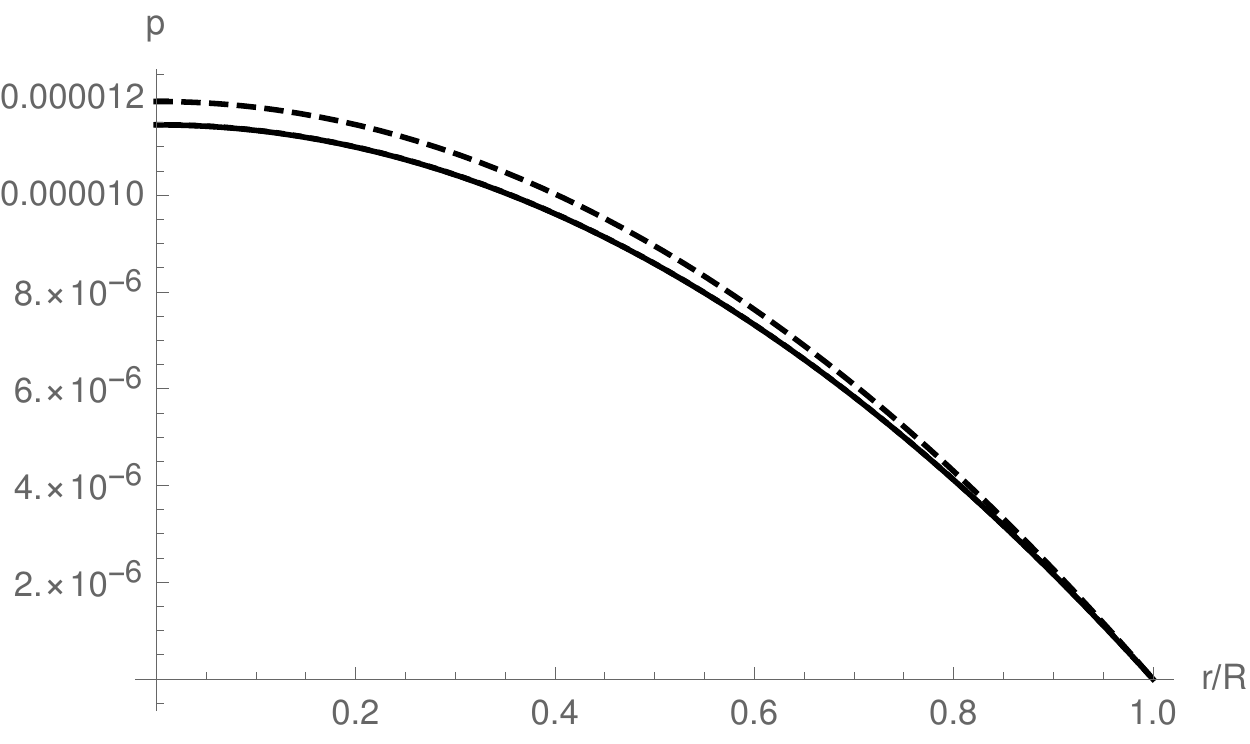}
$\ $
\includegraphics[width=5cm,height=4cm]{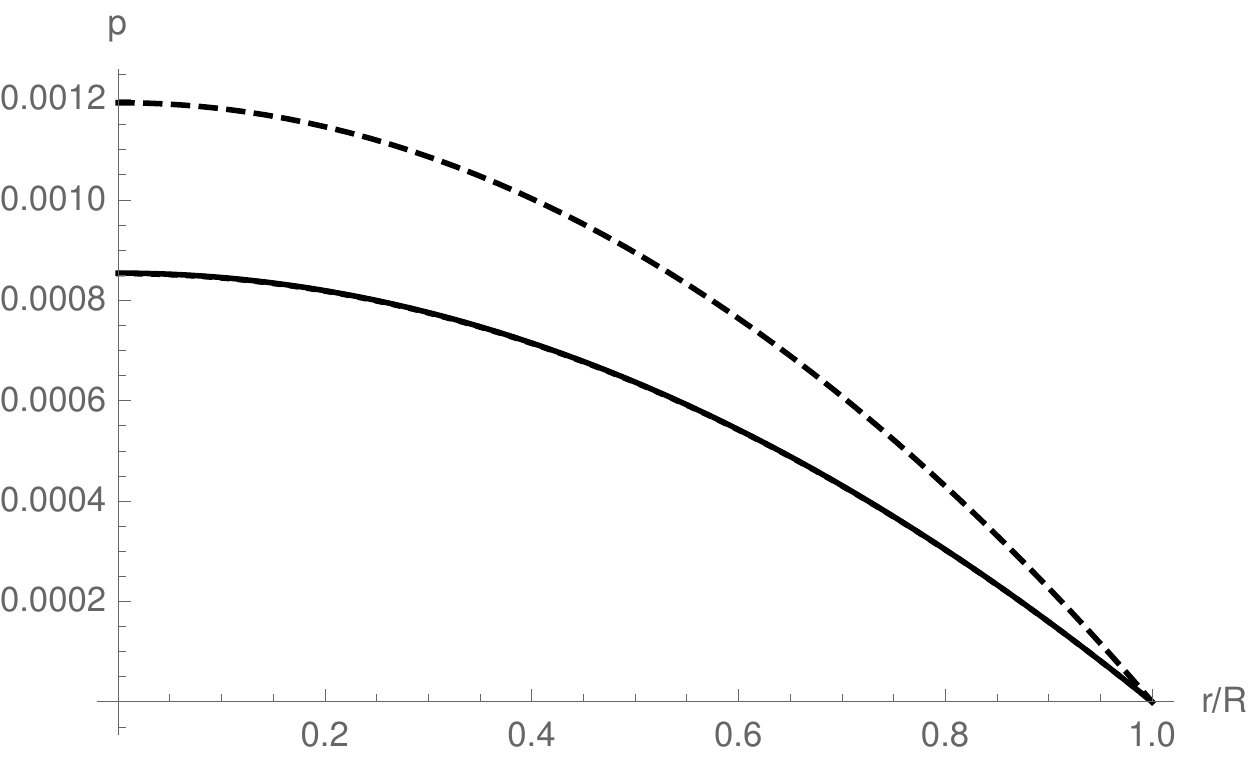}
$\ $
\includegraphics[width=5cm,height=4cm]{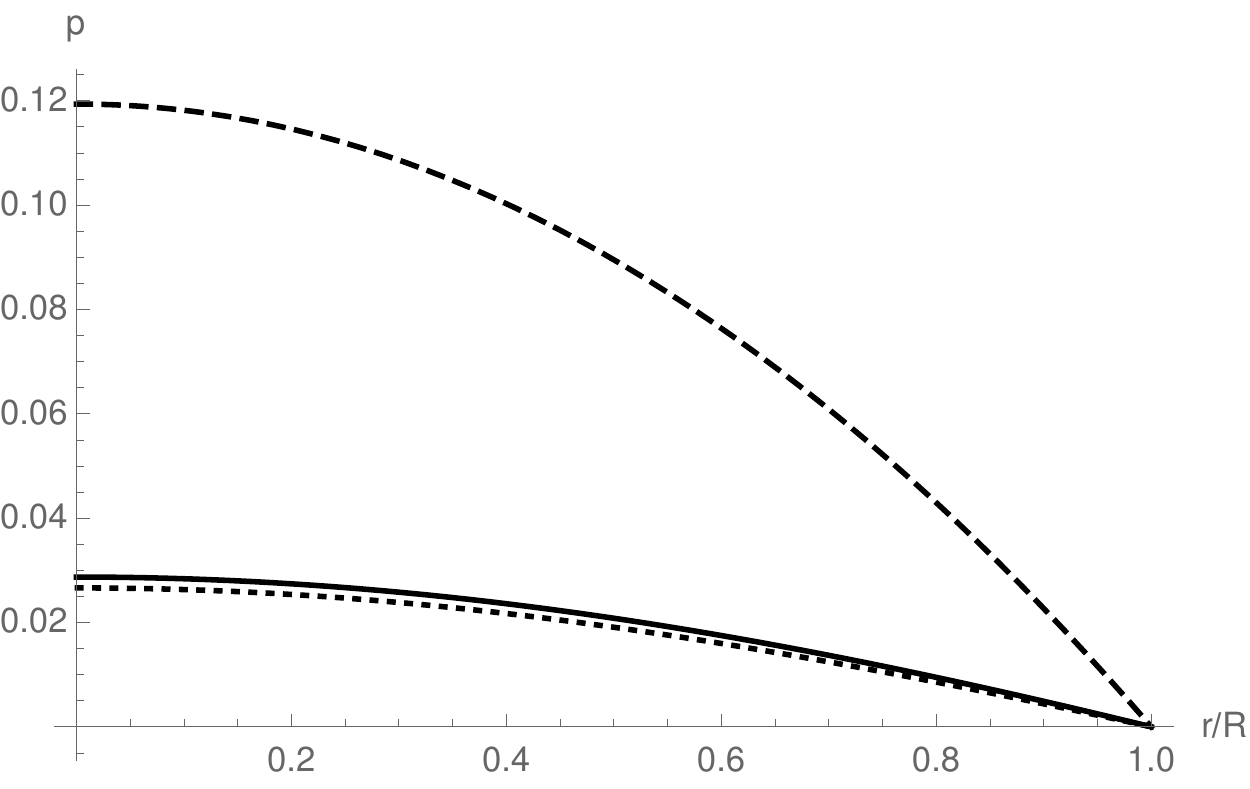}
\caption{Pressure obtained from the expansion~\eqref{Vs0}
(solid line) {\em vs\/} numerical pressure (dotted line) {\em vs\/}
Newtonian pressure~\eqref{pressureNewt} (dashed line),
for $\gn\,M/R =1/100$  (left panel), $\gn\,M/R = 1/10$ (center panel)
and $\gn\,M/R = 1$ (right panel).
}
\label{pressuresmall}
\end{figure}
Finally, it is important to remark that, as opposed to what was done in Ref.~\cite{BootN}, 
the pressure now acts as a source and can be consistently evaluated with the help of
Eqs.~\eqref{pressure} and~\eqref{Vs0}.
The plots in Fig.~\ref{pressuresmall} clearly show that the pressure can be well approximated
by the Newtonian formula in the regime of low compactness, to wit
\be
\label{pressureNewt}
p 
\simeq
\frac{3\, \gn \, M^2\, (R^2 - r^2)}{8\, \pi \, R^6}
\ ,
\ee
again in qualitative agreement with Ref.~\cite{BootN}.
Nevertheless, the same plots indicate that it rapidly departs from the Newtonian expression 
when we approach the regime of intermediate compactness, while remaining 
almost identical to the numerical approximation.
\subsubsection{Large compactness}
\label{ss:largeC}
For $\gn\,M/R\gg 1$, rather than employing a Taylor expansion like we did for small compactness,
it is more convenient to fully rely on comparison methods~\cite{BVPexistence,BVPexistence1,ode}
and start from the exact solution of the simpler equation
\be
\psi''
=
\frac{3\,\gn\,M_0}{R^3}\,e^{V_R-\psi}
\ ,
\ee 
which is given by
\be
\psi(r;A,B)
=
-A\left(B+\frac{r}{R}\right)
+
2\,\ln\left[1+\frac{3\,\gn\,M_0}{2\,A^2\,R}\,e^{A\,(B+r/R)+V_R}\right]
\ ,
\label{psiAB}
\ee
where the constants $A$, $B$ and $M_0$ can be fixed (for any value of $R$) by imposing
the boundary conditions~\eqref{b0}, \eqref{bR} and~\eqref{dbR}.
Regularity at $r=0$ in particular yields
\be
M_0
=
\frac{2\,A^2\,R}{3\,\gn}\,e^{-A\,B-V_R}
\ .
\label{M00}
\ee
Eq.~\eqref{dbR} for the continuity of the derivative across $r=R$ then reads
\be
A\,\tanh(A/2)
=
R\,V_R'
\ .
\label{eqA}
\ee
For large compactness, $R\,V_R'\sim (\gn\,M/R)^{2/3}\gg 1$, and we can 
approximate the above equation as
\be
A
\simeq
R\,V_R'
\ .
\ee
The continuity Eq.~\eqref{bR} for the potential finally reads
\be
2\,\ln\left(1+e^{R\,V_R'}\right)
-R\,V_R'\,(1+B)
=
V_R
\ ,
\ee
and can be used to express $B$ in terms of $M$ and $R$.
Putting everything together, we obtain
\be
\psi(r;M,R)
&\!\!\simeq\!\!&
\frac{1}{4}
\left\{
1
-
\frac{1+({2\,\gn\,M}/{R})
\left(1+{2\,r}/{R}\right)}
{\left(1+{6\,\gn\,M}/{R}\right)^{1/3}}
+
8\,
\ln\left[
\frac{1+e^{\frac{\gn\,M\,r/R^2}{\left(1+6\,\gn\,M/R\right)^{1/3}}}}
{1+e^{\frac{\gn\,M/R}{\left(1+6\,\gn\,M/R\right)^{1/3}}}}
\right]
\right\}
\nonumber
\\
&\!\!\simeq\!\!&
\frac{1}{2}
\left(\frac{\gn\,M}{\sqrt{6}\,R}\right)^{2/3}
\left(\frac{2\,r}{R}-5\right)
\ ,
\label{psi0}
\ee 
and
\be
\frac{M_0}{M}
&\!\!\simeq\!\!&
\frac{\gn\,M/R}
{3\left(1+6\,\gn\,M/R\right)^{2/3}
\left\{1+\cosh\left[\frac{\gn\,M/R}{\left(1+6\,\gn\,M/R\right)^{1/3}}\right]\right\}}
\nonumber
\\
&\!\!\simeq\!\!&
\frac{1}{3}
\left(\frac{2\,\gn\,M}{9\,R}\right)^{1/3}
e^{-\left(\frac{\gn\,M}{\sqrt{6}\,R}\right)^{2/3}}
\ ,
\label{M0psi}
\ee
in which we showed the leading behaviours for $\gn\,M\gg R$.
It is important to remark that the condition~\eqref{b0} is not apparently satisfied
by the above approximate expressions, although it was imposed from the very beginning,
which shows once more how complex is to obtain analytical approximations for the
problem at hand.
\par
The solutions to the complete equation~\eqref{eomV} could then be written as
\be
V_{\rm in}
=
f(r;A,B)\,\psi(r;A,B)
\ ,
\ee
where $A$, $B$ and $M_0$ should again be computed from the three boundary
conditions, so that $V_{\rm in}$ eventually depends only on the parameters $M$ and $R$.
Since solving for $f=f(r)$ is not any simpler than the original task, we shall instead just
find lower and upper bounds, that is constants $C_\pm$ such that
\be
C_-<f(r)<C_+
\ ,
\ee
in the whole range $0\le r\le R$.
In particular, we consider the bounding functions
\be \label{VPM}
V_\pm
=
C_\pm\,\psi(r;A_\pm,B_\pm)
\ ,
\ee
where $A_\pm$, $B_\pm$ and $C_\pm$ are constants computed by imposing the boundary
conditions~\eqref{b0}, \eqref{bR} and~\eqref{dbR} and such that $E_+(r)<0$ and $E_-(r)>0$ for $0\le r\le R$.
\par
In details, we first determine a function $V_C=C\,\psi(r;A,B)$ which satisfies the three boundary conditions for
any constant $C$.
Eq.~\eqref{b0} yields the same expression~\eqref{M00}, whereas the l.h.s.~of Eq.~\eqref{eqA} is just rescaled
by the factor $C$ and continuity of the derivative therefore gives the approximate solution
\be
C\,A
\simeq
R\,V_R'
\ .
\label{dbRapp}
\ee
Eq.~\eqref{bR} for the continuity of the potential likewise reads
\be
2\,C\,\ln\left(1+e^{R\,V_R'/C}\right)
-R\,V_R'\,(1+B)
=
V_R
\ ,
\ee
Upon solving the above equations one then obtains $V_C=C\,\psi(r;A(M,R,C),B(M,R,C))$ and
$M_0=M_0(M,R,C)$.
For fixed values of $R$ and $M$, one can then numerically determine a constant $C_+$ such that
$E_+<0$ and a constant $C_-<C_+$ such that $E_->0$.
\begin{figure}[t]
\centering
\includegraphics[width=8cm]{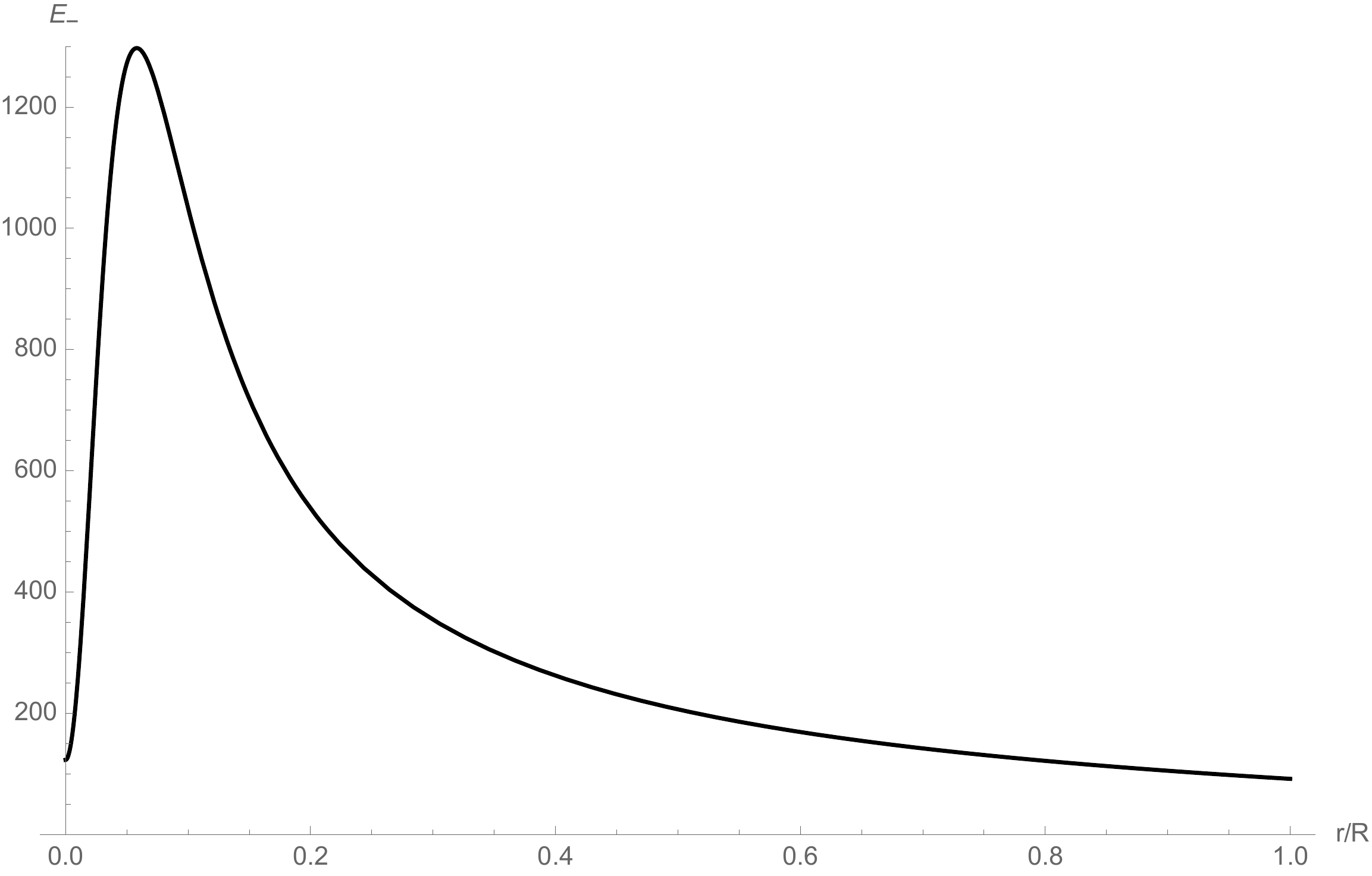}
$\ $
\includegraphics[width=8cm]{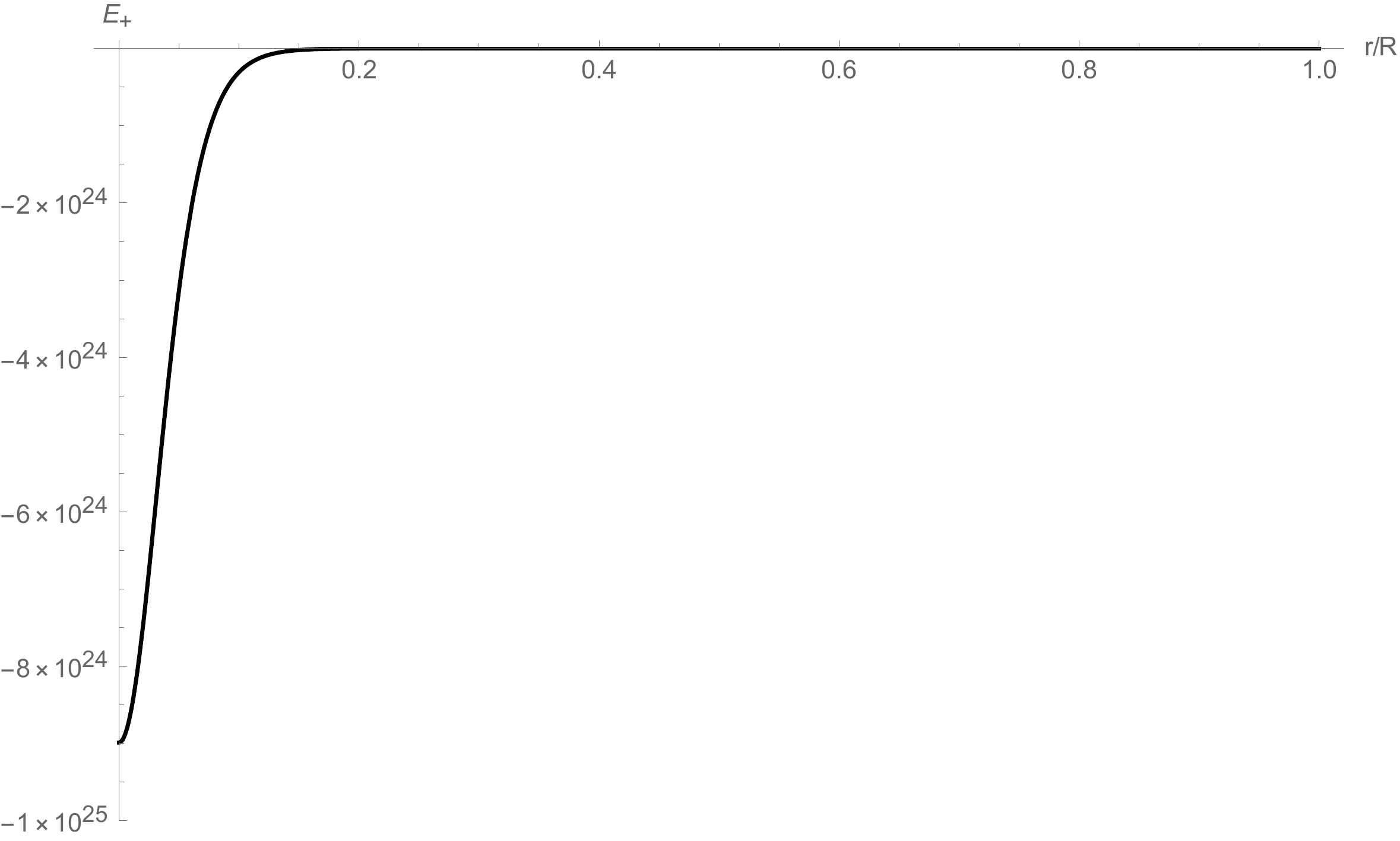}
\caption{Left panel: $E_-$ for $C_-=1$.
Right panel: $E_+$ for $C_+=1.6$.
Both plots are for $\gn\,M/R=10^3$.
}
\label{Epm}
\end{figure}
\begin{figure}[h]
\centering
\includegraphics[width=8cm]{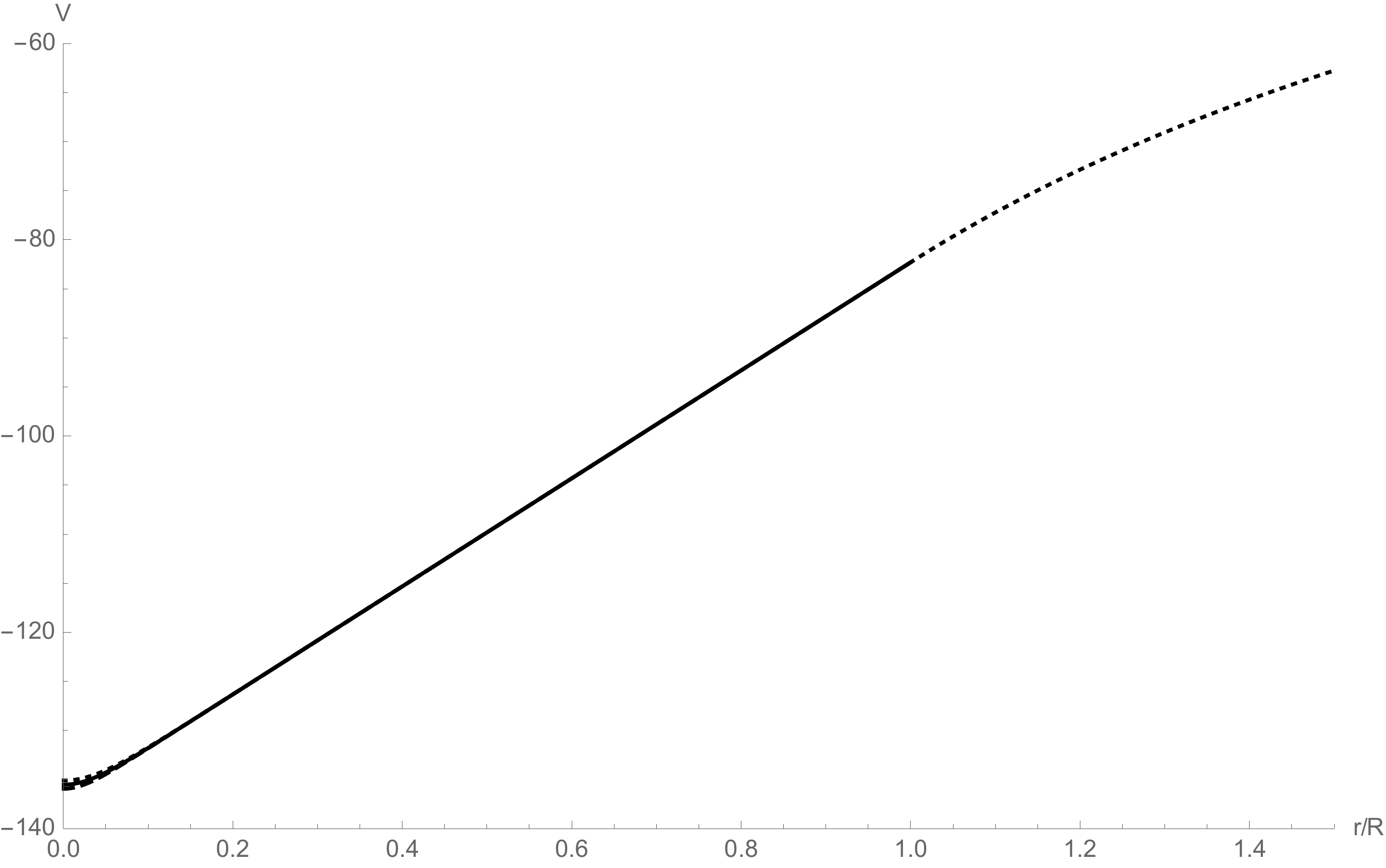}
$\ $
\includegraphics[width=8cm]{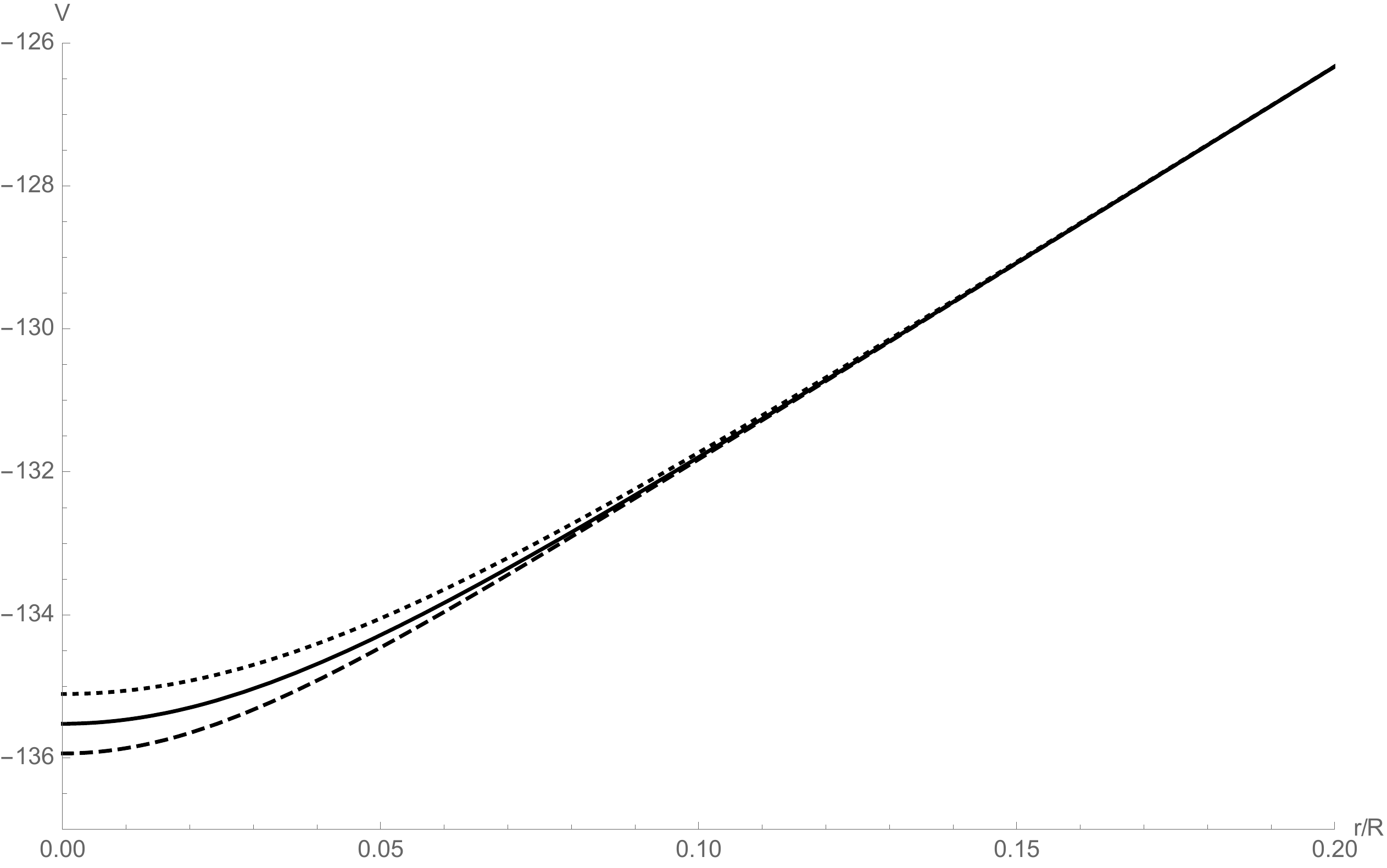}
\caption{Left panel: approximate inner potentials $V_-$ (dashed line), $\tilde V$ (solid line) 
and $V_+$ (dotted line) for $0\le r\le R$ and exact outer potential $V_{\rm out}$ (dotted line)
for $r>R$.
Right panel: approximate inner potentials $V_-$ (dashed line), $\tilde V$ (solid line) 
and $V_+$ (dotted line) for $0\le r\le R/5$. 
Both plots are for $\gn\,M/R=10^3$.
}
\label{Vpm}
\end{figure}
\begin{figure}[h]
\centering
\includegraphics[width=8cm]{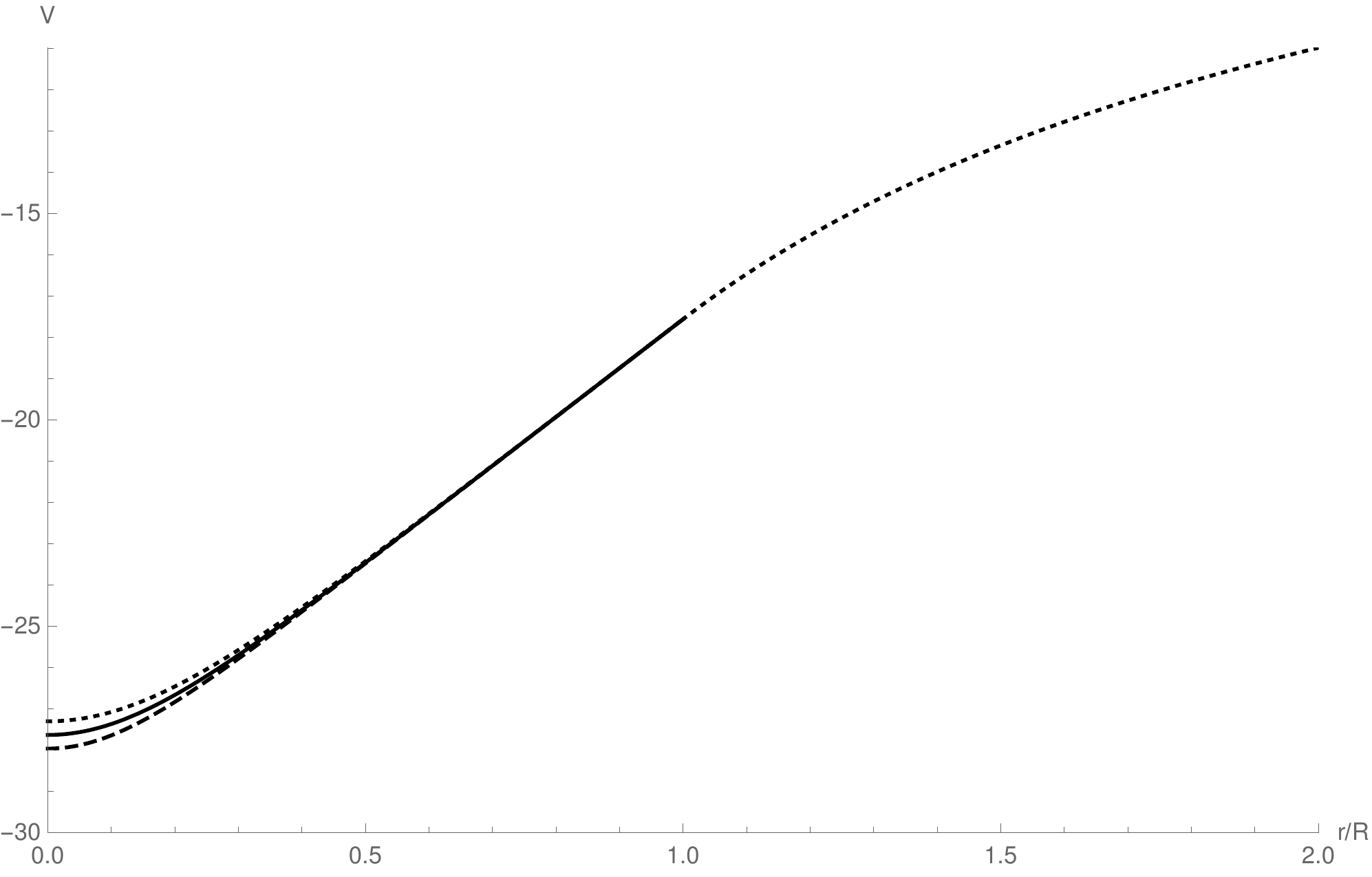}
$\ $
\includegraphics[width=8cm]{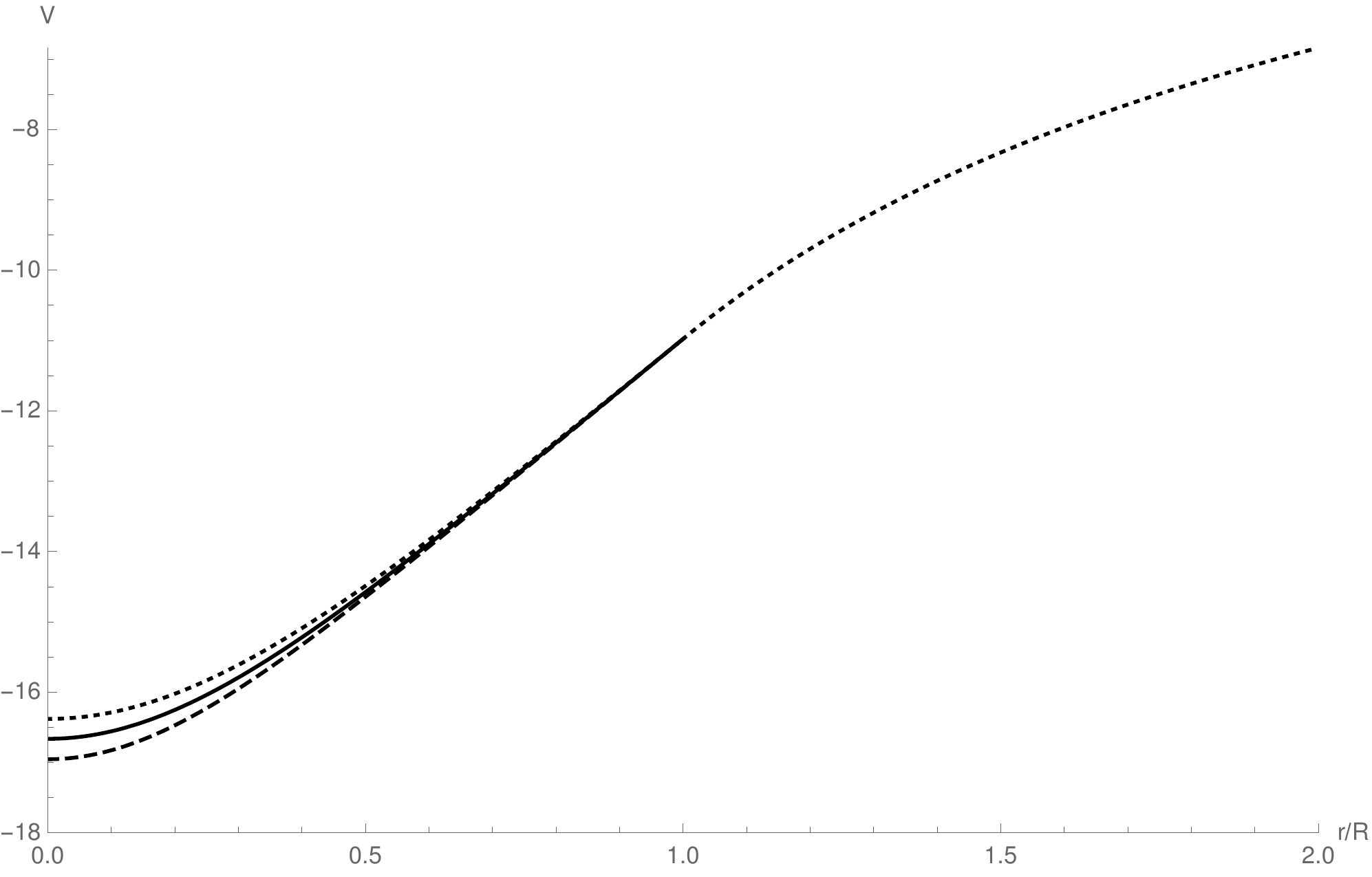}
\caption{Approximate inner potentials $V_-$ (dashed line), $\tilde V$ (solid line) 
and $V_+$ (dotted line) for $0\le r\le R$ and exact outer potential $V_{\rm out}$ (dotted line)
for $r>R$ and for $\gn\,M/R=10^2$ (left panel, with $C_-=1.042$ and $C_+=1.52$)
and $\gn\,M/R=50$ (right panel, with $C_-=1.073$ and $C_+=1.5$)
}
\label{Vpm1}
\end{figure}
\begin{figure}[t]
\centering
\includegraphics[width=8cm]{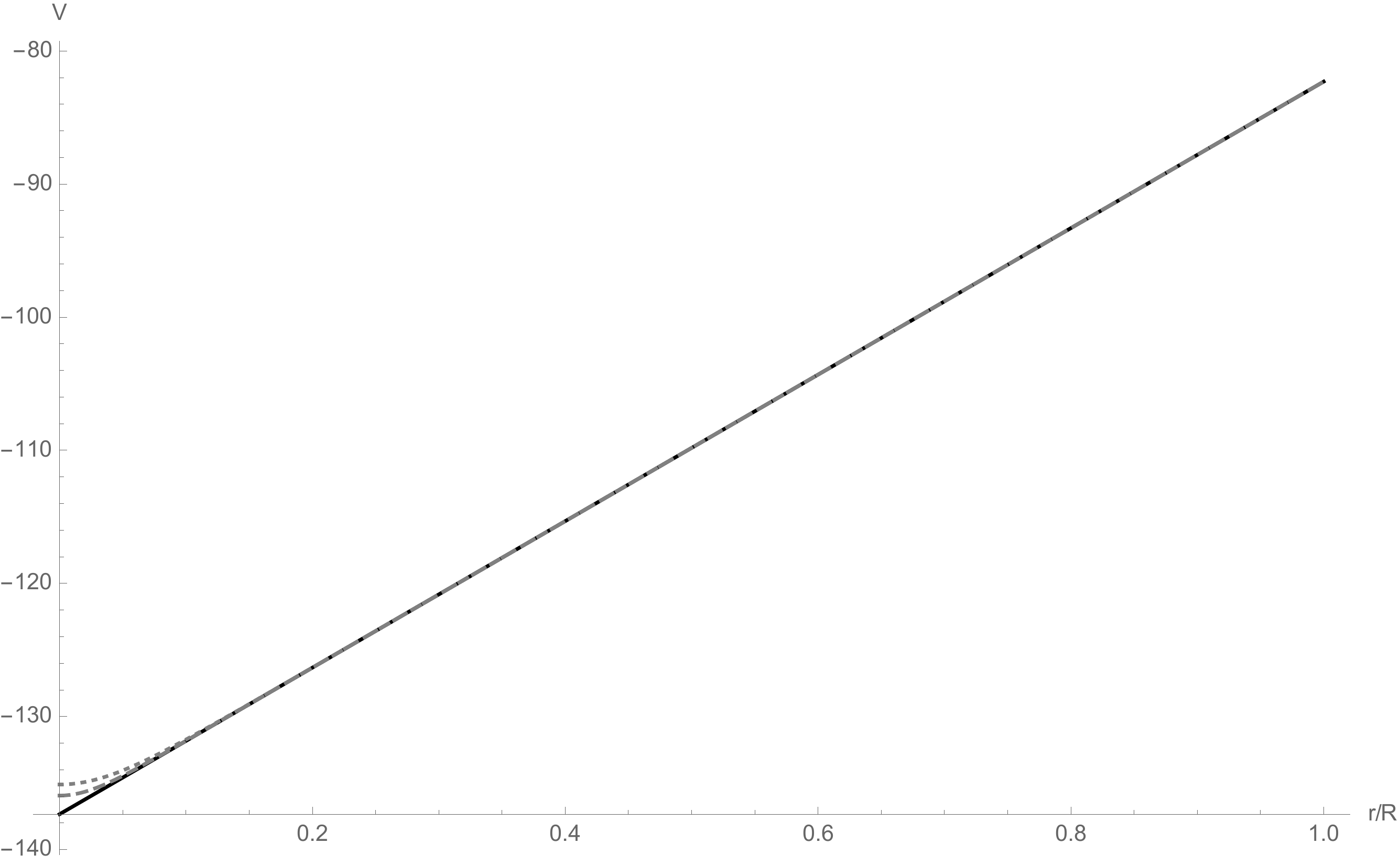}
\caption{Approximate inner potentials $V_-$ (dashed line), $V_{\rm lin}$ (solid line) 
and $V_+$ (dotted line) for $0\le r\le R$. 
Both plots are for $\gn\,M/R=10^3$.
}
\label{Vlin}
\end{figure}
\par
For example, for the compactness $\gn\,M/R=10^3$, we can use $C_-\simeq 1$ and $C_+\simeq 1.6$,
and the plots of $E_-$ and $E_+$ are shown in Fig.~\ref{Epm}.
In particular, the minimum value of $|E_+|\simeq 14$.
The corresponding potentials $V_\pm$ along with $\tilde V=\tilde C\,\psi$, where $\tilde C=(C_++C_-)/2$,
are displayed in Fig.~\ref{Vpm}.
It is easy to see that the three approximate solutions essentially coincide almost everywhere,
except near $r=0$ where they start to fan out, albeit still very slightly (the right panel of Fig.~\ref{Vpm}
shows a close-up of this effect).
A similar behaviour is obtained for larger values of $\gn\,M/R$.
For smaller values of the compactness up to $\gn\,M/R\simeq 50$, the approximation~\eqref{dbRapp} is still quite accurate 
(see Fig.~\ref{Vpm1}), even if the smaller the compactness the bigger the difference between $V_\pm$. 
Actually, the error in the derivative of the potential at $r=R$ is of the order of $0.01\,\%$ and
$0.6\,\%$ for $\gn\,M/R=10^2$ and $\gn\,M/R=50$, respectively.
In order to obtain a comparable precision for lower compactness, the approximate expression~\eqref{dbRapp}
should be improved, but we do not need to do that given how accurate is the perturbative expansion
employed in Section~\ref{SS:lowC}.
\par
From the left panel of Fig.~\ref{Vpm}, it is clear that for $\gn\,M/R=10^3$ the potential $V_{\rm in}$
is practically linear, except near $r=0$ where it turns into a quadratic shape, in order to ensure the
regularity condition~\eqref{b0}.
An approximate expression for the source proper mass $M_0$ in terms of $M$ can then be obtained
from the simple linear approximation
\be
V_{\rm lin}
\simeq
V_R
+
V'_R\left(r-R\right)
\ ,
\label{eVlin}
\ee
where $V_R$ and $V_R'$ are given by the usual expressions~\eqref{VR} and \eqref{dVR},
and which is shown in Fig.~\ref{Vlin} for $\gn\,M/R=10^3$.
Upon replacing the approximation~\eqref{eVlin} into the equation~\eqref{eomV} for $r=R$,
we obtain
\be
\frac{M_0}{M}
\simeq
\frac{2\left(1+5\,\gn\,M/R\right)}
{3\left(1+6\,\gn\,M/R\right)^{4/3}}
\ .
\label{M0M}
\ee
\begin{figure}[t]
\centering
\includegraphics[width=5cm,height=4cm]{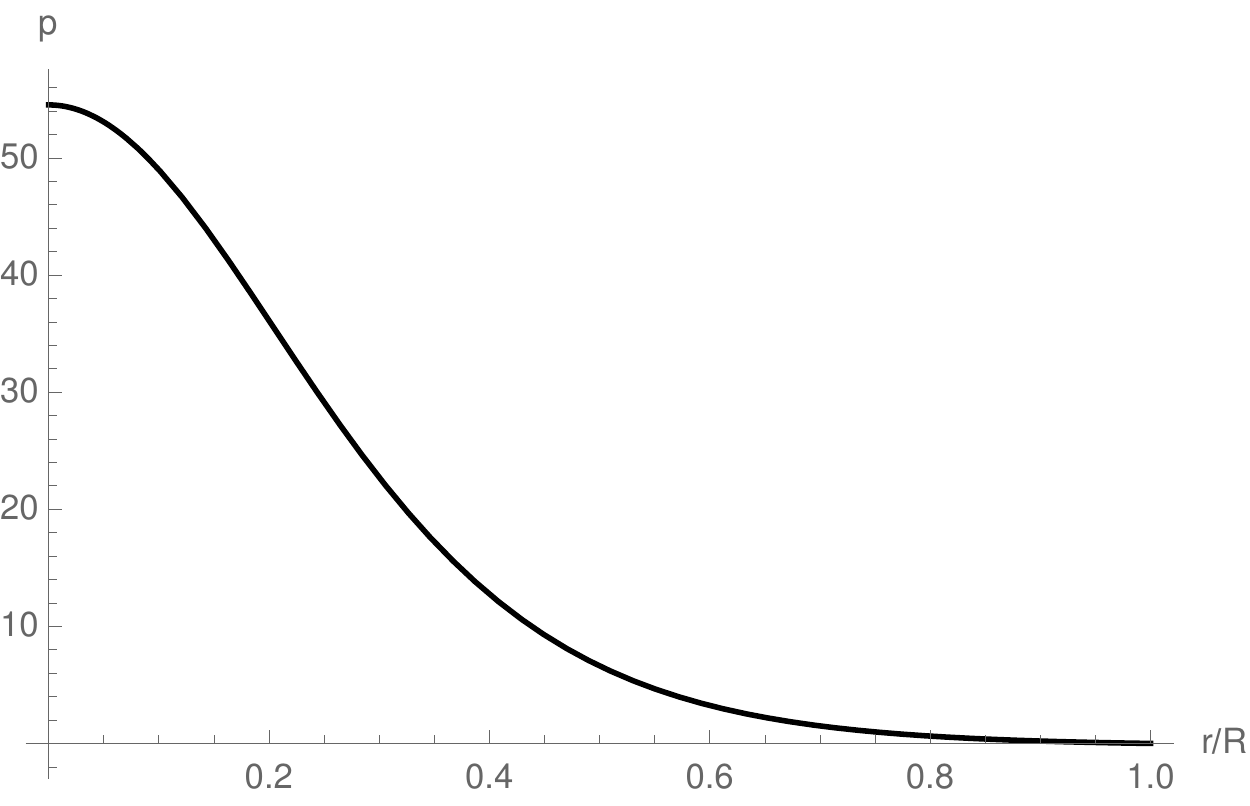}
$\ $
\includegraphics[width=5cm,height=4cm]{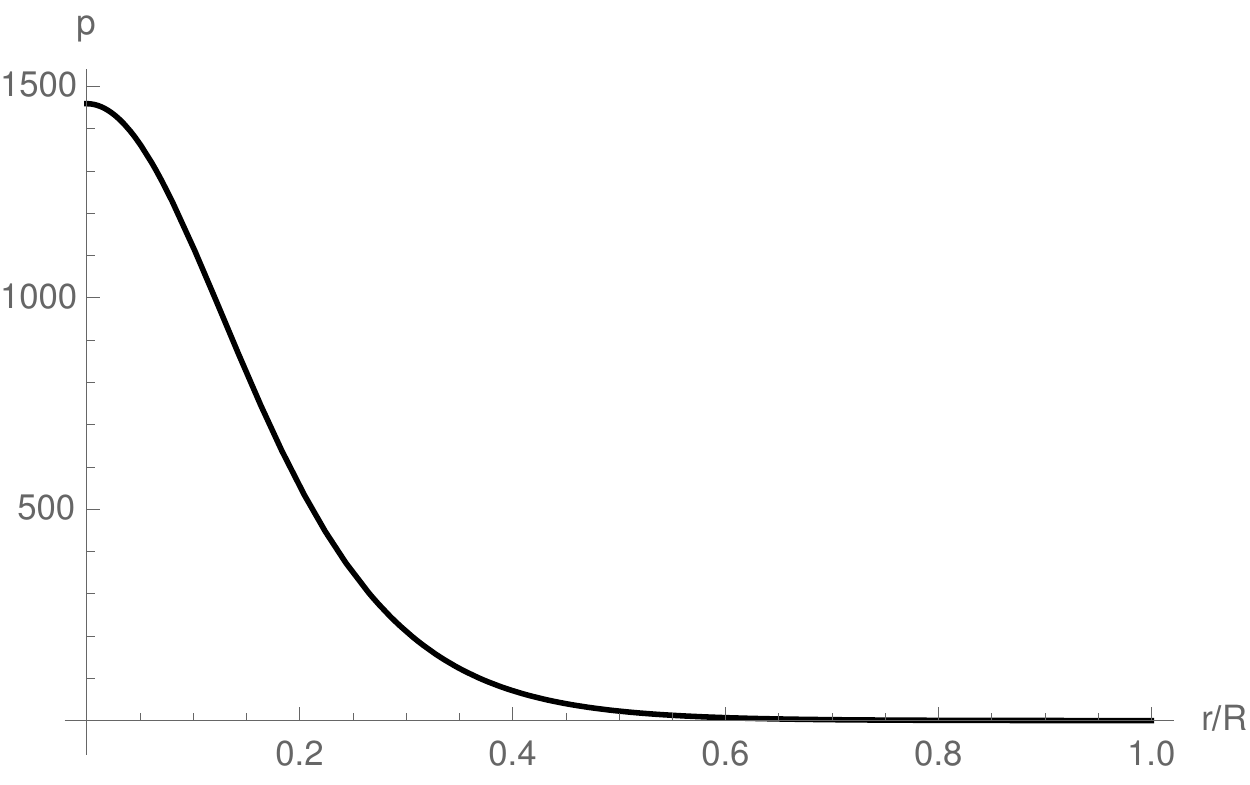}
$\ $
\includegraphics[width=5cm,height=4cm]{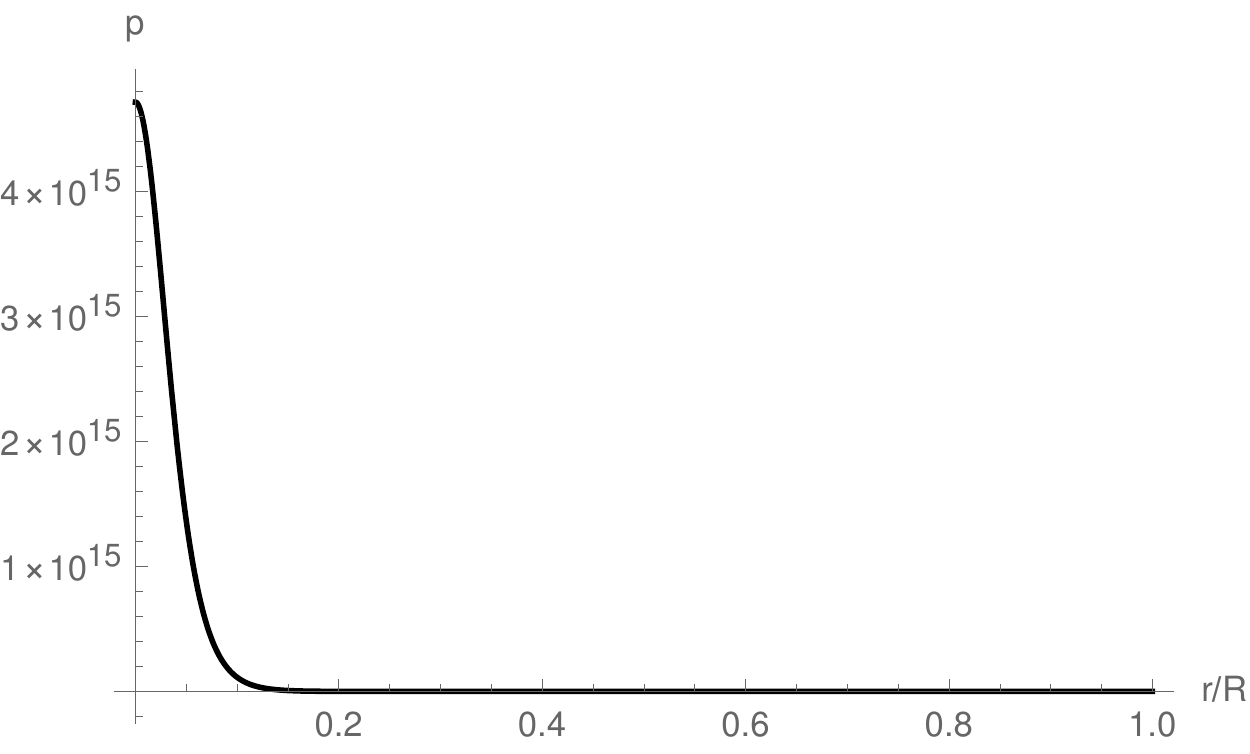}
\caption{Pressure evaluated using the approximation $\tilde V= \tilde C\,\psi$ for $\gn\,M/R =50$ 
(left panel), $\gn\,M/R = 100$ (center panel) and $\gn\,M/R = 1000$ (right panel). 
The constant $\tilde C=(C_+ + C_-)/2$, where $C_+$ and $C_-$ are the 
same as in Figs.~\ref{Vpm} and~\ref{Vpm1} for the corresponding cases. 
}
\label{pressurebig}
\end{figure}
\par
\begin{figure}[t]
\centering
\includegraphics[width=5cm,height=4cm]{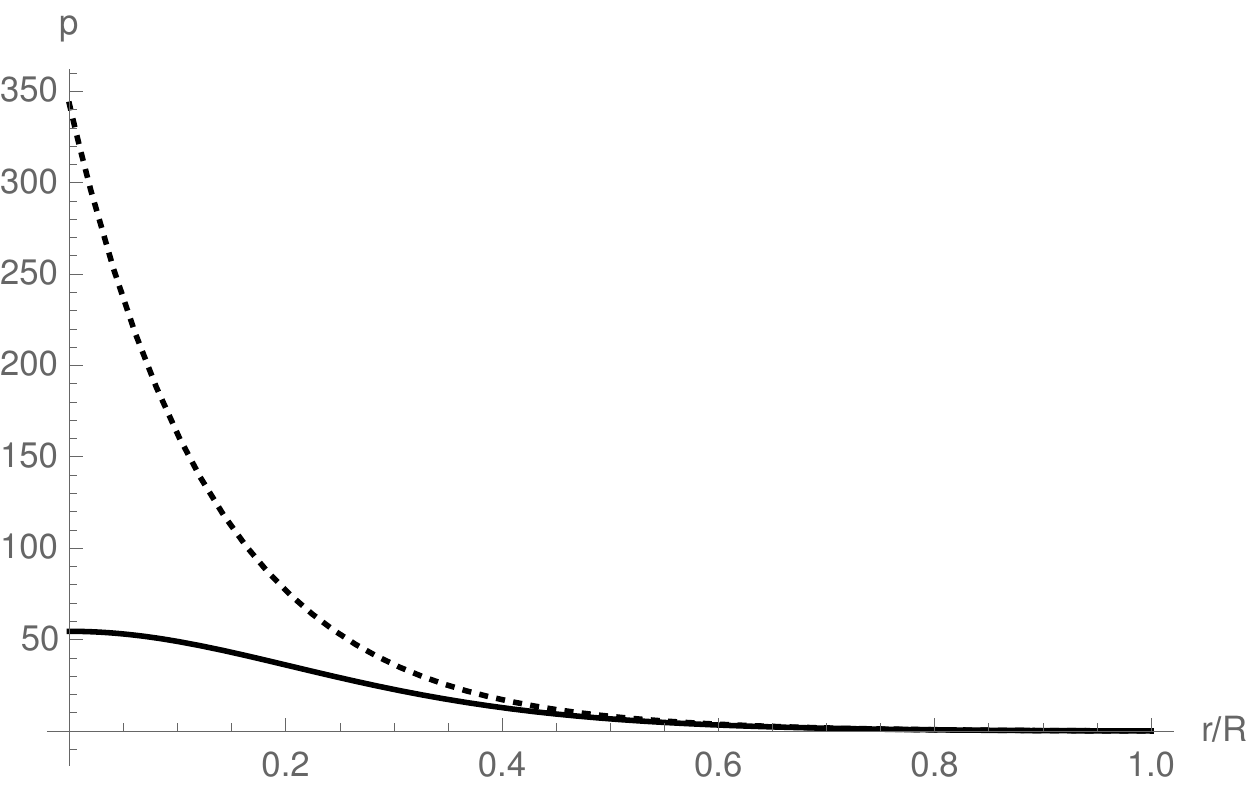}
$\ $
\includegraphics[width=5cm,height=4cm]{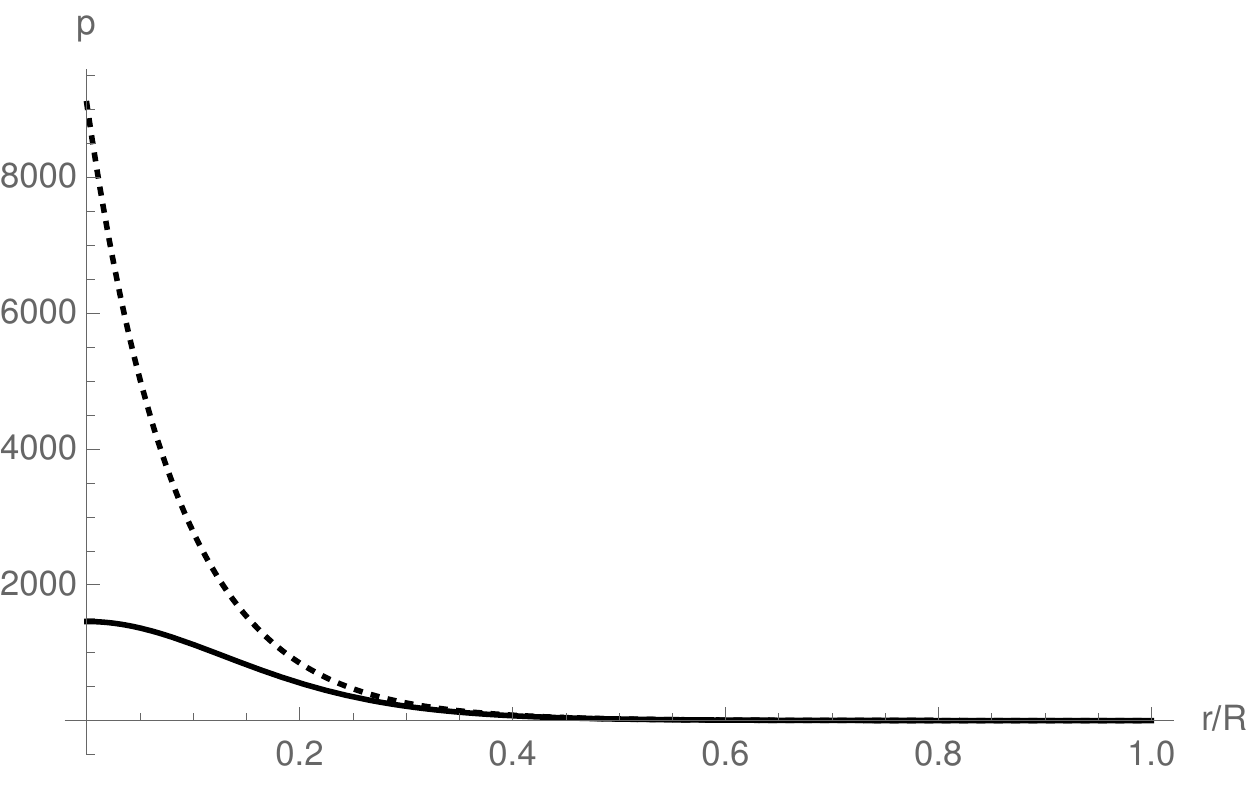}
$\ $
\includegraphics[width=5cm,height=4cm]{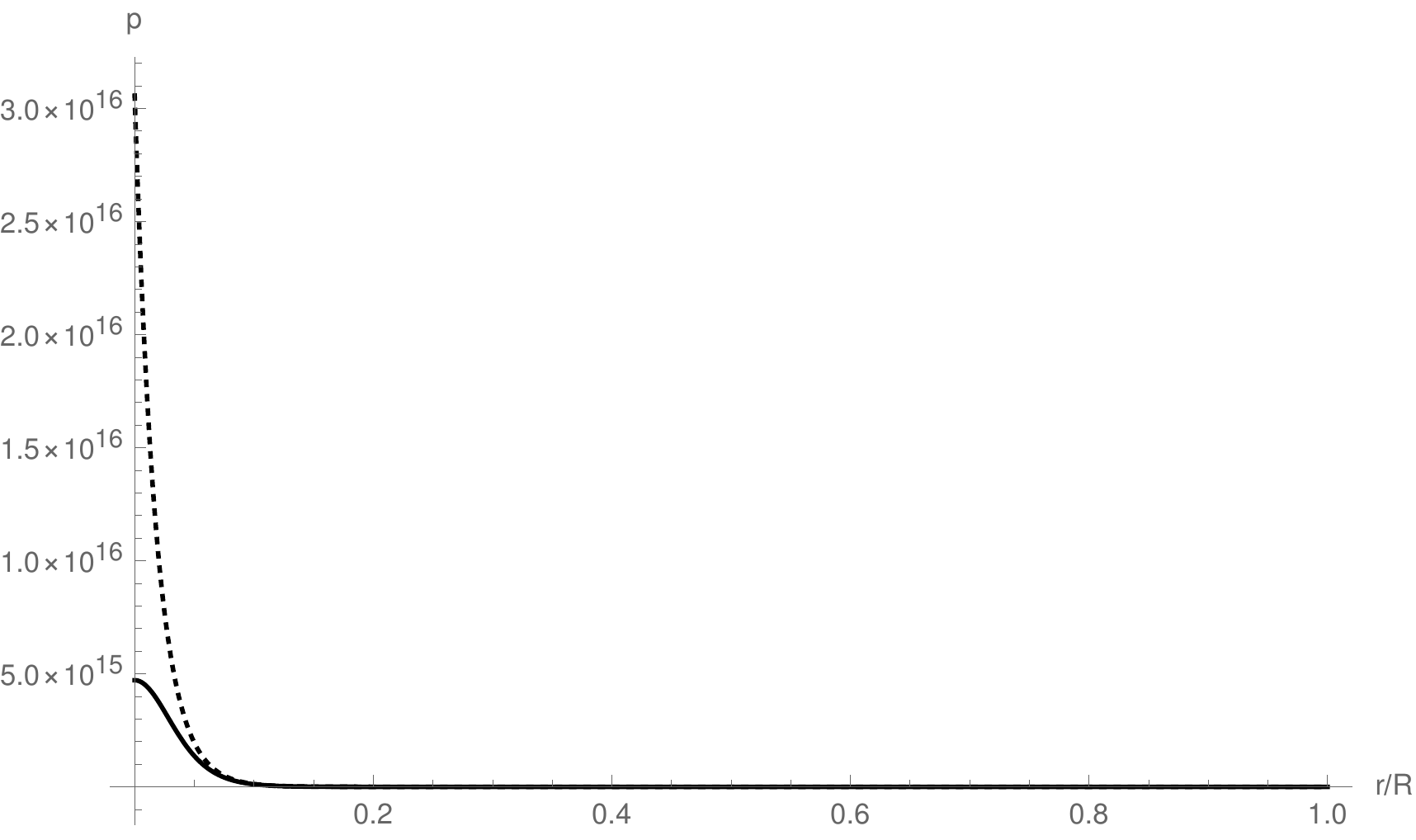}
\\
\includegraphics[width=5cm,height=4cm]{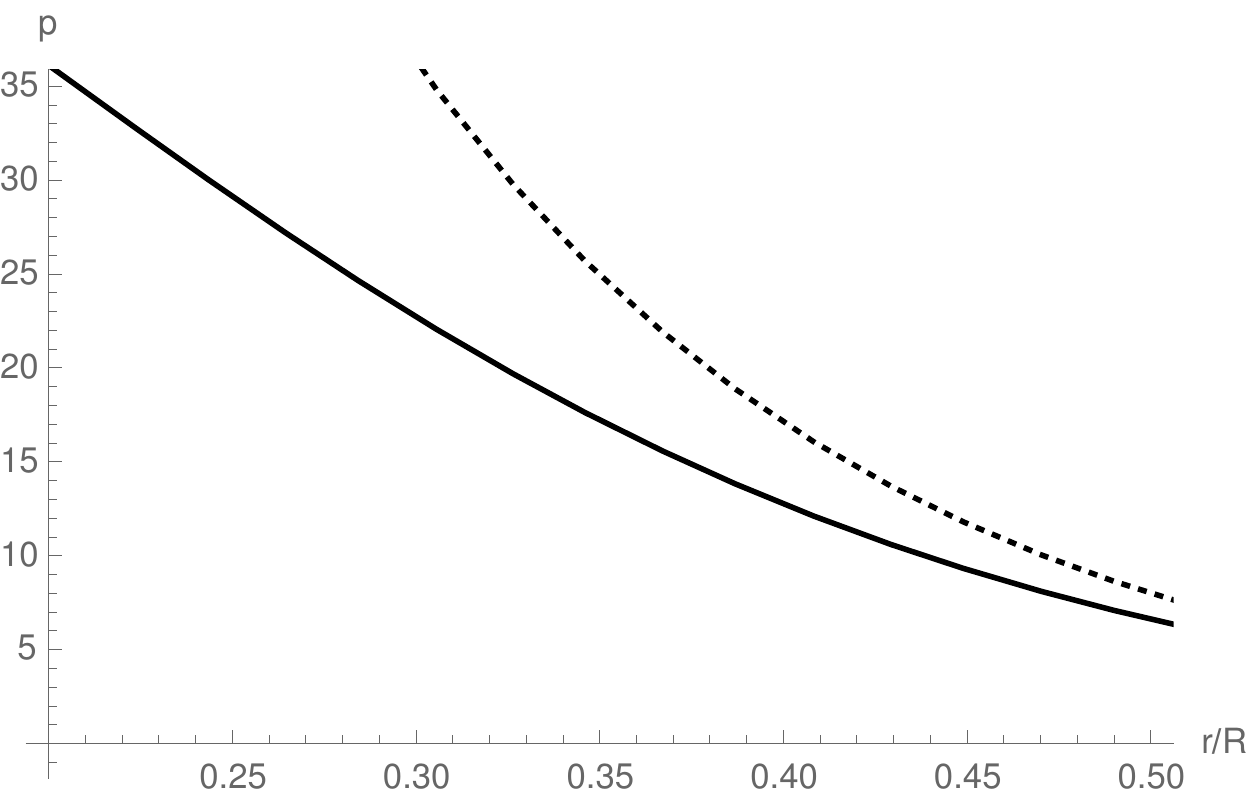}
$\ $
\includegraphics[width=5cm,height=4cm]{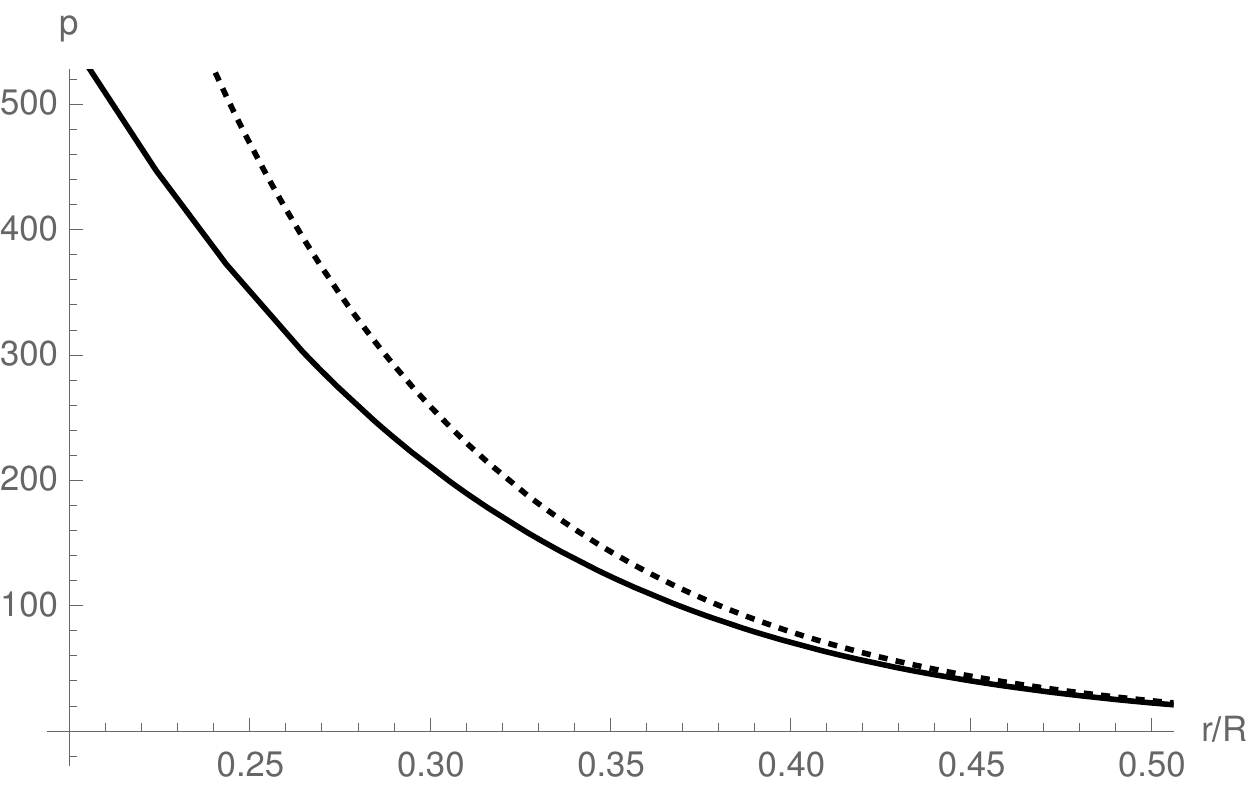}
$\ $
\includegraphics[width=5cm,height=4cm]{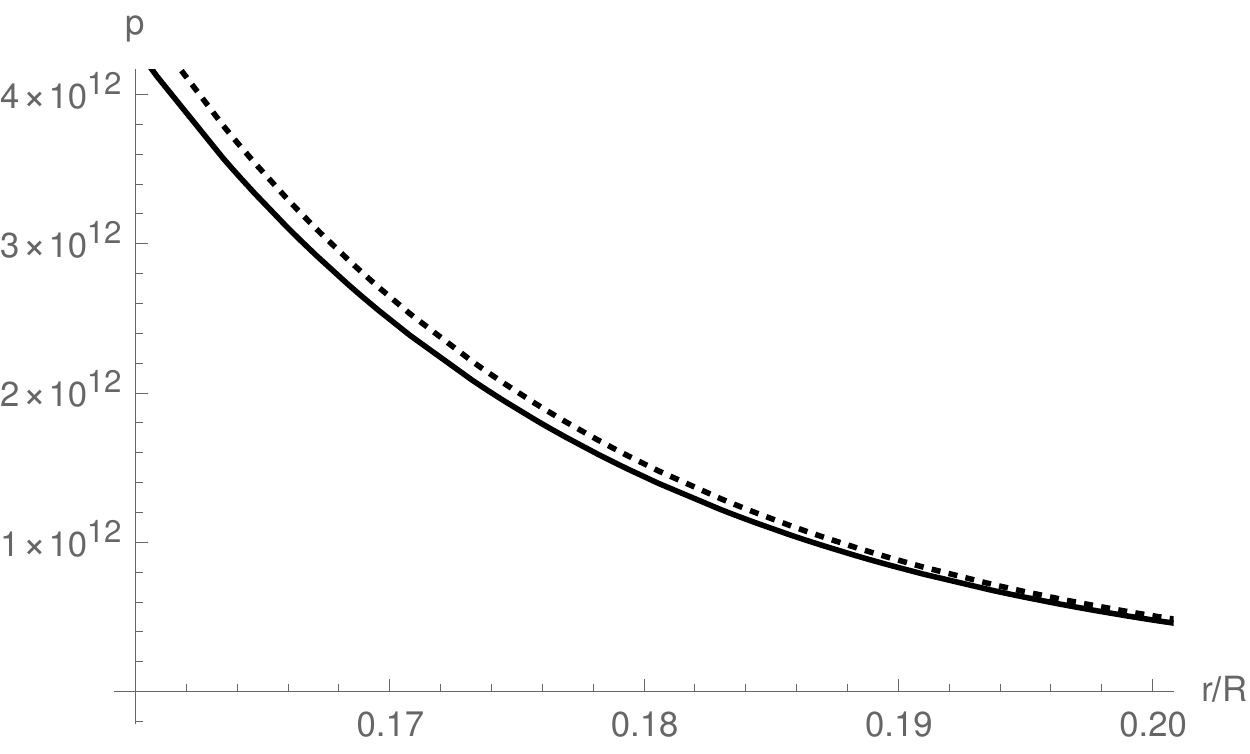}
\caption{Comparison between the approximate pressure~\eqref{pressurebig} (dotted line)
{\em vs\/} solution of Eq.~\eqref{pressB} with $\tilde V=\tilde C\,\psi$ and $\tilde C=(C_+ +C_-)/2$
(solid line) for $\gn\,M /R = 50$ (top left panel),
$\gn\,M / R= 100$ (top central panel),
$\gn\,M / R= 1000$ (top right panel),
and the corresponding close-ups in the bottom panels.
}
\label{pinVSpsimpl}
\end{figure}
\begin{figure}[t]
\centering
\includegraphics[width=5cm,height=4cm]{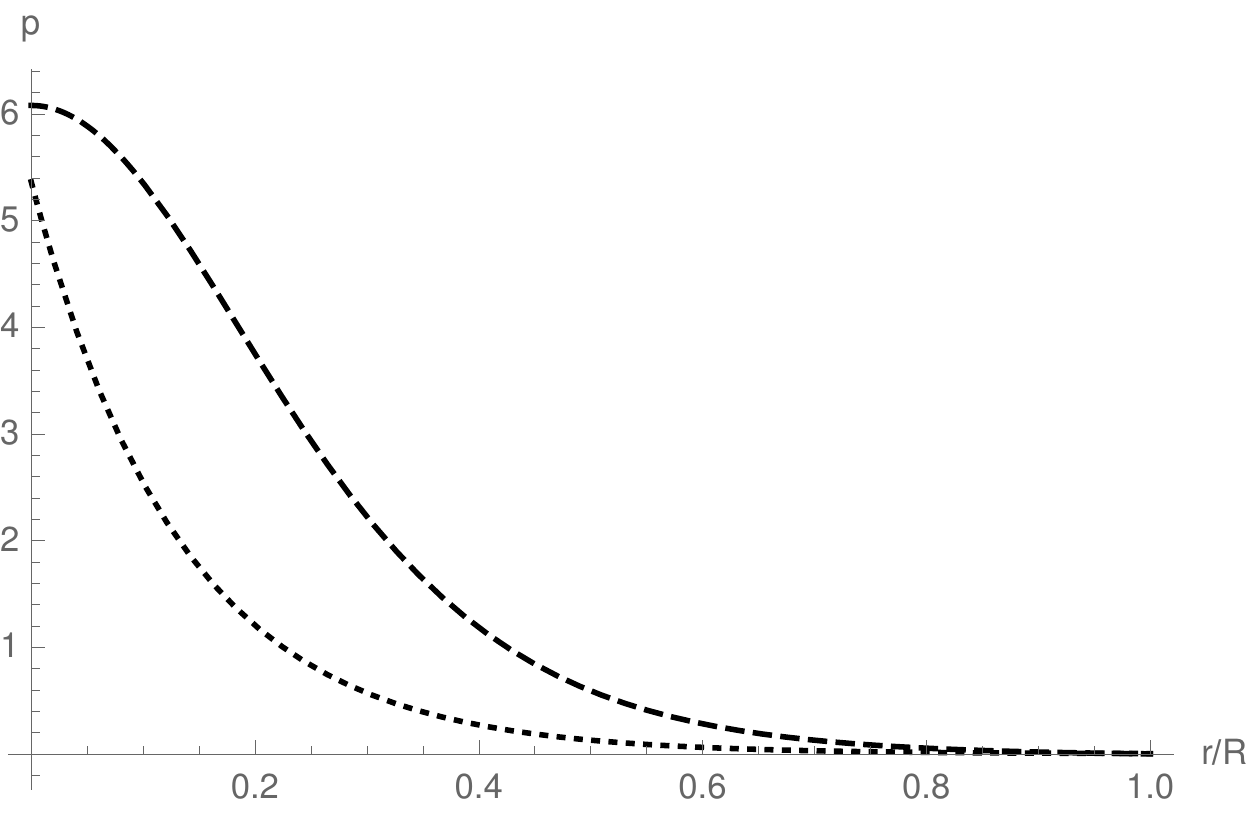}
$\ $
\includegraphics[width=5cm,height=4cm]{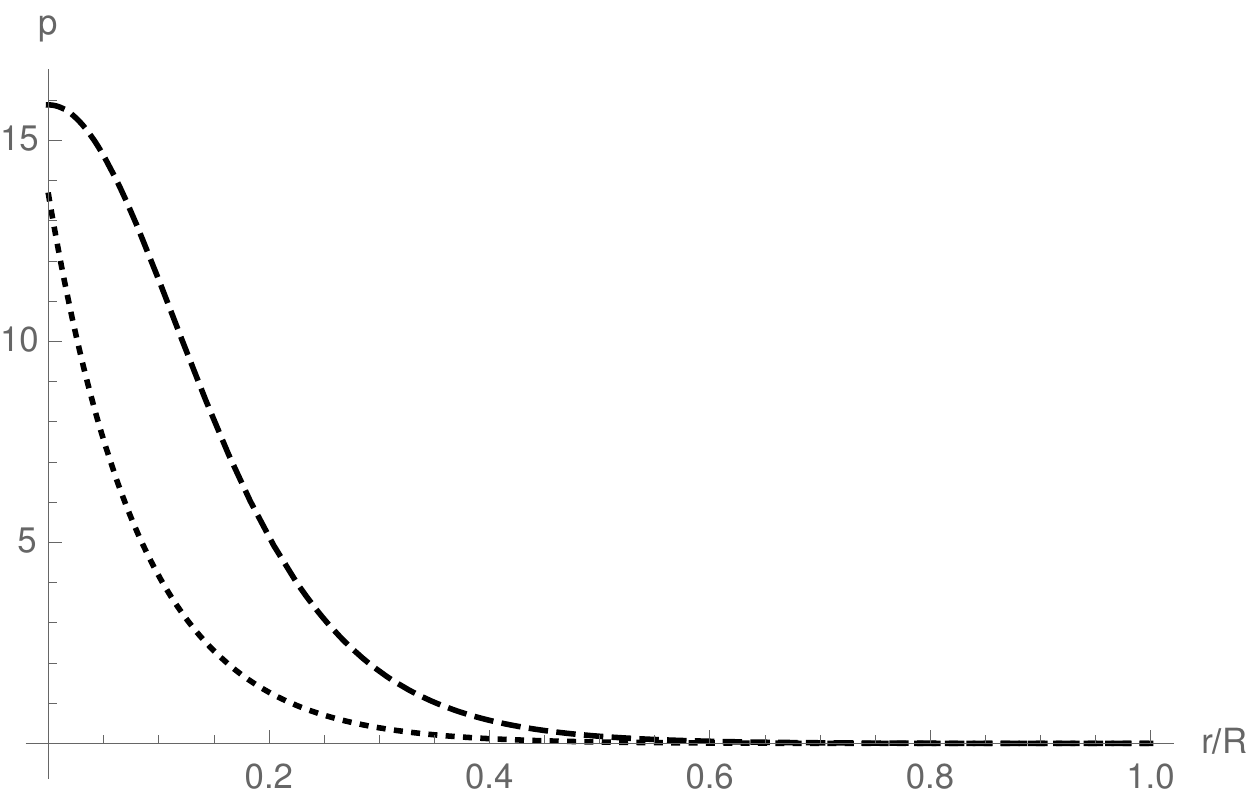}
$\ $
\includegraphics[width=5cm,height=4cm]{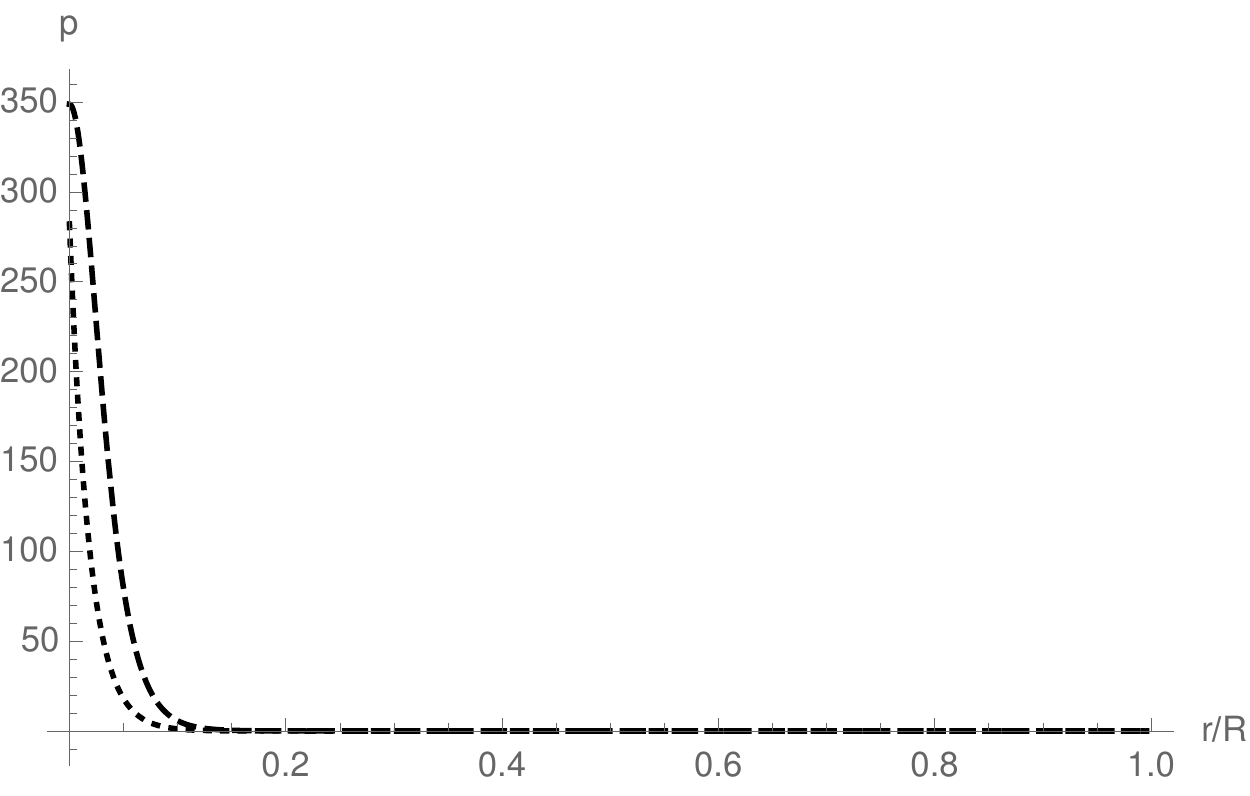}
\caption{Pressure evaluated from $V_-=C_-\,\psi$ (dashed line) {\em vs\/}
pressure evaluated from $\tilde V=\tilde C\,\psi$ (dotted line) 
for $\gn\,M/R =50$ (left panel), $\gn\,M/R = 100$ (center panel) and $\gn\,M/R = 1000$ (right panel).
}
\label{pmin}
\end{figure}
\begin{figure}[t]
\centering
\includegraphics[width=5cm,height=4cm]{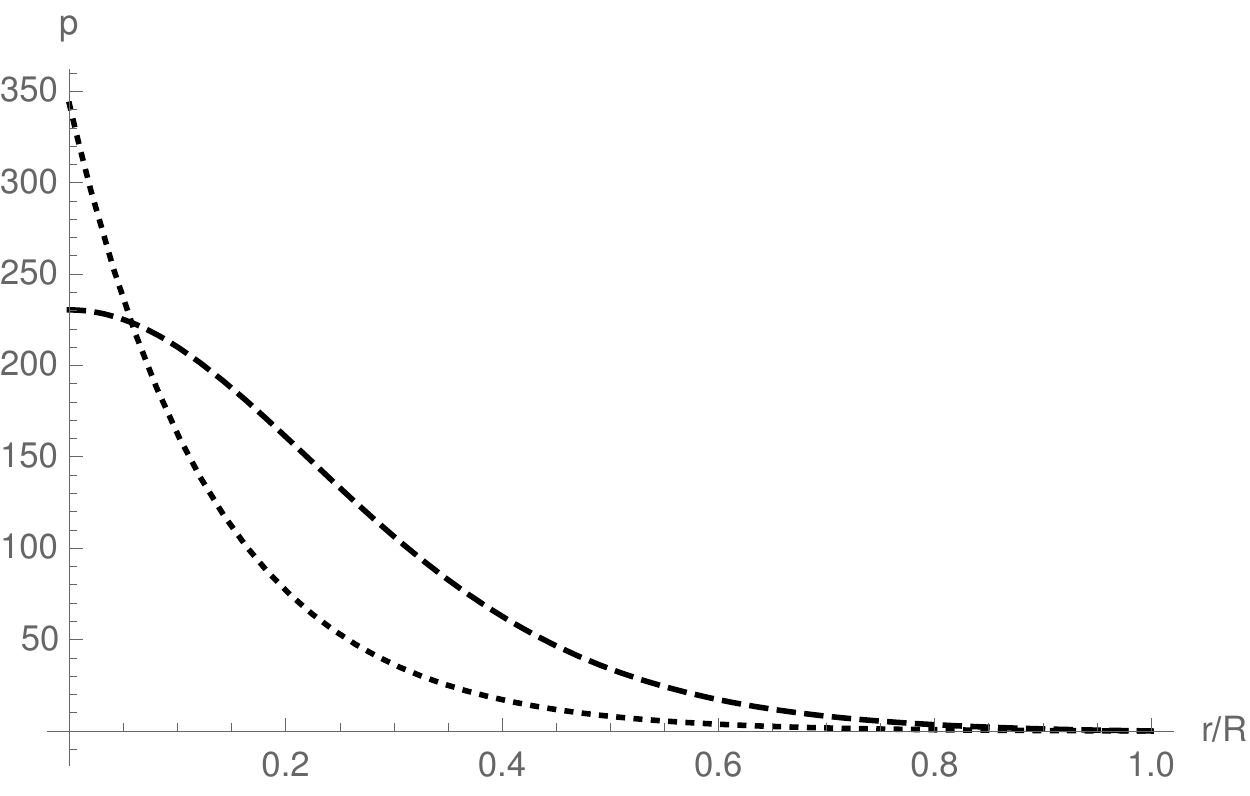}
$\ $
\includegraphics[width=5cm,height=4cm]{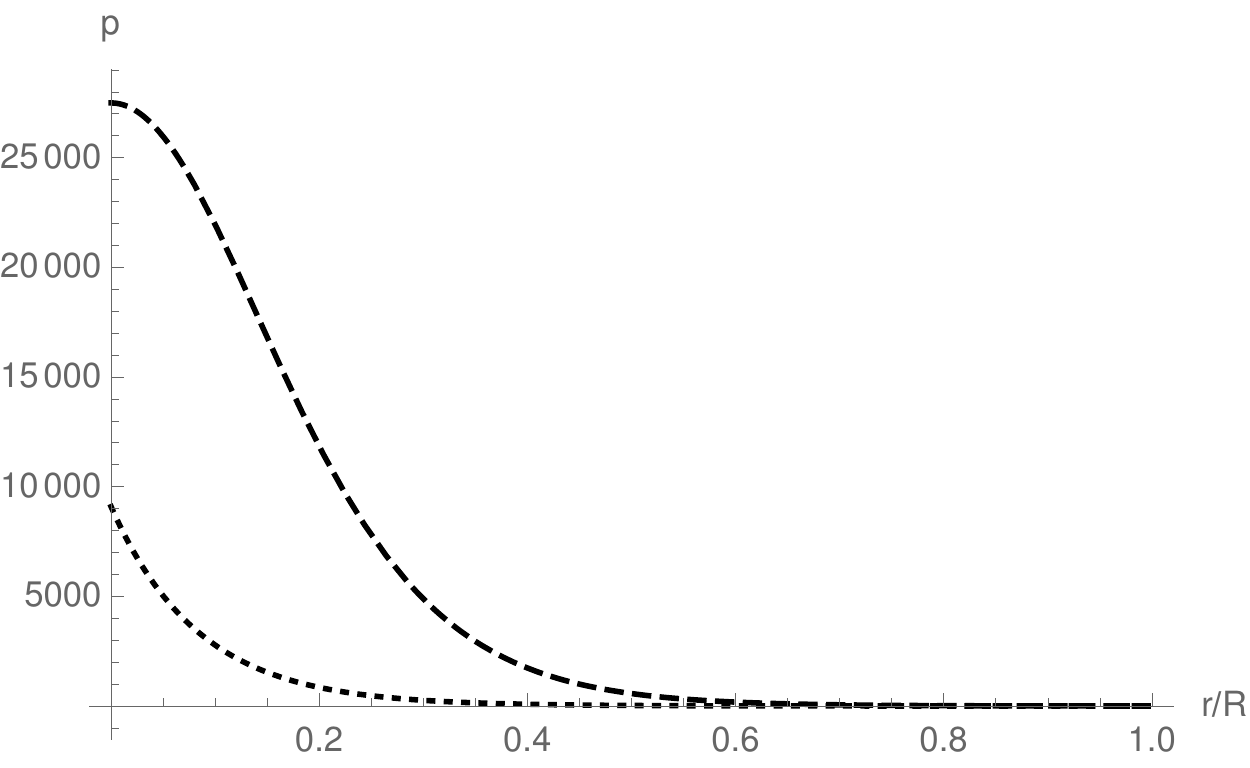}
$\ $
\includegraphics[width=5cm,height=4cm]{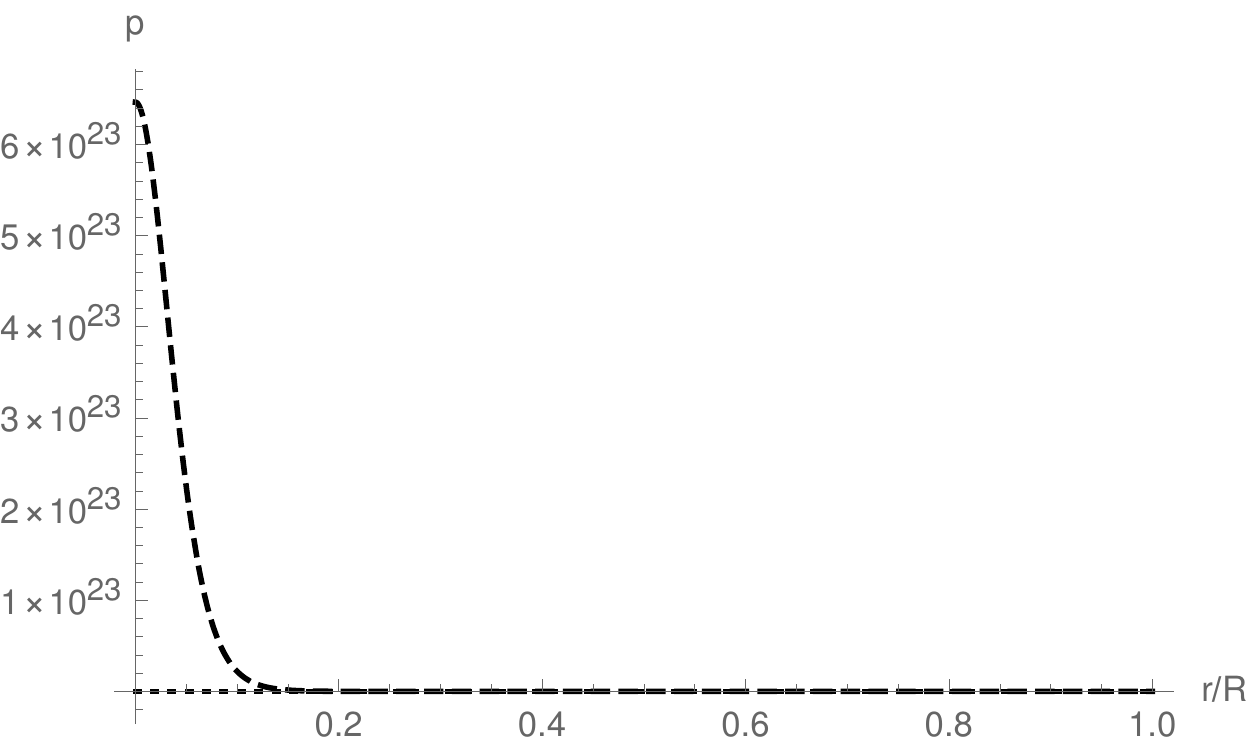}
\caption{Pressure evaluated from $V_+=C_+\,\psi$ (dashed line) {\em vs\/} 
pressure evaluated from $\tilde V=\tilde C\,\psi$ with $\tilde C=(C_+ +C_-)/2$ (dotted line)  
for $\gn\,M/R =50$ (left panel), $\gn\,M/R = 100$ (center panel) and $\gn\,M/R = 1000$ (right panel).
}
\label{pmax}
\end{figure}

The linear approximation is not very useful when it comes to evaluate the maximum 
value of the pressure, which we expect to occur in the origin at $r=0$, precisely where
this approximation must fail.
We therefore consider again the approximation  $\tilde V=\tilde C\,\psi$, which replaced into
Eq.~\eqref{pressure} gives rise to the pressure shown in Fig.~\ref{pressurebig}.
Since the full expression is very cumbersome, we just show the leading order contribution
for large compactness
\be
p
\simeq
\frac{\gn\,M^2\,e^{\frac{1}{2}\left(\frac{\gn\,M}{\sqrt{6}\,R}\right)^{2/3}\left(3- \frac{5}{\tilde C}\right)}}
{2\,\pi\,\tilde C^2\,R^4\,(6\,\gn\,M/R)^{2/3}}
\left[e^{\left(\frac{\gn\,M}{\sqrt{6}\,R}\right)^{2/3}\left(1- \frac{r}{R}\right)}-1\right]
\ ,
\label{pressB}
\ee
which yields
\be
p(0)
\simeq
\frac{\gn\,M^2\,e^{\frac{5}{2}\left(\frac{\tilde C-1}{\tilde C}\right)\left(\frac{\gn\,M}{\sqrt{6}\,R}\right)^{2/3}}}
{2\,\pi\,\tilde C^2\,R^4\,(6\,\gn\,M/R)^{2/3}}
\ ,
\ee
where we find that $\tilde C>1$ for $\gn\,M/R\gg 1$.
It is clear from this expression and Fig.~\ref{pressurebig} how rapidly 
the pressure grows near the origin when the compactness increases,
but still remaining finite and regular everywhere even for very large compactness. 
In Fig.~\ref{pinVSpsimpl} we can see the comparison of the above approximate 
expression with the graphs shown in Fig.~\ref{pressurebig}. 
Of course the biggest the compactness the more rapidly the approximation~\eqref{pressB}
approaches the results of Fig.~\ref{pressurebig}.
In Figs.~\ref{pmin} and~\ref{pmax} we instead plot the comparison between the 
approximation~\eqref{pressB} with $\tilde C=(C_+ +C_-)/2$ and the pressure
evaluated from Eq.~\eqref{pressure} and $V_\pm=C_\pm\,\psi$. 
The values of $C_-$ and $C_+$ are the same as in Figs.~\ref{Vpm} and~\ref{Vpm1} 
for the corresponding compactness.
\section{Horizon and gravitational energy}
\setcounter{equation}{0}
\label{S:horizon}
The approach we used so far completely neglects any geometrical aspect of gravity.
In particular, it is well known that collapsing matter is responsible for the emergence 
of black hole geometries, providing us with the associated Schwarzschild radius~\eqref{def:Rh}.
In general relativity, this marks the boundary between sources which we consider as stars
($R\gg\Rh$) and black holes ($R\lesssim\Rh$).
Moreover, if the pressure is isotropic, stars must have a radius $R>(9/8)\,\Rh$,
known as the Buchdahl limit~\cite{buchdahl}, otherwise the necessary pressure diverges.
\par
\begin{figure}[t]
\centering
\includegraphics[width=8cm]{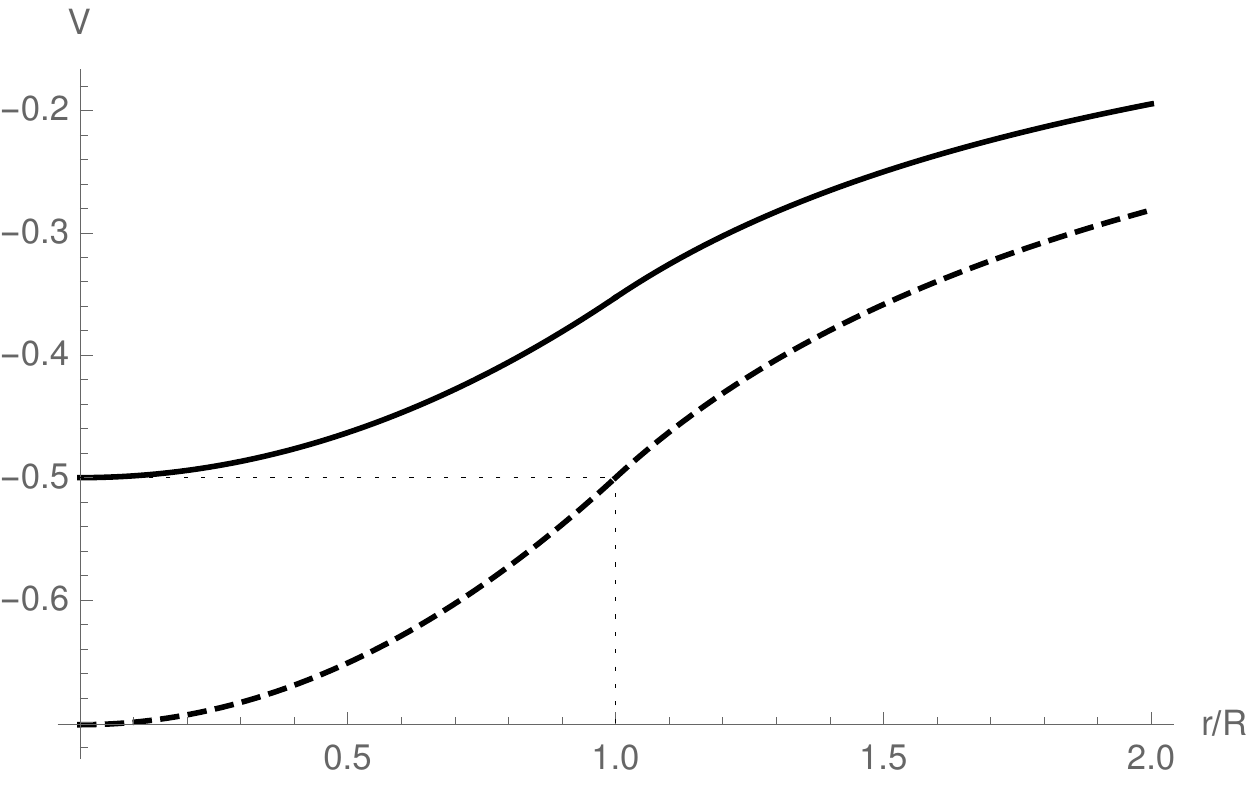}
\caption{Potentials corresponding to $\rh=0$ (solid line) and $\rh=R$ (dashed line).}
\label{horizons}
\end{figure}

We found that the pressure is always finite in our bootstrapped picture, hence there is no analogue
of the Buchdahl limit.
This means that the source can have arbitrarily large compactness, including $R<\Rh$.
Lacking precise geometrical quantities, we will follow a Newtonian argument and define the
horizon as the value $\rh$ of the radius at which the escape velocity of test particles equals the speed 
of light, namely
\be\label{Vrh}
2\,V(r_{\mathrm{H}})= -1
\ ,
\ee
as in Ref.~\cite{BootN}.
Of course, when the source is diluted no horizon should exist and the above definition correctly
reproduces this expectation, since that condition is never fulfilled for small compactness 
(see Figs.~\ref{VinVSVzoomSmall} and~\ref{VinVSVzoomInter}).
In fact, we can find a limiting lower value for the compactness at which Eq.~\eqref{Vrh} has a solution,
by requiring
\be
2\, V_{\rm in}(\rh=0) 
=
-1
\ ,
\ee
which gives  $\gn\,M/R \simeq 0.46$ if we use $V(0) = V_0$ from Eq.~\eqref{V0si}.
Upon increasing the compactness, the horizon radius $\rh$ will increase and eventually approach the 
radius $R$ of the matter source, which occurs when
\be
2\, V_{\rm in}(\rh=R)
=
2\, V_{\rm out}(R)
=
-1
\ ,
\ee
where $V_{\rm out}(R) = V_R$ is given by the exact expression in Eq.~\eqref{VR}.
This yields the compactness $\gn\,M/R \simeq 0.69$ and $\rh \simeq R \simeq 1.43\, \gn\,M$.
For even larger values of the compactness, the horizon radius will always appear in the
outer potential~\eqref{sol0} and therefore remain fixed at this value in terms of $M$.
We can summarise the situation as follows
\be \label{horizon}
\left\{
\begin{array}{ll}
{\rm no\ horizon}
&
{\rm for}
\ \
\gn\,M/R
\lesssim
0.46
\\
\\
0< r_{\mathrm{H}}\leq R
\simeq
1.4\,\gn\,M
&
{\rm for}
\ \
0.46 \lesssim \gn\,M/R \leq 0.69
\\
\\
r_{\mathrm{H}}
\simeq
1.4\,\gn\,M
&
{\rm for}
\ \
\gn\,M/R \gtrsim 0.69
\ .
\end{array}
\right.
\ee
The above values of the compactness further correspond to proper masses
\be
\frac{M_0}{M}
\simeq
\left\{
\begin{array}{ll}
0.56
&
{\rm for}
\ \
\gn\,M/R
\simeq
0.46
\\
\\
0.47
&
{\rm for}
\ \
\gn\,M/R \simeq 0.69
\ ,
\end{array}
\right.
\ee
so that, when the horizon is precisely at the surface of the source, we have 
\be
\rh
\simeq
1.4\,\gn\,M
\simeq
3\,\gn\,M_0
\ .
\ee
It is also important to remark that the horizon $\rh$ lies inside the source for a relatively narrow
range of the compactness (see Fig.~\ref{horizons} for the corresponding potentials).
\par
We can next estimate the gravitational potential energy $U_{\rm G}$ from the effective
Hamiltonian~\eqref{HamV} (with $q_{\Phi}=1$).
For calculation and conceptual purposes, it is convenient to separate $U_{\rm G}$ into three different parts:
the ``baryon-graviton'' contribution, for which the radial integral has only support inside
the matter source, given by
\be
\label{UBG}
U_{\rm BG}
&\!\!=\!\!&
4\,\pi
\int_0^\infty
r^2\,\d r
\left(\rho+p\right) V\left(1-2\,V\right)
=
\frac{3\, M_0}{R^3}\int_0^R
r^2\,\d r\,
e^{V_{R}-V_{\rm in}} V_{\rm in}\left(1-2\,V_{\rm in}\right) 
\ ,
\ee
where we employed Eq.~\eqref{pressure};
the ``graviton-graviton'' contribution due to the potential self-interaction inside the source
\be
\label{UGGin}
U_{\rm GG}^{\rm in} 
&\!\!=\!\!&
\frac{1}{2\,\gn}
\int_0^R
r^2\,\d r\,
\left(V_{\rm in}'\right)^2
\left(1-4\,V_{\rm in}\right)
\ee
and outside the source 
\be
\label{UGGout}
U_{\rm GG}^{\rm out} 
&\!\!=\!\!&
\frac{1}{2\,\gn}
\int_R^\infty
r^2\,\d r\,
\left(V_{\rm out}'\right)^2
\left(1-4\,V_{\rm out}\right)
\ ,
\ee
While the contribution from the outside is exactly given by
\be
U_{\rm GG}^{\rm out} 
= 
\frac{\gn\,M^2}{2\,R}
\ .
\ee
the inner contributions $U_{\rm BG}$ and $U_{\rm GG}^{\rm in}$ can only be evaluated within the approximations
for the potential employed in the previous sections.
\par
The energy contributions for objects of low compactness $\gn\,M/R\ll 1$ can be evaluated
straightforwardly.
Starting from the approximate expression in~\eqref{M0small} and~\eqref{Vins1} the total energy
is calculated to be 
\be
U_{\rm G} 
=
U_{\rm BG}
+
U_{\rm GG}^{\rm in} 
+
U_{\rm GG}^{\rm out} 
\simeq
-\frac{3\, \gn\, M^2}{5\,R}
+
\frac{9\, \gn^2\, M^3}{7\,R^2}\ ,
\label{UGlow}
\ee
where we immediately notice the usual newtonian term at the lowest order.
\par
\begin{figure}[t]
\centering
\includegraphics[width=5cm,height=4cm]{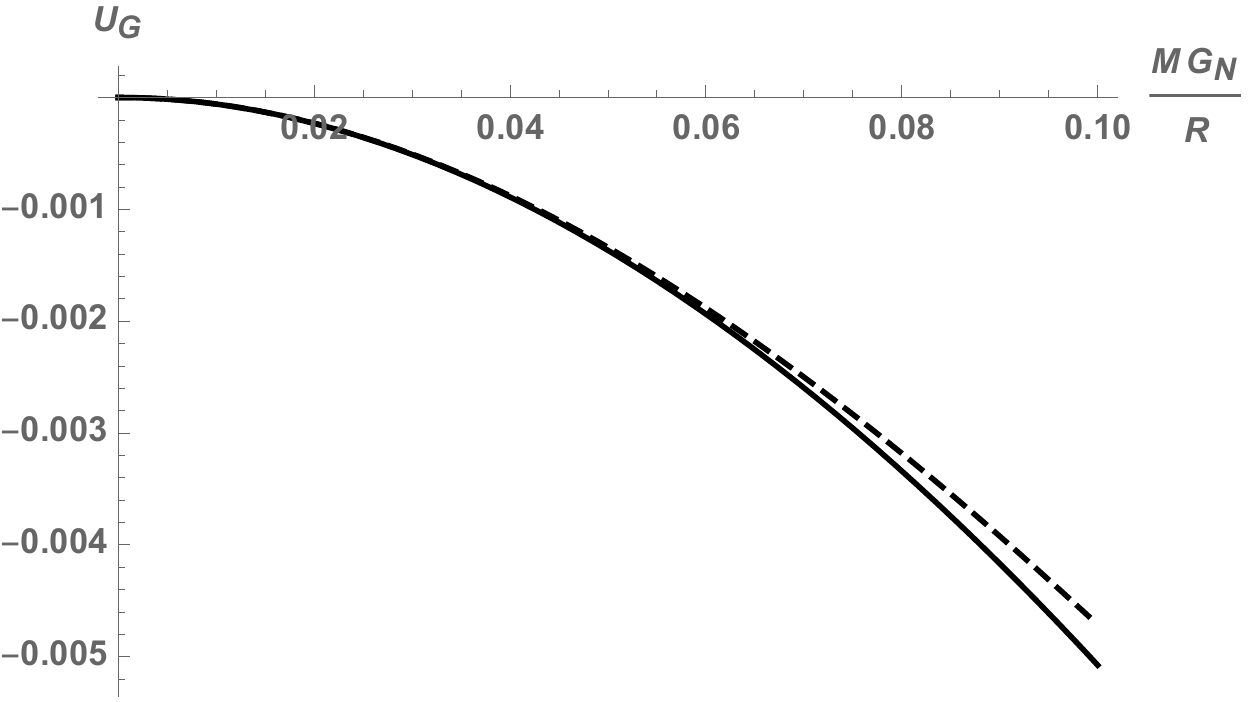}
$\ $
\includegraphics[width=5cm,height=4cm]{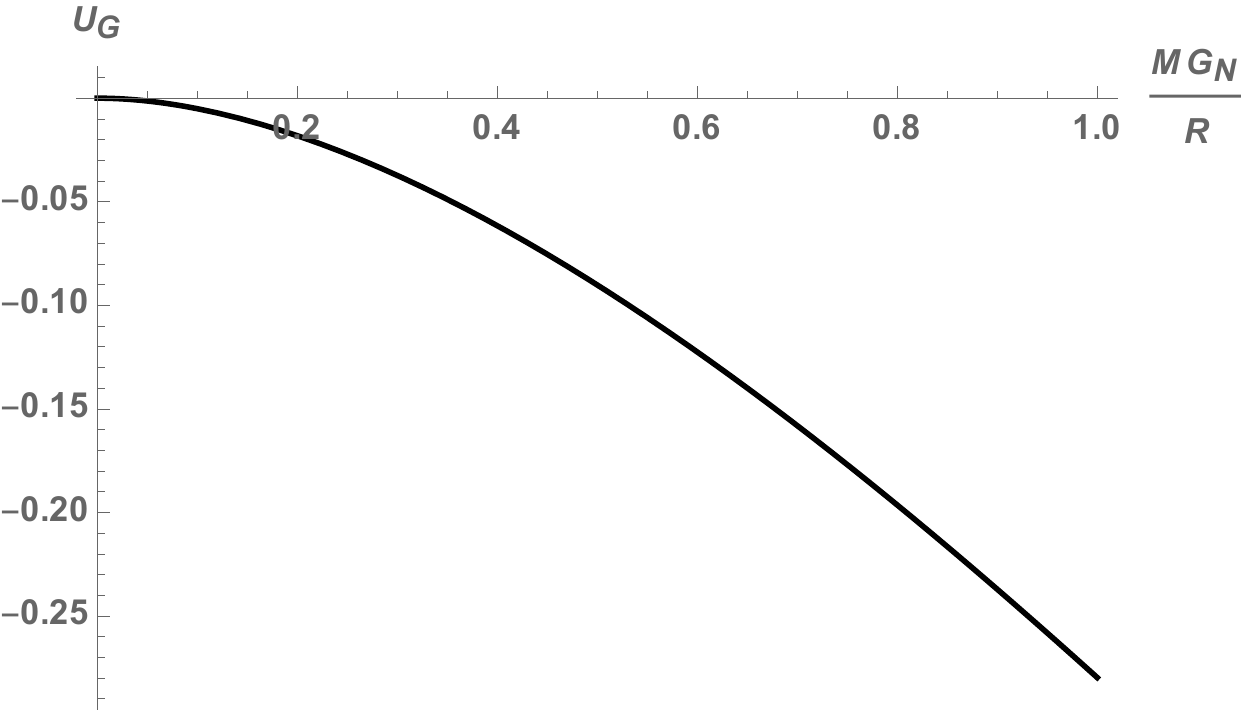}
$\ $
\includegraphics[width=5.4cm,height=4cm]{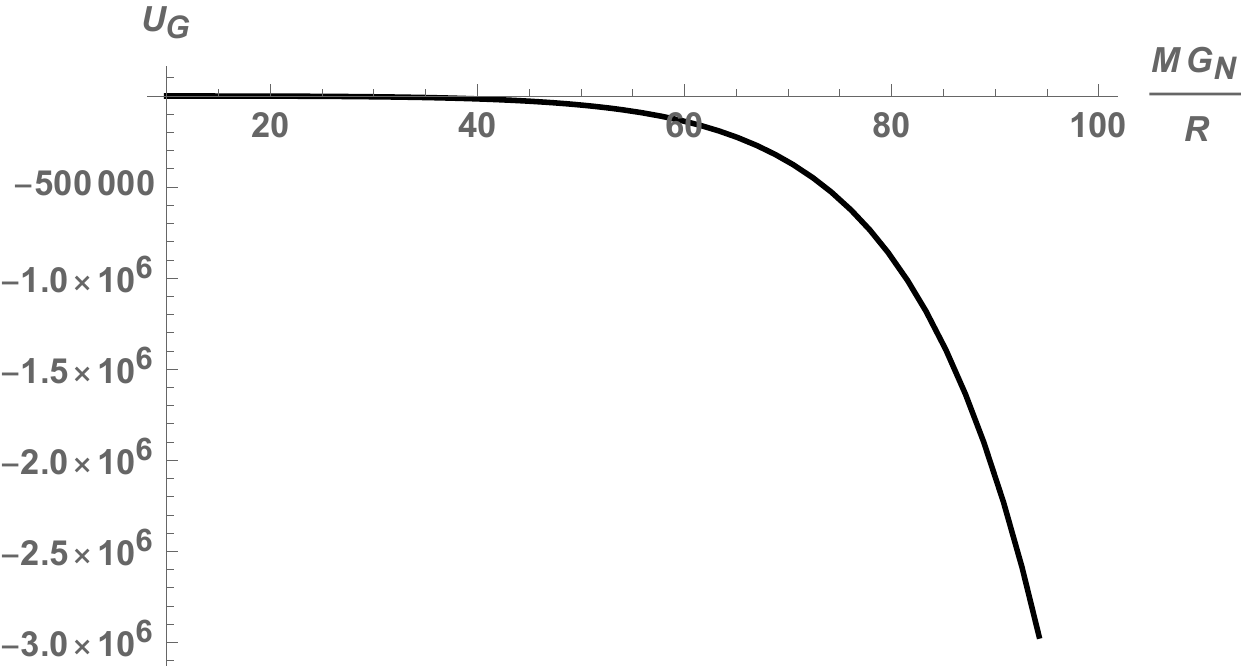}
\caption{Total gravitational potential energy $U_{\rm G}$.
Left panel: $U_{\rm G}$ in the low compactness regime
from the analytic approximations valid in the low and intermediate regime (continuous line)
{\em vs\/} $U_{\rm G}$ from Eq.~\eqref{UGlow} (dashed line).
Center panel: $U_{\rm G}$ in the low and intermediate compactness regime. 
Right panel: $U_{\rm G}$ in the high compactness regime. 
}
\label{UG}
\end{figure}

One can also calculate the three components of the gravitational potential energy in the regime
of  intermediate compactness $\gn\,M/R\sim 1$, but the explicit expressions would be too cumbersome
to display.
Instead, the left panel of Fig.~\ref{UG} shows a comparison in the regime of low compactness between
the above expression for $U_{\rm G}$ and the one obtained starting from the analytic
approximations from Eqs.~\eqref{M0} and \eqref{Vintermediate}, which are valid both for sources
of low and intermediate compactness.
It can be seen that the two approximations lead to similar results for objects that have low compactness.
The center panel also shows the behaviour of $U_{\rm G}$ for objects of intermediate compactness.
As expected, the gravitational potential energy becomes more and more negative as the density of the
source increases.
\par
We conclude with the high compactness regime, in which the increase in modulus of the
negative gravitational potential energy is even more dramatic, as shown in the right panel
of Fig.~\ref{UG}. 
To make things easier, we are going to evaluate the contributions~\eqref{UBG} 
and~\eqref{UGGin} in the limit $\gn\,M/R\gg 1$, with the help of the linear
approximation~\eqref{eVlin} and~\eqref{M0M}.
The leading order in $\gn\,M/R\gg 1$ then reads
\be
U_{\rm BG} 
\simeq
-\frac{125\,R}{3\,\gn}
e^{\left(\frac{\gn\,M}{\sqrt{6}\,R}\right)^{2/3}}
\label{UbgL}
\ee
and 
\be
U_{\rm GG}^{\rm in} 
\simeq
\frac{5\,\gn\,M^2}{36\,R}\ .
\ee
One expects that this negative and large potential energy $U_{\rm G}$ is counterbalanced by
the positive energy~\eqref{Ubp} associated with the pressure~\eqref{pressB} inside the matter source.
\par
Of course, the total energy of the system should still be given by the ADM-like mass $M$,
which must therefore equal the sum of the matter proper mass $M_0$ and the energy
associated with the pressure (see Appendix~\ref{A:energy} for more details about the energy
balance). 
\section{Conclusions and (quantum) outlook}
\setcounter{equation}{0}
\label{S:conc}
In this work we have fully developed a bootstrapped model of isotropic and homogeneous stars,
in which the pressure and density both contribute to the potential describing the gravitational pull  
on test particles.
No equivalent of the Buchdahl limit was found, and the matter source can therefore be kept in
equilibrium by a sufficiently large (and finite) pressure for any (finite) value of the compactness $\gn\,M/R$.
When the compactness of the source exceeds a value of order $0.46$, a horizon appears
inside the source and its radius $\rh\simeq R$ for $\gn\,M/R\simeq 0.69$.
For larger values of the compactness, the source is entirely inside $\rh$ and we can consider
cases with $\rh\gtrsim R$ as representing bootstrapped Newtonian black holes. 
\par
When the matter is collapsed further inside the horizon, that is for larger compactness
such that $\rh\gg R$, the gravitational potential energy grows even more negative, and a correspondingly
very large pressure $p$ is required in order to support the matter core.
In fact, if we assume that black holes have regular inner cores of finite proper mass $M_0$ and
thickness $R$, from Eq.~\eqref{M0M} we obtain
\be
\frac{\gn\,M_0}{R}
\sim
\left(\frac{\gn\,M}{R}\right)^{2/3}
\ ,
\ee
so that $M_0/M\sim (R/\gn\,M)^{1/3}$ for  $\gn\,M/R\gg 1$.
This means that most of the matter energy must be accounted for by the interactions that give rise
to the pressure in very compact sources.
Such a huge pressure $p\gg\rho$ violates the dominant energy condition~\cite{Wald} and
could only be of purely quantum nature, thus requiring a 
quantum description of the matter in the source.
\par
Correspondingly, the regular potential we obtained in the present work should be viewed as the
mean field description of the quantum state of the (off-shell) gravitons in a (regular~\footnote{For
a review, see Ref.~\cite{nicolini}})
black hole when $R\lesssim \rh$.
It will be therefore a natural development to investigate the quantum features of this potential,
as it affects both the quantum state of matter inside the black hole (or falling into it) and
the dynamics of the gravitons themselves.
Eventually, one would also like to identify the fully quantum state that generates this potential,
like it was done for the Newtonian potential in Refs.~\cite{Casadio:2017cdv,Mueck},
or in Refs.~\cite{Casadio:2014vja,QHBH}.
Finally, we would like to remark that, although we found that $M\gg M_0$ for very large compactness,
and one could thus infer that matter become almost irrelevant inside a black hole~\cite{DvaliGomez},
the above picture inherently requires the presence of matter,
whose role in black hole physics we believe needs more
investigations~\cite{Foit:2015wqa,kuhnelBaryons,Dvali:2016mur,Kuhnel}.
\section*{Acknowledgments}
R.C.~and M.L.~are partially supported by the INFN grant FLAG.
The work of R.C.~has also been carried out in the framework
of activities of the National Group of Mathematical Physics (GNFM, INdAM)
and COST action {\em Cantata\/}. 
O.M.~is supported by the grant Laplas~VI of the Romanian National Authority for Scientific
Research. 
\appendix
\section{Gravitational current}
\setcounter{equation}{0}
\label{A:J_V}
We present here an alternative derivation of the gravitational current leading to the same Lagrangian~\eqref{LagrV}
of Section~\ref{S:action}.
The starting point will now be the Newtonian energy evaluated on-shell inside a sphere of radius $r$, that is
\be
U_{\rm N}(r)
&\!\!=\!\!&
2\,\pi
\int_0^r 
{\bar r}^2\,\d {\bar r}
\,\rho(\bar r)\, V(\bar r)
\nonumber
\\
&\!\!=\!\!&
\frac{1}{2\,\gn}\,
\int_0^r 
{\bar r}^2 \,\d {\bar r}\,
V(\bar r)\,\triangle V(\bar r)
\ ,
\ee
in which we do not perform any integration by parts.
We can then define a current $\tilde J_V$ proportional to the energy density by deriving $U_{\rm N}(r)$
with respect to the volume $\mathcal{V}$, which yields
\be
\tilde J_V
\simeq
2\,\frac{\d U_{\rm N}}{\d \mathcal{V}} 
=
\frac{V(r)\,\triangle V(r)}{4\,\pi\,\gn}
\ .
\label{J'V}
\ee
One can immediately notice that we chose to have a different numerical factor in front of $\tilde J_V$ from the one
in $J_V$ of Eq.~\eqref{JV} in order to keep the same coupling parameter $\tilde q_V=q_V$.
It is now easy to see that by adding all other sources described in Section~\ref{S:action} together with~\eqref{J'V},
we end up with the same Lagrangian~\eqref{LagrV}, 
\be
\tilde L[V]
=
L_{\rm N}[V]
-4\,\pi
\int_0^\infty
r^2\,\d r
\left[
q_V\,\tilde J_V\,V
+
q_{\rm B}\,J_{\rm B}\,V
+
q_\rho\,J_\rho\left(\rho+p\right)
\right]
=
L[V]
\ ,
\ee
where we discarded vanishing boundary terms.
In fact, we have
\be
\int_0^\infty
r^2\,\d r\,J_V\,V
=
2\,\int_0^\infty
r^2\,\d r\,\tilde J_V\,V
+
\left[r^2\,V^2\,V'\right]_{r=0}^{r\to\infty}
\ ,
\ee
and the second term in the right hand side vanishes because of the boundary conditions at $r\to\infty$
and Eq.~\eqref{b0} at $r=0$.
\section{Newtonian solution}
\setcounter{equation}{0}
\label{A:newton}
We recall that the Newtonian solution of the Poisson equation~\eqref{EOMn} with a homogeneous
source of mass $M_0$ and radius $R$, 
\be
\triangle V_{\rm N}
=
\frac{3\, \gn\,M_0}{R^3}\,\Theta(R-r)
\ ,
\label{tEOMVn}
\ee
is given by
\be
V_{\rm N}
=
\left\{
\begin{array}{lrl}
\strut\displaystyle\frac{\gn\, M_0}{2\,R^3}\left(r^2 - 3\,R^2\right)
&
{\rm for}
&
0\le  r<R 
\\
\\
-\strut\displaystyle\frac{\gn\,M_0}{r}
&
{\rm for}
&
r>R
\ ,
\end{array}
\right.
\label{Vn}
\ee
which is continuous, with continuous first derivative across $r=R$.
We also remark that there is one and the same mass parameter $M_0=M$ in the
interior and exterior part of the potential.
\section{Comparison method}
\setcounter{equation}{0}
\label{A:comparison}
We have shown in Section~\ref{SS:innerV} that a solution to Eq.~\eqref{eomV} satisfying Eq.~\eqref{vminvmax}
exists by employing comparison functions~\cite{BVPexistence, BVPexistence1} and we recall the
fundamentals of this method here for the sake of convenience.
\par
Let us consider an equation of the form
\be
\label{BVP}
u''(r) = F(r, u(r), u'(r))
\ ,
\ee
where $F$ is a real function of its arguments, $r$ varies in the finite interval $[r_1,r_2]$ and a prime denotes
the derivative with respect to $r$.
We want to find a solution which further satisfies the general boundary conditions
\be
\label{BV1}
&&
a_1\, u(r_1) - a_2\, u'(r_1)
=
A_0
\ , 
\\
\label{BV2}
&&
b_1\, u(r_2) + b_2\, u'(r_2)
=
B_0
\ ,
\ee
with $A_0$, $B_0$, $a_1$, $b_1$ real numbers and $a_2$, $b_2$ non negative real numbers satisfying
$a_1^2 + a_2^2 > 0$ and $b_1^2 + b_2^2 > 0$.     
The theorems in Refs.~\cite{BVPexistence, BVPexistence1} guarantee that such a solution $u \in C^2 ([r_1,r_2])$ 
exists under the following three conditions:
\begin{enumerate}
\item
we can find a lower bounding function
\be
&&
u_-''(r)
\geq
F(r, u_-(r), u_-'(r))
\\
&&
a_1\, u_-(r_1) - a_2\,u_-'(r_1)
\leq
A_0
\\ 
&&
b_1\, u_-(r_2) + b_2\,u_-'(r_2)
\leq
B_0
\ ,
\ee
and an upper bounding function
\be
&&
u_+''(r)
\leq
F(r, u_+(r), u_+'(r))
\\
&&
a_1\, u_+(r_1) - a_2\,u_+'(r_1)
\geq
A_0
 \\ 
&&
b_1\,u_+(r_2) + b_2\,u_+'(r_2)
\geq
B_0
\ ;
\ee
\item
the function $F$ is continuous on the domain
$D = \{(r, u, u')\in [r_1,r_2]\times \mathbb{R}^2\, |\, u_-\leq u \leq u_+\}$;
\item
the function $F$ satisfies a {\em Nagumo condition\/}:
there exists a continuous and positive function $\phi$ such that 
\be
\int_0^{\infty}
\frac{s\, \d s}{\phi(s)}
=
\infty
\ee
and, $\forall (t, u, u')\in D$, 
\be 
|F(r, u(r), u'(r))|
\leq
\phi(|u'|)
\ .
\ee
\end{enumerate}
Moreover, the solution $u$ will satisfy 
\be
u_- (t)\leq u(t) \leq u_+ (t)
\ .
\ee
\par
We can now apply the above general theorem to our problem inside the source, for which
$r_1=0$ and $r_2=R$.
We first rewrite Eq.~\eqref{eomV} as 
\be \nonumber
V''
&\!\!=\!\!&
\frac{3\,\gn\,M_0}{R^3}\,e^{V_R-V}
+
\frac{2\left(V'\right)^2}
{1-4\,V}
-
\frac{2\, V'}{r}
\\
\label{F}
&\!\!\equiv\!\!&
F(r, V, V')\ ,
\ee
and recall the boundary conditions~\eqref{b0} and~\eqref{bR}, that is
\be
&&
V'(0) = 0
\label{b00}
\\
&&
V(R) = V_{R}
\ .
\ee
We can now verify all the requirements of the theorem, and will do so for the
case of large compactness analysed in Section~\ref{ss:largeC}.
The upper and lower bounding functions are therefore $V_\pm$ given in Eq.~\eqref{VPM}
and the domain
\be
D = \{(r, V, V') \in [0, R]\times\mathbb{R}^2\, |\, V_{-} \leq V \leq V_{+}\,\}
\ .
\ee
Continuity of $F$ on $D$ is easily verified.
In fact, the first term on the right hand side of Eq.~\eqref{F} 
is an exponential of $V$ which is always regular in $D$.
The same is true for the second term considering that $V_\pm < 0$, thus $V<0$ as well. 
The last term could be tricky but the boundary condition~\eqref{b00} require that $V'$ vanishes 
at $r=0$ at least as fast as $r$ [see the expansion around $r = 0$ in Eq.~\eqref{Vs0}]
so that this is also regular in $D$.
Finally, we can choose
\be
\phi
=
\underset{D}
{\rm max}(F)
\ ,
\ee
which must be finite given that $F$ is continuous in $D$.
\par
All of the hypotheses of the theorem hold and a solution to Eq.~\eqref{eomV} therefore exists
and satisfies Eq.~\eqref{vminvmax}.
By imposing the remaining boundary condition~\eqref{dbR}, one can then obtain a relation 
between $M_0$, which appears in the equation~\eqref{eomV}, and $M$, which appears in the
boundary conditions~\eqref{bR} and \eqref{dbR}, for any given value of $R$. 
\section{Energy balance}
\label{A:energy}
\setcounter{equation}{0}
In Section~\ref{S:horizon}, we only computed the gravitational energy from the
Hamiltonian~\eqref{HamV}. 
The purely baryonic contribution will be given by the proper mass $M_0$ and the
pressure energy contribution found again from the newtonian argument~\eqref{JP},
whereby
\be
U_{\rm B}(R)
=
D(M,R) 
-
4\,\pi
\int_0^R r^2\,\d r\, p(r)
\ .
\ee
In the newtonian regime, the integration constant $D(M,R)$ can be fixed so as
to guarantee that the work done by gravity is equal and opposite to the work done
by the forces responsible for the pressure $p$.
In other words, in that case we find $D(M,R)$ by requiring that the gravitational
force is conservative. 
This will also ensure that the total energy related to the Hamiltonian 
constraint equals the ADM-like mass $M$ of the system, that is
\be
\label{Etot}
E
=
M_0 + U_{\rm G} + U_{\rm B}
=
M
\ .
\ee
Of course, in the Newtonian case Eq.~\eqref{Etot} simply reads $E=M_0\equiv M$,
as shown in Ref.~\cite{BootN}. 
\par
In the bootstrapped picture, gravity is not a linear interaction any more and it is not
at all obvious that it will still be conservative.
A precise energy estimate would therefore require a complete knowledge of the
dynamical process which led to the formation of the equilibrium configuration
of given ADM-like mass $M$ and radius $R$.
Without that knowledge, we can only assume that the total energy of the equilibrium
configuration equals $M$ and fix $D(M,R)$ so that the Hamiltonian constraint~\eqref{Etot}
is satisfied.
\par
With that prescription, we can now evaluate the baryonic contributions.
In the low compactness case, we expand all the terms in Eq.~\eqref{Etot} to order $M^3$,
namely
\be
M_0
\simeq
M 
-
\frac{5\,\gn\,M^2}{2\,R}
+
\frac{81\,\gn^2\,M^3}{8\,R^2}\ ,
\ee
and the pressure energy
\be
U_{\rm B} 
\simeq
D_s(M,R)
-
\frac{\gn\,M^2}{5\,R}
+
\frac{61\,\gn^2\,M^3}{70\,R^2}
\ .
\ee
Eq.~\eqref{Etot} is then satisfied for 
\be
D_s(M,R)
\simeq
-\frac{33\,\gn\,M^2}{10\,R}
+\frac{3439\,\gn^2\,M^3}{280\,R^2}
\ee
so that
\be
U_{\rm B}
\simeq
\frac{31\,\gn\,M^2}{10\,R}
\left(1
-
\frac{6390\,\gn\,M}{1736\,R}
\right)
\ ,
\ee
which is positive only for small compactness, as its approximation requires.
\par
The high compactness regime of course yields quite different results. 
To make things easier, we again look at the limiting case of very high compactness,
where the linear approximation~\eqref{eVlin} holds, and consider the Hamiltonian
constraint~\eqref{Etot} only at leading order in $M$.
The proper mass in Eq.~\eqref{M0M} can be simplified further to give
\be
M_0
\simeq
\frac{5\,M}{9\,(6\,\gn\,M/R)^{1/3}}
\ ,
\ee
while the pressure energy can be written as
\be
U_{\rm B}
\simeq
D_b(M,R)
-
\frac{20\,R^3}{\gn^3\,M^2}
\left(\frac{\gn\,M}{\sqrt{6}\,R} \right)^{2/3}
e^{\left(\frac{\gn\,M}{\sqrt{6}\,R} \right)^{2/3}}
\ .
\ee
Again, we just impose Eq.~\eqref{Etot} and find
\be
D_b(M,R)
&\!\!\simeq\!\!&
M
+
\frac{20\,R^3}{\gn^3\,M^2}
\left(\frac{\gn\,M}{\sqrt{6}\,R} \right)^{2/3}
e^{\left(\frac{\gn\,M}{\sqrt{6}\,R} \right)^{2/3}}
+
\frac{125\,R}{3\,\gn}
e^{\left(\frac{\gn\,M}{\sqrt{6}\,R}\right)^{2/3}}\\
&&
-
\frac{7\,\gn\,M^2}{36\,R}
-
\frac{5\,M}{9\,(6\,\gn\,M/R)^{1/3}}
\ ,
\ee
so that
\be
U_{\rm B}
&\!\!\simeq\!\!&
\frac{125\,R}{3\,\gn}
e^{\left(\frac{\gn\,M}{\sqrt{6}\,R}\right)^{2/3}}
\ ,
\label{Ubp}
\ee
which is positive as it should, and precisely counterbalances Eq.~\eqref{UbgL}.
\end{document}